\documentclass[onecolumn]{aa}

\usepackage{txfonts} 
\usepackage{amsmath, amssymb}
\usepackage{bm}
\usepackage[english=british]{csquotes}
\usepackage[T1]{fontenc}
\usepackage[utf8]{inputenc}
\usepackage{lmodern}
\usepackage{graphicx}
\usepackage{natbib}
 \bibpunct{(}{)}{;}{a}{}{,}
\usepackage{hyperref}
\usepackage{verbatim}
\usepackage{physics} 
\usepackage{siunitx} 
\usepackage{multirow}
\usepackage{tabularx}
\usepackage{tikz-cd}
\usepackage[margin=10pt,font=small,labelfont=bf]{caption}

\usepackage{subcaption}

\makeatletter
\renewcommand*\aa@pageof{page \thepage{} of \pageref*{LastPage}}
\makeatother

\usepackage{hyperref} 
\hypersetup{
colorlinks   = true, 
urlcolor     = blue, 
linkcolor    = blue, 
citecolor   = blue 
}

\defcitealias{Heydenreich2022}{Paper I}

\newcommand{\Omm}{\Omega_\mathrm{m}}
\newcommand{\Map}{M_\mathrm{ap}}
\newcommand{\MapEst}{\hat{M}_\mathrm{ap}}

\newcommand{\MapMapMap}{\expval{\Map^3}}

\newcommand{\MapMapMapEst}{\MapEst^3}

\newcommand{\astroang}[1]{\ang[angle-symbol-over-decimal]{#1}}
\newcommand{\ellvec}{\bm{\ell}}
\newcommand{\dirac}{\delta_\mathrm{D}}
\newcommand{\kronecker}[2]{\delta_{#1,#2}}
\newcommand{\Npix}{N_\mathrm{pix}}

\newcommand{\varthetavec}{\bm{\vartheta}}
\newcommand{\varthetamax}{\vartheta_\mathrm{max}}
\newcommand{\I}{\mathrm{i}}
\newcommand{\E}{\mathrm{e}}
\newcommand{\etavec}{\vec{\eta}}
\newcommand{\alphavec}{\vec{\alpha}}
\newcommand{\qvec}{\vec{q}}
\newcommand{\realspace}{\mathbb{R}}

\newcommand{\svec}{\vec{s}}

\newcommand{\myexpval}[1]{\Big\langle #1 \Big\rangle}
\newcommand{\thetavec}{\vec{\theta}}

\newcommand{\TI}{T_{PPP, 1}}
\newcommand{\TII}{T_{PPP, 2}}
\newcommand{\TIV}{T_{BB}}
\newcommand{\TV}{T_{PT, 1}}
\newcommand{\TVI}{T_{PT, 2}}
\newcommand{\TVII}{T_{P_6}}

\begin{document}

\title{A roadmap to cosmological parameter analysis with third-order shear statistics II: Analytic covariance estimate\footnote{ The modelling code is available at \url{https://github.com/sheydenreich/threepoint/releases}. }}
\titlerunning{Covariance of third-order aperture statistics}

\author{Laila Linke \inst{1},
        Sven Heydenreich \inst{1},
        Pierre A. Burger \inst{1},
        Peter Schneider \inst{1}
      }
\authorrunning{Linke et al.}

\institute{
      Argelander-Institut f\"ur Astronomie, Auf dem H\"ugel 71, 53121 Bonn, Germany 
      \\ \email{llinke@astro.uni-bonn.de}
      }
\date{Version \today; received xxx, accepted yyy} 

\abstract
{Third-order weak lensing statistics are a promising tool for cosmological analyses since they extract cosmological information in the non-Gaussianity of the cosmic large-scale structure. However, such analyses require precise and accurate models for the covariance of the statistics.}{In this second paper of a series on third-order weak lensing statistics, we derive and validate an analytic model for the covariance of the third-order aperture statistics $\MapMapMap$.} {We derive the covariance model from a real-space estimator for $\MapMapMap$, including the Gaussian and non-Gaussian parts. We validate the model by comparing it to estimates from simulated Gaussian random fields (GRFs) and two sets of $N$-body simulations. Finally, we perform mock cosmological analyses with the model covariance and the simulation estimate to compare the resulting parameter constraints.}{We find good agreement between the analytic model and the simulations, both for the GRFs and the $N$-body simulations. The figure of merit in the $S_8$-$\Omega_\mathrm{m}$ plane from our covariance model is within 3\% of the one obtained from the simulated covariances. We also show that our model, which is based on an estimator using convergence maps, can be used to obtain upper and lower bounds for the covariance of an estimator based on three-point shear correlation functions. This second estimator is required to measure $\MapMapMap$ in realistic survey data. In our derivation, we find that the covariance of $\MapMapMap$ cannot be obtained from the bispectrum covariance and that it includes several `finite-field terms' that do not scale with the inverse survey area. 
}{Our covariance model is sufficiently accurate for analysing stage III surveys. Covariances for statistics in Fourier space cannot always be straightforwardly converted into covariance for real-space statistics.}


\maketitle

\section{Introduction}
\label{sec: introduction}
Using weak gravitational lensing of the cosmic large-scale structure (LSS) has become a precise and accurate method to infer cosmological parameters (\citealp{Heymans2021}, \citealp{Hikage2019}, \citealp{Abbott2022}). Weak-lensing analyses have mainly concentrated on second-order statistics, such as shear two-point correlation functions or galaxy-galaxy-lensing. However, since the cosmic LSS is non-Gaussian, second-order statistics do not contain its total information content. To access the additional information, higher-order statistics (HOS) are needed. These HOS depend on cosmological parameters differently than second-order statistics, so they can be used to break degeneracies (\citealt{Takada2004}, \citealt{Kilbinger2005}, \citealt{Kayo2013}) or constrain nuisance parameters such as galaxy bias or intrinsic alignment (\citealt{Pyne2021}, \citealt{Troxel2012}).

Many HOS have been suggested to complement second-order analyses, for example, peak statistics (\citealt{Kacprzak2016}, \citealt{Zuercher2021} \citealt{Martinet2021}), density split statistics (\citealt{Friedrich2018}, \citealt{Gruen2018}, \citealt{Burger2022}), or persistent homology \citep{Heydenreich2021, Heydenreich2022a}. Most of these statistics, though, cannot be modelled from analytical theories and instead require time-consuming realistic $N$-body simulations. Additionally, for many HOS, the measurement requires converting weak lensing shear estimates to mass convergence maps. This conversion becomes complicated in the presence of masks and for finite survey areas, which can lead to biased estimators \citep{Seitz1996}.

In this series of papers, we are considering the third-order aperture mass $\MapMapMap$, which is not affected by these problems. It is linearly related to the third-order shear correlation functions $\Gamma_i$ (\citealt{Jarvis2004}, \citealt{Schneider2005}), which can be directly measured in shear catalogues and converted into $\MapMapMap$. 
This was demonstrated recently by \citet{Secco2022}, who measured $\MapMapMap$ with high significance from the Dark Energy Survey Year 3 shear catalogues. Moreover, in contrast to many other HOS, $\MapMapMap$ and $\Gamma_i$ can be modelled analytically without directly relying on simulations. 

There are several advantages to using $\MapMapMap$ instead of $\Gamma_i$, among them that $\MapMapMap$ compress the information in the $\Gamma_i$ from a data vector with thousands of entries to one containing only tens of entries in a non-tomographic setup, and that the $\MapMapMap$ have a clear separation into $E$- and $B$-modes, allowing for the detection of untreated systematic effects. Our goal is to lay the groundwork for a cosmological analysis with $\MapMapMap$. Such an analysis requires two ingredients: a model of $\MapMapMap$ and an estimate of its covariance matrix. The first ingredient is the subject of \citet[][\citetalias{Heydenreich2022} hereafter]{Heydenreich2022}, where we present and test an analytic model of $\MapMapMap$ based on the \texttt{BiHalofit} bispectrum model \citep{Takahashi2020}. Here, we are concerned with the second ingredient, which is the covariance.

The covariance of the estimator of a statistic can be obtained in three ways. The first possibility is applying the estimator to a large set of (quasi-)independent cosmological $N$-body simulations and using the sample covariance of the estimates. However, an unbiased and precise estimate requires many more realisations than entries in the data vector. For example, for a tomographic analysis of $\MapMapMap$ with four redshift bins and four different scale radii, the data vector contains 400 entries, so hundreds of simulations are needed for a simulated covariance estimate. Such an estimate would be very time-consuming, as both the creation of this many independent simulations and the measurement of $\MapMapMap$ in them is computationally demanding. The second possibility is to estimate the covariance directly from the data using jackknife resampling or bootstrap methods. For this, the survey area is divided into smaller patches, the statistics are estimated separately on each of these patches, and the sample covariance of these individual estimates is taken as the covariance estimate. However, this method has two main disadvantages. First, information on correlations larger than the patch size is lost. Since the number of patches must be at least as large as the length of the data vector, the required number of patches becomes large and their size small. Second, the method implicitly assumes that the individual patches are independent of each other. This assumption is not true, in particular, if small, neighbouring patches are considered. This causes biases in the covariance estimate. Third, triplets covering two or three patches need to be neglected for the estimation of the data vector. Otherwise, the covariance from jackknife or bootstrap resampling would overestimate the true covariance of the data.

Instead, cosmic shear covariances can also be calculated analytically for a given estimator. This was done, for example, for two-point statistics \citep{Joachimi2008} or the matter bispectrum \citep{Joachimi2009}. While the derivation of an expression for a covariance can be tedious, once it is done, evaluating them for the characteristics of a specific survey is usually accurate and quick, at least compared to running hundreds of cosmological simulations. It is also easier to infer the dependence of the covariance on survey area and geometry from an analytic expression.  Consequently, we follow this approach and derive an analytical expression for the covariance of $\MapMapMap$.  We test our result by comparing it with the covariance measured in simulated mock data, both simple Gaussian shear fields and full cosmological simulations, and investigate the influence of different terms in the covariance by performing mock cosmological analyses.

In our derivation, we make five crucial findings. First, the covariance consists of six terms, two of which belong to the Gaussian part and four to the non-Gaussian part. These terms all depend on the survey area and window function (Sect.~\ref{sec: Derivation subsec: Gaussian} and \ref{sec: Derivation subsec: Non-Gaussian}).
Second, the individual covariance terms can be estimated from correlation functions of the aperture mass map. These correlation functions need to be known only on scales inside the survey area (Sect.~\ref{sec: Derivation subsec: Correlation Functions}). Third, under the `large-field approximation', which assumes a broad survey window function, two covariance terms vanish, while the others scale inversely with the survey area. The vanishing terms decrease faster than $1/A$ with survey area $A$. We refer to them as finite-field terms. One of these terms is already present for Gaussian fields (Sect.~\ref{sec: Derivation subsec: Large-field approximation}). Fourth, the covariance of $\MapMapMap$ cannot be obtained from the bispectrum covariance unless the large-field approximation is valid.  Using a linear transformation of the covariance of a Fourier space quantity to obtain the covariance for a real-space observable has been done for second-order statistics \citep{Joachimi2021, Friedrich2021}. However, for $\MapMapMap$ this approach already fails for Gaussian fields if the large-field approximation is not used (Sect.~\ref{sec: Derivation From Bispec}). Finally, our covariance model, based on an estimator using convergence maps, can be used to obtain upper and lower limits on the covariance for an estimator based on third-order shear correlation functions (Sect.~\ref{sec: Validation subsec: Results subsubsec: N-body}).

This paper is structured as follows. In Sect.~\ref{sec: Theory} we give a short overview of third-order shear statistics, and we introduce our notation, the third-order aperture mass $\MapMapMap$, and its estimator. We derive the covariance of this estimator in Sect.~\ref{sec: Derivation}. In Sect.~\ref{sec: Derivation From Bispec} we show that an alternative derivation of the $\MapMapMap$ covariance from the bispectrum covariance is not correct. We validate the model by comparing its predictions to simulated data in Sect.~\ref{sec: Validation}. Finally, in Sect.~\ref{sec: MCMC}, we perform mock cosmological analyses to compare the results from the validation data to our model and investigate the impact of the individual covariance terms. We conclude with a summary and discussion in Sect.~\ref{sec: Discussion}. Throughout this paper, we assume a spatially flat universe. Our covariance modelling code is publically available \footnote{\url{https://github.com/sheydenreich/threepoint/releases}}.

\section{Theoretical background}
\label{sec: Theory}
In this section, we briefly review some fundamental weak lensing quantities relevant to third-order shear statistics and the covariance calculation in the remainder of this work. We refer to \citet{Bartelmann2001, Hoekstra2008} or \citet{Bartelmann2010} for in-depth reviews on weak lensing.

One of the fundamental quantities in weak gravitational lensing is the convergence $\kappa(\varthetavec)$, which is the normalised surface mass density at angular position $\varthetavec$. In a flat universe, it is related to the density contrast $\delta(\chi\varthetavec, \chi)$ at angular position $\varthetavec$ and comoving distance $\chi$ via
\begin{equation}
    \kappa(\varthetavec) = \frac{3H_0^2\Omm}{2c^2}\int_0^\infty \dd{\chi}\; q(\chi)\,\frac{\delta(\chi\varthetavec, \chi)}{a(\chi)}\;, \textrm{where }  q(\chi) = \int_\chi^\infty \dd{\chi'}\, p(\chi')\, \frac{\chi'-\chi}{\chi'}\;,
\end{equation}
with the Hubble constant $H_0$, the matter density parameter $\Omega_\mathrm{m}$, the speed-of-light $c$, the cosmic scale factor $a(\chi)$ at $\chi$, normalised to unity today, and
the probability distribution $p(\chi)\,\dd{\chi}$ of source galaxies in comoving distance.

We treat $\kappa(\varthetavec)$ as a homogenous and isotropic random field, which is characterised by the full set of its polyspectra $\mathcal{P}_n$, defined by
\begin{equation}
    \label{eq: polyspectra_real}
    \expval{\tilde{\kappa}(\ellvec_1)\dots\tilde{\kappa}(\ellvec_n)}_\mathrm{c} = \mathcal{P}_n(\ellvec_1,\dots,\ellvec_n)\,(2\pi)^2\, \dirac(\ellvec_1+\dots+\ellvec_n)\;,
\end{equation}
where $\tilde{\kappa}(\ellvec)$ is the Fourier transform of $\kappa$ and the $\expval{\dots}_\mathrm{c}$ are connected correlation functions.
Interesting for us in this work is the powerspectrum $P$, bispectrum $B$, trispectrum $T$, and the pentaspectrum $P_6$, defined as
\begin{align}
    P(\ell) = P_2(\ellvec, -\ellvec)\,, \quad    B(\ellvec_1, \ellvec_2) = \mathcal{P}_n(\ellvec_1, \ellvec_2, -\ellvec_1-\ellvec_2)\,, \quad
    T(\ellvec_1, \ellvec_2, \ellvec_3) = \mathcal{P}_n(\ellvec_1, \ellvec_2, \ellvec_3, -\ellvec_1-\ellvec_2-\ellvec_3)\,,
\end{align}
and
\begin{align}
    &P_6(\ellvec_1, \ellvec_2, \ellvec_3, \ellvec_4, \ellvec_5)= \mathcal{P}_n(\ellvec_1, \ellvec_2, \ellvec_3, \ellvec_4, \ellvec_5, -\ellvec_1-\ellvec_2-\ellvec_3-\ellvec_4-\ellvec_5)\;.
\end{align}
The $\kappa$-polyspectra of are related to the polyspectra $\mathcal{P}_n^\mathrm{(3d)}$ of the three-dimensional density contrast $\delta$, using the Limber-approximation \citep{Kaiser1997, Kayo2013},
\begin{align}
\label{eq: limber}
    \mathcal{P}_n(\ellvec_1,\dots,\ellvec_n)&=\left(\frac{3H_0^2\Omm}{2c^2}\right)^n\,\int_0^\infty \dd{\chi}\; \frac{q^n(\chi)}{\chi^{n-2}\, a(\chi)^n}\, \mathcal{P}_n^\mathrm{(3d)}(\ellvec_1/\chi,\dots,\ellvec_n/\chi, \chi)\;.
\end{align}
We model the three-dimensional power spectrum with the revised \verb|Halofit| prescription by \citet{Takahashi2012} and the three-dimensional bispectrum with \verb|BiHalofit| \citep{Takahashi2020}. For the tri- and pentaspectrum, we use the halo model \citep{Cooray2002} with a Sheth--Tormen halo mass function and halo bias \citep{Sheth1999} and Navarro--Frenk--White-halo profiles. To simplify our calculation, we use only the 1-halo term for the trispectrum, which we expect to be dominant at the considered $\ell$-scales. For the pentaspectrum, we use the 1- and part of the 2-halo term (see Sect.~\ref{sec: Derivation subsec: Large-field approximation} and Appendix~\ref{app: TVII approximation}) 

The convergence cannot be directly observed, but it is related to the weak lensing shear $\gamma$, which describes the change in the observed ellipticity of source galaxies due to the lensing effect. For third-order shear statistics, we are interested in the three-point correlation function of the shear, whose so-called natural components $\Gamma_i$ were derived by \citet{Schneider2003}. However, these components are challenging to handle in practice since they require a large number of bins (typically of the order of $10^3$) and are complex to model because they relate to the matter bispectrum via multi-dimensional integrals involving oscillating Bessel functions. As shown in \citetalias{Heydenreich2022}, more practical quantities with similar information content are the third-order aperture statistics $\MapMapMap$.

Aperture statistics are moments of the aperture mass $\Map$, a smoothed convergence map, which depends on its position $\alphavec$ and a characteristic smoothing scale $\theta$,
\begin{equation}
\label{eq: Definition Map}
    \Map(\alphavec, \theta)=\int \dd[2]{\vartheta}\; U_\theta(|\alphavec - \varthetavec|)\, \kappa(\varthetavec)\;, \textrm{ with }     U_\theta(\vartheta) = \frac{1}{\theta^2}\, u\left(\frac{\vartheta}{\theta}\right)\;,
\end{equation}
where $U_\theta$ is a filter function with aperture scale radius $\theta$. As long as this is a compensated filter function, meaning $
\int \dd{\vartheta} \vartheta\, U_\theta(\vartheta)=0\;,$
the aperture mass can be estimated from the tangential shear $\gamma_\mathrm{t}$ instead of the convergence \citep{Schneider1996}, 
\begin{equation}
    \label{eq: map_from_gamma_via_convolution}
    \Map(\alphavec, \theta)=\int \dd[2]{\vartheta}\; Q_\theta(|\alphavec - \varthetavec|)\, \gamma_\mathrm{t}(\alphavec;\varthetavec)\;, \quad \textrm{with}\quad Q_\theta(\vartheta)=\frac{2}{\vartheta^2}\, \int_0^\vartheta \dd{\vartheta'}\; \vartheta' U_\theta(\vartheta') - U_\theta(\vartheta)\;. 
\end{equation}

Here, we are interested in the covariance of the third-order aperture statistics $\MapMapMap$, given by
\begin{equation}
\MapMapMap(\theta_1, \theta_2, \theta_3)= \expval{\Map(\alphavec, \theta_1)\,\Map(\alphavec, \theta_2)\,\Map(\alphavec, \theta_3)}\,,
\end{equation}
which is our main observable for third-order shear statistics. $\MapMapMap$ is related to the convergence bispectrum by 
\begin{align}
\label{eq: Map3 from Bispec}
    \MapMapMap(\theta_1,\theta_2,\theta_3) &= \int \frac{\dd[2]{\ell_1}}{(2\pi)^2} \int \frac{\dd[2]{\ell_2}}{(2\pi)^2}\; \tilde{u}(\ell_1\,\theta_1)\,\tilde{u}(\ell_2\,\theta_2)\,\tilde{u}\left(|\ellvec_1+\ellvec_2|\,\theta_3\right)\, B(\ellvec_1,\ellvec_2)\;.
\end{align}

For the comparison of the analytical covariance to mock data in Sect.~\ref{sec: Validation} and the estimate of the cosmological parameter analysis in Sect.~\ref{sec: MCMC} we choose the exponential filter function from \citet{Crittenden2002},
\begin{equation}
    \label{eq: exponential filter}
    u(x) = \frac{1}{2\pi} \, \left(1-\frac{x^2}{2}\right)\, \exp(-\frac{x^2}{2})\;.
\end{equation}
The normalisation of the aperture filter functions is chosen such that $\int d^2 \vartheta Q(\vec{\vartheta}) =1$.

\begin{figure}
\begin{minipage}[c]{0.49\linewidth}
\centering
\begin{tikzpicture}
    \draw (0,0) rectangle (5,5);
    \draw (5,0) node[anchor=south east]{$A'$};
    \draw [fill=pink](1,1) rectangle (4,4);
    \draw (4,1) node[anchor=south east]{$A$};
    
    \filldraw [black] (1.5, 2) circle (2pt);
    \draw (1.5, 2) node[anchor=west]{$\alphavec$};
    \draw (1.5, 2) circle (1);
    \draw (1.5, 2) -- (0.5,2);
    \draw (0.8, 2) node[anchor=north]{$\theta$};

    \filldraw [black] (2.5, 4.5) circle (2pt);
    \draw (2.5, 4.5) node[anchor=west]{$\alphavec'$};
    \draw (2.5, 4.5) circle (1);
    \draw (2.5, 4.5) -- (2.5, 3.5);
    \draw (2.5, 3.75) node[anchor=west]{$\theta$};
    \end{tikzpicture}
\end{minipage}
\begin{minipage}[t]{0.49\linewidth}
\caption{Illustration of aperture mass estimation. The area $A'$ is the size of the full convergence field, on which we place apertures with scale radius $\theta$, illustrated by the circles, to obtain $\Map(\alphavec, \theta)$. Apertures centred on positions $\alphavec$ within the smaller area $A$ lie completely within $A'$, while apertures centred on positions $\alphavec'$ outside of $A$ extend outside of $A'$, so $\Map(\alphavec', \theta)$ is biased.}
\label{fig: aperture mass estimation}
\end{minipage}

\end{figure}
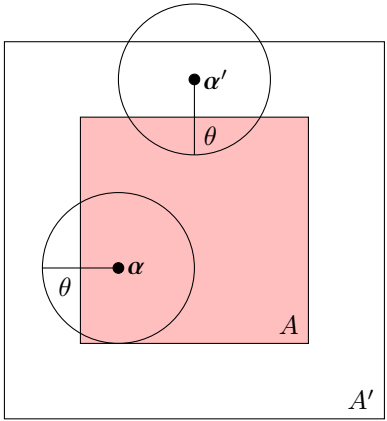

Based on Eq.~\eqref{eq: Definition Map}, $\Map(\alphavec, \theta)$ can be estimated from convergence fields using
\begin{equation}
    \MapEst(\alphavec, \theta) = \int_{A'}\dd[2]{\vartheta} U_\theta(|\alphavec-\varthetavec|)\, \kappa(\varthetavec)\;,
\end{equation}
where $A'$ is the full area of the convergence field $\kappa$. This estimator can be visualised as placing an aperture of radius $\theta$ centred at position $\alphavec$ and averaging the convergence values within the aperture weighted by the filter function $U_\theta$ (see Fig.~\ref{fig: aperture mass estimation}). The $\MapMapMap(\theta_1, \theta_2, \theta_3)$ can then be estimated by averaging the product of the $\MapEst(\alphavec, \theta)$ for the three filter radii over the aperture positions $\alphavec$. However, for $\alphavec$ close to the border of the survey area, $\MapEst(\alphavec, \theta)$ is biased because the aperture includes regions outside the survey where $\kappa$ is not known. Consequently, for an unbiased estimate of $\MapMapMap$, we only average over an area $A$ smaller than $A'$, which leads to the estimator
\begin{align}
    \MapMapMapEst(\theta_1, \theta_2, \theta_3)
    &= \frac{1}{A}\int_A \dd[2]{\alpha}\, \prod_{i=1}^3 \int_{A'} \dd[2]{\varthetavec_i} U_{\theta_i}(|\alphavec-\varthetavec_i|)\,\kappa(\varthetavec_i)\,.
\end{align}
We have assumed here that apertures are placed densely so that the separation between aperture centres is much smaller than the aperture radius $\theta$, to use a continuous integral over $\alphavec$ instead of a sum. If $A'$ is large enough that all $U_\theta$ centred on points within $A$ vanish outside of $A'$, we can replace $A'$ with the full $\realspace^2$. We also introduce the survey window function $W_A(\alphavec)$, which is unity for $\alphavec$ inside $A$ and zero otherwise. With $W$, we convert integrals over $A$ into integrals over $\realspace^2$, which we denote as two-dimensional integrals without integration borders in the following. The estimator for $\MapMapMap$ becomes
\begin{align}
    \label{eq: EstimatorMap^3_fft}
    \MapMapMapEst(\theta_1, \theta_2, \theta_3)
    &= \frac{1}{A}\,\int \dd[2]{\alpha}\,W_A(\alphavec)\, \left[\prod_{i=1}^3 \int \dd[2]{\varthetavec_i} U_{\theta_i}(|\alphavec-\varthetavec_i|)\,\kappa(\varthetavec_i)\right]\;.
\end{align}
This estimator is easily applied to simulations, where convergence maps without masks are available (see Sect.~\ref{sec: Validation subsec: Measurement subsubsec: FFT}). However, this estimator should not be used for survey data that includes masked areas. In that case, one would instead measure the correlation functions $\Gamma_i$ and convert these into $\MapMapMap$ \citepalias[see the discussion in Section 5.3 of][]{Heydenreich2022}. In this approach, one does not need to decrease the survey area from $A'$ to $A$ because no apertures are laid down. Therefore, the covariance for this estimator is smaller than the covariance for the $\MapMapMapEst$ in Eq.~\eqref{eq: EstimatorMap^3_fft}. However, the unbiased construction of the aperture mass on the whole area $A'$ requires shear estimates outside of $A'$ as well. Therefore, even $\Gamma_i$ measured from shear estimates on all of $A'$ cannot include the full information that would be contained in an unbiased estimate of $\MapMapMapEst$ on $A'$. Consequently, in the absence of masked regions, we expect the magnitude of the covariance for the estimator based on the $\Gamma_i$ to be between the covariance for $\MapMapMapEst$ for a survey area of $A$ and a survey area of $A'$. We verify this expectation in Sect.~\ref{sec: Validation}.

\section{Derivation of aperture statistics covariance from the real-space estimator}
\label{sec: Derivation}
\begin{figure}
\centering
    \begin{tikzcd}[column sep = 0.1em]
         & C_{\MapMapMapEst} & = & \myexpval{\MapMapMapEst\MapMapMapEst}\arrow[dlll]\arrow[dl]\arrow[dr]\arrow[drrr] & - & \myexpval{\MapMapMapEst}\myexpval{\MapMapMapEst} &  && \\
         \myexpval{\MapMapMapEst\MapMapMapEst}_\mathrm{G}\arrow[d]\arrow[dr] && \myexpval{\MapMapMapEst\MapMapMapEst}_\mathrm{BB}\arrow[d]\arrow[dr] && \myexpval{\MapMapMapEst\MapMapMapEst}_\mathrm{PT}\arrow[d]\arrow[dr] && \myexpval{\MapMapMapEst\MapMapMapEst}_\mathrm{P_6}\arrow[d] \\
        \TI\textrm{, Eq.~\eqref{eq: T1 final}}\arrow[d,dashed] & \TII\textrm{, Eq.~\eqref{eq: T2 final}}\arrow[d,dashed] & \myexpval{\MapMapMapEst}\myexpval{\MapMapMapEst} & \TIV\textrm{, Eq.~\eqref{eq: T4 final}}\arrow[d,dashed] & \TV\textrm{, Eq.~\eqref{eq: T5 final}}\arrow[d,dashed]&\TVI\textrm{, Eq.~\eqref{eq: T6 final}}\arrow[d,dashed] & \TVII\textrm{, Eq.~\eqref{eq: T7 final}}\arrow[d,dashed] \\
        \TI^\infty\textrm{, Eq.~\eqref{eq: TI for infinite fields}} & 0 &  & \TIV^\infty\textrm{, Eq.~\eqref{eq: TIV for infinite fields}} & \TV^\infty\textrm{, Eq.~\eqref{eq: TV for infinite fields}} & 0 & \TVII^\infty\textrm{, Eq.~\eqref{eq: TVII for infinite fields}} \\
    \end{tikzcd}
    \begin{tikzpicture}[mybox/.style={draw, thick, rounded corners, inner sep=5pt}]
        \node[mybox] (box){%
            \begin{tikzcd}[    ,row sep = 0ex
    ,/tikz/column 1/.append style={anchor=base east}
    ,/tikz/column 2/.append style={anchor=base west}]
                \arrow[r] & \textrm{Decomposition into summands (see Sect.~\ref{sec: Derivation subsec: Gaussian} and \ref{sec: Derivation subsec: Non-Gaussian})} &
                \arrow[r,dashed] & \textrm{Large-field approximation (see Sect.~\ref{sec: Derivation subsec: Large-field approximation})}
            \end{tikzcd}
        };
    \end{tikzpicture}
\caption{Schematic representation of the calculation of the covariance $C_{\MapMapMapEst}$. The covariance is given by the difference between $\expval{\MapMapMapEst\,\MapMapMapEst}$ and $\expval{\MapMapMapEst}\,\expval{\MapMapMapEst}$, the first of which can be decomposed (indicated by solid arrows) into one Gaussian (denoted by G) and three non-Gaussian parts (denoted by BB, PT, and $\mathrm{P}_6$). We discuss the Gaussian part in Sect.~\ref{sec: Derivation subsec: Gaussian} and the non-Gaussian parts in Sect.~\ref{sec: Derivation subsec: Non-Gaussian}. These parts can be further decomposed into terms depending on different permutations of the aperture scale radii, called $\TI$ to $\TVII$. For large survey areas, the $T_i$ can be approximated, indicated by dashed arrows, as shown in Sect.~\ref{sec: Derivation subsec: Large-field approximation}. Under this approximation, two terms vanish, which is why we term them `finite-field terms'. Also shown are equation numbers for the relevant expressions.}
\label{fig: covariance scheme}
\end{figure}
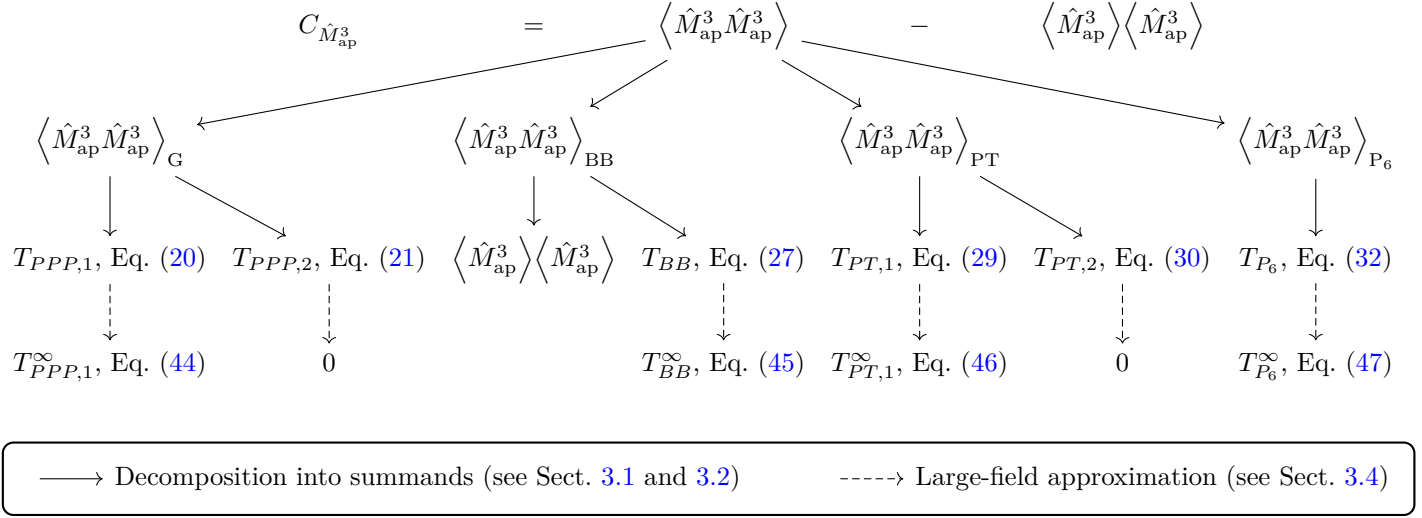

We now derive the covariance $C_{\MapMapMapEst}$ of $\MapMapMapEst$. An overview of our calculation is given in Fig.~\ref{fig: covariance scheme}, with references to the main equations in our derivation. 
In the derivation, we find that the covariance comprises several terms, including various permutations of scale radii. For simplicity, we write here `Perm.' to indicate permutations; the complete list of permutations for all terms is given in Appendix~\ref{app: permutations}. We are not explicitly addressing the effect of shape noise in this section, but shape noise can be easily included in the derived expressions by replacing the power spectrum $P$ with $P+{\sigma_\epsilon^2}/{2n}$, where $n$ is the galaxy number density and $\sigma^2_\epsilon$ the two-component ellipticity dispersion. We show the validity of this treatment in Appendix~\ref{app: shape noise}.

The covariance is
\begin{align}
\label{eq: DefinitionCovariance}
    C_{\MapMapMapEst}(\Theta_1, \Theta_2) &= \expval{\MapMapMapEst\,\MapMapMapEst}(\Theta_1, \Theta_2)- \expval{\MapMapMapEst(\Theta_1)}\, \expval{\MapMapMapEst(\Theta_2)}\;,
\end{align}
where $\Theta_1=(\theta_1, \theta_2, \theta_3)$ and $\Theta_2=(\theta_4, \theta_5, \theta_6)$. With Eq.~\eqref{eq: EstimatorMap^3_fft} and $\kappa_i:=\kappa(\varthetavec_i)$,
\begin{align}
\label{eq: cov_from_estimator}
\notag\expval{\MapMapMapEst\, \MapMapMapEst}(\Theta_1, \Theta_2)&= \frac{1}{A^2}\int \dd[2]{\alpha_1} \int \dd[2]{\alpha_2} W_A(\alphavec_1)\,W_A(\alphavec_2)\,\Bigg[\prod_{i=1}^3 \int \dd[2]{\vartheta_i} U_{\theta_i}(|\alphavec_1-\varthetavec_i|)\Bigg]\,\Bigg[\prod_{j=4}^6 \int \dd[2]{\vartheta_j} U_{\theta_j}(|\alphavec_2-\varthetavec_j|)\Bigg]\\
&\times\,\expval{\kappa_1\,\kappa_2\,\kappa_3\,\kappa_4\,\kappa_5\,\kappa_6}.
\end{align}
The six-point correlation can be written in terms of connected correlation functions, denoted by $\expval{}_\mathrm{c}$. This, together with $\expval{\kappa(\varthetavec)}=0$, leads to
\begin{align}
	\label{eq: cov_from_estimator_decomposed}
\notag\expval{\MapMapMapEst\, \MapMapMapEst}(\Theta_1, \Theta_2)  &= \frac{1}{A^2}\int \dd[2]{\alpha_1} \int \dd[2]{\alpha_2} W_A(\alphavec_1)\, W_A(\alphavec_2)\, \Bigg[\prod_{i=1}^3 \int \dd[2]{\vartheta_i} U_{\theta_i}(|\alphavec_1-\varthetavec_i|)\Bigg]\,\Bigg[\prod_{j=4}^6 \int \dd[2]{\vartheta_j} U_{\theta_j}(|\alphavec_2-\varthetavec_j|)\Bigg]\\
	&\quad \times \Big[ \left(\expval{\kappa_1\,\kappa_2}_\mathrm{c}\,\expval{\kappa_3\,\kappa_4}_\mathrm{c}\,\expval{\kappa_5\,\kappa_6}_\mathrm{c} +     14\,\mathrm{Perm.}\right) + \left(     \expval{\kappa_1\,\kappa_2\,\kappa_3}_\mathrm{c}\,\expval{\kappa_4\,\kappa_5\,\kappa_6}_\mathrm{c} + 9\,\mathrm{Perm.}\right)\\
    &\notag \qquad +  \left(   \expval{\kappa_1\,\kappa_2}_\mathrm{c}\,\expval{\kappa_3\,\kappa_4\,\kappa_5\,\kappa_6}_\mathrm{c} + 14\,\mathrm{Perm.} \right) +    \expval{\kappa_1\,\kappa_2\,\kappa_3\,\kappa_4\,\kappa_5\,\kappa_6}_\mathrm{c} \Big]\\
    &\notag= 	\expval{\MapMapMapEst\, \MapMapMapEst}_\textrm{G}(\Theta_1, \Theta_2) + \expval{\MapMapMapEst\, \MapMapMapEst}_{BB}(\Theta_1, \Theta_2)  +\expval{\MapMapMapEst\, \MapMapMapEst}_{PT}(\Theta_1, \Theta_2)\\
    &\notag\quad + \expval{\MapMapMapEst\, \MapMapMapEst}_{P_6}(\Theta_1, \Theta_2) \;.
\end{align}
The first term $\expval{\MapMapMapEst\, \MapMapMapEst}_\textrm{G}$ is the Gaussian part of the covariance -- the only part that is non-zero for Gaussian random fields. It depends only on the two-point correlation of $\kappa$, or, equivalently, the matter power spectrum $P(\ell)$. The other terms comprise the non-Gaussian part of the covariance. They depend on higher-order polyspectra, namely the bi-, tri- and the pentaspectrum.

\subsection{Gaussian part}
\label{sec: Derivation subsec: Gaussian}
We first concentrate on the Gaussian part $\expval{\MapMapMapEst\MapMapMapEst}_\mathrm{G}$, given by
\begin{align}
	\label{eq: cov_from_estimator_gaussian}
	\notag\expval{\MapMapMapEst\MapMapMapEst}_\mathrm{G}(\Theta_1, \Theta_2) &=  \frac{1}{A^2}\int \dd[2]{\alpha_1} \int \dd[2]{\alpha_2} W_A(\alphavec_1)\, W_A(\alphavec_2)\,\Bigg[\prod_{i=1}^3 \int \dd[2]{\vartheta_i} U_{\theta_i}(|\alphavec_1-\varthetavec_i|)\Bigg]\,\Bigg[\prod_{j=4}^6 \int \dd[2]{\vartheta_j} U_{\theta_j}(|\alphavec_2-\varthetavec_j|)\Bigg]\\
	&\quad \times  \left[\expval{\kappa_1\,\kappa_2}_\mathrm{c}\,\expval{\kappa_3\,\kappa_4}_\mathrm{c}\,\expval{\kappa_5\,\kappa_6}_\mathrm{c} +     14\,\mathrm{Perm.}\right]\;.
\end{align}
We split the permutations in Eq.~\eqref{eq: cov_from_estimator_gaussian} into two groups, as
\begin{align}
	\label{eq: cov_from_estimator_gaussian_split}
   \notag \expval{\MapMapMapEst\MapMapMapEst}_\mathrm{G}(\Theta_1, \Theta_2) &= \frac{1}{A^2}\int \dd[2]{\alpha_1} \int \dd[2]{\alpha_2} W_A(\alphavec_1)\, W_A(\alphavec_2)\, \Bigg[\prod_{i=1}^3 \int \dd[2]{\vartheta_i} U_{\theta_i}(|\alphavec_1-\varthetavec_i|)\Bigg]\,\Bigg[\prod_{j=4}^6 \int \dd[2]{\vartheta_j} U_{\theta_j}(|\alphavec_2-\varthetavec_j|)\Bigg]\\
	&\qquad \times  \left[\expval{\kappa_1\,\kappa_4}_\mathrm{c}\,\expval{\kappa_3\,\kappa_5}_\mathrm{c}\,\expval{\kappa_2\,\kappa_6}_\mathrm{c} +     5\,\textrm{Perm.}\right]\\
	&\notag\quad+ \frac{1}{A^2}\int \dd[2]{\alpha_1} \int \dd[2]{\alpha_2} W_A(\alphavec_1)\, W_A(\alphavec_2)\, \Bigg[\prod_{i=1}^3 \int \dd[2]{\vartheta_i} U_{\theta_i}(|\alphavec_1-\varthetavec_i|)\Bigg]\,\Bigg[\prod_{j=4}^6 \int \dd[2]{\vartheta_j} U_{\theta_j}(|\alphavec_2-\varthetavec_j|)\Bigg]\\
	&\notag\qquad \times  \left[\expval{\kappa_1\,\kappa_2}_\mathrm{c}\,\expval{\kappa_3\,\kappa_4}_\mathrm{c}\,\expval{\kappa_5\,\kappa_6}_\mathrm{c} + 8\,\textrm{Perm.}\right]\\
	&\notag= 	\TI(\Theta_1, \Theta_2) + 	\TII(\Theta_1, \Theta_2)\;.
\end{align}
The first group consists of six terms, in which for all three $ \expval{\kappa_i\,\kappa_j}_\mathrm{c}$, the first index $i \in \lbrace 1, 2, 3 \rbrace$, and the second index $j \in \lbrace 4, 5, 6 \rbrace$. The second group consists of the nine other permutations. To evaluate $\TI$ and $\TII$, we use Eq.~\eqref{eq: polyspectra_real} for $n=2$ and 
\begin{equation}
    \int \dd[2]{\vartheta}\, U_\theta(|\alphavec-\varthetavec|) \, \E^{-\I\ellvec\cdot\varthetavec} = \tilde{u}(\ell\,\theta)\, \E^{-\I\ellvec\cdot\alphavec}\;,
\end{equation}
and find 
\begin{align}
	\TI(\Theta_1, \Theta_2)
	\notag&= \frac{1}{A^2} \left[\prod_{i=1}^3 \int \frac{\dd[2]{\ell_i}}{(2\pi)^2} P(\ell_i) \right] \, \left[\tilde{u}(\ell_1\,\theta_1)\,\tilde{u}(\ell_2\,\theta_2)\,\tilde{u}(\ell_3\,\theta_3)\,\tilde{u}(\ell_1\,\theta_4)\,\tilde{u}(\ell_2\,\theta_5)\,\tilde{u}(\ell_3\,\theta_6) + \textrm{5 Perm.}\right]\\
 &\label{eq: T1 final}\quad \times \int \dd[2]{\alpha_1} \int \dd[2]{\alpha_2} W_A(\alphavec_1)\, W_A(\alphavec_2)\, \E^{-\I(\alphavec_1-\alphavec_2)\,(\ellvec_1+\ellvec_2+\ellvec_3)} \\
 	&\notag= \left[\prod_{i=1}^3 \int \frac{\dd[2]{\ell_i}}{(2\pi)^2} P(\ell_i) \right] \, G_A(\ellvec_1+\ellvec_2+\ellvec_3)\,\left[ \tilde{u}(\ell_1\,\theta_1)\,\tilde{u}(\ell_2\,\theta_2)\,\tilde{u}(\ell_3\,\theta_3)\,\tilde{u}(\ell_1\,\theta_4)\,\tilde{u}(\ell_2\,\theta_5)\,\tilde{u}(\ell_3\,\theta_6)\,  + \textrm{5 Perm.}\right]\;,
\end{align}
and
\begin{align}
	\TII(\Theta_1, \Theta_2)
	&\notag= \frac{1}{A^2} \left[\prod_{i=1}^3 \int \frac{\dd[2]{\ell_i}}{(2\pi)^2} P(\ell_i) \right] \,\left[\tilde{u}(\ell_1\,\theta_1)\,\tilde{u}(\ell_1\,\theta_2)\,\tilde{u}(\ell_2\,\theta_3)\,\tilde{u}(\ell_2\,\theta_4)\,\tilde{u}(\ell_3\,\theta_5)\,\tilde{u}(\ell_3\,\theta_6)+\textrm{8 Perm.}\right]\\
	&\label{eq: T2 final}\quad \times \int \dd[2]{\alpha_1} \int \dd[2]{\alpha_2} W_A(\alphavec_1)\, W_A(\alphavec_2)\, \E^{-\I(\alphavec_1-\alphavec_2)\cdot\ellvec_2} \\
 	&\notag= \left[\prod_{i=1}^3 \int \frac{\dd[2]{\ell_i}}{(2\pi)^2} P(\ell_i) \right] \, G_A(\ellvec_2)\,\left[\tilde{u}(\ell_1\,\theta_1)\,\tilde{u}(\ell_1\,\theta_2)\,\tilde{u}(\ell_2\,\theta_3)\,\tilde{u}(\ell_2\,\theta_4)\,\tilde{u}(\ell_3\,\theta_5)\,\tilde{u}(\ell_3\,\theta_6) + \textrm{8 Perm.}\right]\;,
\end{align}
where the geometry factor $G_A(\ellvec)$, defined by
\begin{equation}
\label{eq: G_A}
    G_A(\ellvec) = \frac{1}{A^2}\int \dd[2]{\alpha_1} \int \dd[2]{\alpha_2}\;W_A(\alphavec_1)\, W_A(\alphavec_2)\,\E^{-\I(\alphavec_1-\alphavec_2)\cdot\ellvec}\;,
\end{equation}
contains the full dependence of the covariance on the survey area and geometry. For a square survey with side length $\varthetamax=\sqrt{A}$, and $\ellvec=(\ell_{x}, \ell_y)$ the factor is
\begin{equation}
\label{eq: G_A square}
    G_{A, \mathrm{square}}(\ellvec) =  \frac{4\sin^2(\ell_{x}\, \varthetamax/2)}{\ell_{x}^2\, \varthetamax^2}\, \frac{4 \sin^2(\ell_{y}\, \varthetamax/2)}{\ell_{y}^2\, \varthetamax^2}\;.
\end{equation}
The geometry factor $G_A$ is related to a function $E_A(\etavec)$, which, for a point $\alphavec$ within $A$, gives the probability for a second point at $\alphavec+\etavec$ to lie within $A$ as well. It is given by
\begin{equation}
  \label{eq: definition E}
  E_A(\etavec) = \frac{1}{A}\int \dd[2]{\alpha}\, W_A(\alphavec) \, W_A(\alphavec+\etavec)\;, \quad \textrm{so that} \quad  G_A(\ellvec)=\frac{1}{A}\int_A \dd[2]{\eta}\, E_A(\etavec)\, \E^{-\I\etavec\cdot\ellvec}\;.
\end{equation}
For a square area, $E_A(\etavec)$ was determined by \citet{Heydenreich2020}. It is unity for vanishing separation $|\etavec|=0$ and smoothly declines to zero for $\etavec$ outside the square.

\subsection{Non-Gaussian part}
\label{sec: Derivation subsec: Non-Gaussian}
We now derive the non-Gaussian part of the covariance. First, we calculate the covariance part $\expval{\MapMapMapEst\MapMapMapEst}_{BB}$, which depends on third-order correlations of $\kappa$. We divide this term into two parts, as
\begin{align}
	\expval{\MapMapMapEst\MapMapMapEst}_{BB}(\Theta_1, \Theta_2)
	&\notag= \frac{1}{A^2}\int \dd[2]{\alpha_1} \int \dd[2]{\alpha_2}  \Bigg[\prod_{i=1}^3 \int \dd[2]{\vartheta_i} U_{\theta_i}(|\alphavec_1-\varthetavec_i|)\Bigg]\,\left[\prod_{j=4}^6 \int \dd[2]{\vartheta_j} U_{\theta_j}(|\alphavec_2-\varthetavec_j|)\right]\\
	&\notag\quad \times W_A(\alphavec_1)\, W_A(\alphavec_2)\, \left\{ \expval{\kappa_1\,\kappa_2\,\kappa_3}_\mathrm{c}\,\expval{\kappa_4\,\kappa_5\,\kappa_6}_\mathrm{c} 
	+ \left[\expval{\kappa_4\,\kappa_2\,\kappa_3}_\mathrm{c}\,\expval{\kappa_1\,\kappa_5\,\kappa_6}_\mathrm{c} + 8\textrm{ Perm. }\right]\right\}\\
	&= \expval{\MapMapMapEst(\theta_1, \theta_2, \theta_3)}\expval{\MapMapMapEst(\theta_4, \theta_5, \theta_6)} + \TIV(\Theta_1, \Theta_2)\;.
\end{align}
The first term cancels with the second term in the definition of the covariance of the estimator in Eq.~\eqref{eq: DefinitionCovariance}. 
The term $\TIV$ is
\begin{align}
	\TIV(\Theta_1, \Theta_2)
	&=\frac{1}{A^2}\int \dd[2]{\alpha_1} \int \dd[2]{\alpha_2} \Bigg[\prod_{i=1}^3 \int \dd[2]{\vartheta_i} U_{\theta_i}(|\alphavec_1-\varthetavec_i|)\Bigg]\,\left[\prod_{j=4}^6 \int \dd[2]{\vartheta_j} U_{\theta_j}(|\alphavec_2-\varthetavec_j|)\right]\,  W_A(\alphavec_1)\, W_A(\alphavec_2)\,\\
 &\notag\quad \times \left[\expval{\kappa_4\,\kappa_2\,\kappa_3}_\mathrm{c}\,\expval{\kappa_1\,\kappa_5\,\kappa_6}_\mathrm{c}  + \textrm{8 Perm.}\right]\;.
\end{align}
Using Eq.~\eqref{eq: polyspectra_real} for $n=3$ to introduce the bispectrum and the definition of $G_A$ in Eq.~\eqref{eq: G_A} leads to
\begin{align}
	\TIV(\Theta_1, \Theta_2)
\label{eq: T4 final}
	&= 	 \int \frac{\dd[2]{\ell_1}}{(2\pi)^2} \int \frac{\dd[2]{\ell_2}}{(2\pi)^2} \int \frac{\dd[2]{\ell_4}}{(2\pi)^2} \int \frac{\dd[2]{\ell_5}}{(2\pi)^2} \;  B(\ellvec_1, \ellvec_2)\, B(\ellvec_4, \ellvec_5)\,  G_A(\ellvec_1-\ellvec_4) \\
	&\notag\quad \times \left[\tilde{u}(\ell_4\,\theta_1)\,\tilde{u}(\ell_2\,\theta_2)\,\tilde{u}(|\ellvec_1+\ellvec_2|\,\theta_3)\,\tilde{u}(\ell_1\,\theta_4)\,\tilde{u}(\ell_5\,\theta_5)\,\tilde{u}(|\ellvec_4+\ellvec_5|\,\theta_6)+\textrm{8 Perm.}\right]\;.
\end{align}
Second, we calculate the covariance part $\expval{\MapMapMapEst\MapMapMapEst}_{PT}$. We divide this term into two parts as
\begin{align}
\label{eq: cov NG2}
	\expval{\MapMapMapEst\MapMapMapEst}_{PT}(\Theta_1, \Theta_2)
	&\notag= \frac{1}{A^2}\int \dd[2]{\alpha_1} \int \dd[2]{\alpha_2} \Bigg[\prod_{i=1}^3 \int \dd[2]{\vartheta_i} U_{\theta_i}(|\alphavec_1-\varthetavec_i|)\Bigg]\,\left[\prod_{j=4}^6 \int \dd[2]{\vartheta_j} U_{\theta_j}(|\alphavec_2-\varthetavec_j|)\right]\\
	&\notag\quad \times  W_A(\alphavec_1)\, W_A(\alphavec_2)\, \Big[ \expval{\kappa_1\,\kappa_4}_\mathrm{c}\,\expval{\kappa_2\,\kappa_3\,\kappa_5\,\kappa_6}_\mathrm{c} + 8\textrm{ Perm. } 
	+ \expval{\kappa_1\,\kappa_2}_\mathrm{c}\,\expval{\kappa_3\,\kappa_4\,\kappa_5\,\kappa_6}_\mathrm{c} + 5\textrm{ Perm. }\Big]\\
	&= \TV(\Theta_1, \Theta_2) + \TVI(\Theta_1, \Theta_2) \;.
\end{align}
Here, the first permutations contain all terms where the two-point correlation $\expval{\kappa_i\,\kappa_j}$ has $i\in\lbrace 1, 2, 3 \rbrace$ and $j \in \lbrace 4, 5, 6 \rbrace$. The second permutations contain all other terms. We introduce the power- and trispectrum with Eq.~\eqref{eq: polyspectra_real}, and find
\begin{align}
\label{eq: T5 final}
	\TV(\Theta_1, \Theta_2)
	&=\int \frac{\dd[2]{\ell_1}}{(2\pi)^2}\int \frac{\dd[2]{\ell_2}}{(2\pi)^2}\int \frac{\dd[2]{\ell_3}}{(2\pi)^2}\int \frac{\dd[2]{\ell_4}}{(2\pi)^2}\; P(\ell_1)\, T(\ellvec_2, \ellvec_3, \ellvec_4)\, G_A(\ellvec_1+\ellvec_2+\ellvec_3)\\
	&\notag\quad\times \left[ \tilde{u}(\ell_1\,\theta_1)\,\tilde{u}(\ell_2\,\theta_2)\,\tilde{u}(\ell_3\,\theta_3)\,\tilde{u}(\ell_1\,\theta_4)\,\tilde{u}(\ell_4\,\theta_5)\,\tilde{u}(|\ellvec_2+\ellvec_3+\ellvec_4|\,\theta_6) + \textrm{8 Perm.}\right]\;,
\end{align}
and
\begin{align}
\label{eq: T6 final}
	\TVI(\Theta_1, \Theta_2)
	&=\int \frac{\dd[2]{\ell_1}}{(2\pi)^2}\int \frac{\dd[2]{\ell_2}}{(2\pi)^2}\int \frac{\dd[2]{\ell_3}}{(2\pi)^2}\int \frac{\dd[2]{\ell_4}}{(2\pi)^2}\; P(\ell_1)\, T(\ellvec_2, \ellvec_3, \ellvec_4)\, G_A(\ellvec_2)\\
	&\notag\quad\times \left[ \tilde{u}(\ell_1\,\theta_1)\,\tilde{u}(\ell_1\,\theta_2)\,\tilde{u}(\ell_2\,\theta_3)\,\tilde{u}(\ell_3\,\theta_4)\,\tilde{u}(\ell_4\,\theta_5)\,\tilde{u}(|\ellvec_2+\ellvec_3+\ellvec_4|\,\theta_6) + \textrm{5 Perm.}\right]\;.
\end{align}
Finally, we consider $\expval{\MapMapMapEst\MapMapMapEst}_{P_6}$, which is
\begin{align}
	\expval{\MapMapMapEst\MapMapMapEst}_{P_6}(\Theta_1, \Theta_2)
	&\notag= \frac{1}{A^2}\int \dd[2]{\alpha_1} \int \dd[2]{\alpha_2} W_A(\alphavec_1)\, W_A(\alphavec_2)\\
	&\notag\quad \times \Bigg[\prod_{i=1}^3 \int \dd[2]{\vartheta_i} U_{\theta_i}(|\alphavec_1-\varthetavec_i|)\Bigg]\,\Bigg[\prod_{j=4}^6 \int \dd[2]{\vartheta_j} U_{\theta_j}(|\alphavec_2-\varthetavec_j|)\Bigg]\,\expval{\kappa_1\,\kappa_2\,\kappa_3\,\kappa_4\,\kappa_5\,\kappa_6}_\mathrm{c} \\
	&= \TVII(\Theta_1, \Theta_2) \;.
\end{align}
The term $\TVII$ is, with Eq.~\eqref{eq: polyspectra_real} for $n=6$, 
\begin{align}
\label{eq: T7 final}
	\TVII(\Theta_1, \Theta_2)
&=  \int\frac{\dd[2]{\ell_1}}{(2\pi)^2}\, \int\frac{\dd[2]{\ell_2}}{(2\pi)^2}  \int\frac{\dd[2]{\ell_3}}{(2\pi)^2}  \int\frac{\dd[2]{\ell_4}}{(2\pi)^2} \int\frac{\dd[2]{\ell_5}}{(2\pi)^2}\;P_6(\ellvec_1, \ellvec_2, \ellvec_3, \ellvec_4, \ellvec_5)\, G_A(\ellvec_1+\ellvec_2+\ellvec_3)\\
&\notag\quad\times  \tilde{u}(\ell_1\,\theta_1)\,\tilde{u}(\ell_2\,\theta_2)\,\tilde{u}(\ell_3\,\theta_3)\,\tilde{u}(\ell_4\,\theta_4)\,\tilde{u}(\ell_5\,\theta_5)\,\tilde{u}(|\ellvec_1+\ellvec_2+\ellvec_3+\ellvec_4+\ellvec_5|\,\theta_6)\;.
\end{align}

\subsection{Connection of covariance terms to aperture mass correlation functions}
\label{sec: Derivation subsec: Correlation Functions}
We now show that the individual covariance terms are related to correlation functions of the aperture mass. While this might appear like a meaningless algebraic exercise, it has an important corollary: Since the correlation functions can be measured, we can estimate all individual terms of the covariance in the validation data. Thereby, the analytic expressions can be validated. 

We define the $n$-th order aperture mass correlation functions $\Map^{n,m}$, as
\begin{equation}
    \Map^{n,m}(\theta_1,\dots,\theta_m;\theta_{m+1},\dots,\theta_n; \etavec) = \expval{\Map(\varthetavec, \theta_1)\,\dots\,\Map(\varthetavec, \theta_m)\, \Map(\varthetavec+\etavec, \theta_{m+1}) \Map(\varthetavec+\etavec, \theta_n)}\;.
\end{equation}
This function is symmetric in its first $m$ aperture radii and in its second $n-m$ aperture radii. The correlation functions can be expressed in terms of the convergence as
\begin{equation}
\label{eq: Map_nm in terms of kappa}
       \Map^{n,m}(\theta_1,\dots,\theta_m;\theta_{m+1},\dots,\theta_n; \etavec) =  \Bigg[\prod_{i=1}^m\int \dd[2]{\vartheta_i}\, U_{\theta_i}(\vartheta_i)\Bigg]\,  \Bigg[\prod_{j=m+1}^n\int \dd[2]{\vartheta_j}\, U_{\theta_j}(|\varthetavec_j+\etavec|)\Bigg] \expval{\kappa_1\dots\kappa_n}\;,
\end{equation}
where $\kappa_i=\kappa(\varthetavec_i)$ and the average over the convergence contains both connected and unconnected terms. 
Using $n=2$, Eq.~\eqref{eq: cov_from_estimator_gaussian_split} and the definition of $E(\etavec)$ from Eq.~\eqref{eq: definition E}, the first Gaussian covariance term can be written as
\begin{align}
    \TI(\Theta_1, \Theta_2)
        &=\frac{1}{A^2}\int \dd[2]{\alpha_1}\int \dd[2]{\alpha_2}\; W_A(\alphavec_1)\,W_A(\alphavec_2)\, \Map^{2,1}(\theta_1; \theta_4;|\alphavec_1-\alphavec_2|)\,\Map^{2,1}(\theta_2; \theta_5;|\alphavec_1-\alphavec_2|)\\
        &\notag\quad \times \Map^{2,1}(\theta_3; \theta_6;|\alphavec_1-\alphavec_2|) \textrm{+ 5 Perm.}\\
    &=\notag\frac{1}{A}\int_A \dd[2]{\eta}\; E_A(\etavec)\, \Map^{2,1}(\theta_1; \theta_4;\etavec)\,\Map^{2,1}(\theta_2; \theta_5;\etavec)\, \Map^{2,1}(\theta_3; \theta_6;\etavec) \textrm{+ 5 Perm.}
\end{align}
Similarly, $\TII$ can be written as
\begin{align}
    \label{eq: TPPP2 from sims}
    \TII(\Theta_1, \Theta_2)=\Map^2(\theta_1, \theta_2)\, \Map^2(\theta_5, \theta_6)\, \frac{1}{A}\int_A \dd[2]{\eta}\, E_A(\etavec)\, \Map^{2,1}(\theta_3; \theta_4;\etavec) +\textrm{ 8 Perm.}\;,
\end{align}
where $\Map^2(\theta_1, \theta_2) = \Map^{2,0}(\theta_1, \theta_2, \eta=0)$ is the second-order aperture statistic.

The term $\TIV$ becomes
\begin{equation}
    \TIV(\Theta_1, \Theta_2)=\frac{1}{A}\int_A \dd[2]{\eta}\; E_A(\etavec)\, \Map^{3,2}(\theta_1, \theta_2; \theta_4;\etavec)\,\Map^{3,2}(\theta_5, \theta_6; \theta_3;\etavec)+\textrm{ 8 Perm.}
\end{equation}
We relate $\TV$ and $\TVI$ to the correlation functions $\Map^{4,3}$ and $\Map^{4,2}$. We stress that these correlation functions depend on both connected and unconnected correlations of the convergence field. 
Using Eq.~\eqref{eq: cov NG2}, we find
\begin{equation}
    \TV(\Theta_1, \Theta_2)=\frac{1}{A}\int_A \dd[2]{\eta}\; E_A(\etavec)\, \Map^{2,1}(\theta_1; \theta_4;\etavec)\,\Map^{4,2}(\theta_2, \theta_3; \theta_5, \theta_6;\etavec) +\textrm{8 Perm.} - \big[3\,\TI(\Theta_1,\Theta_2)+\TII(\Theta_1,\Theta_2)\big]\;,
\end{equation}
and
\begin{equation}
    \label{eq: TPT2 from sims}
    \TVI(\Theta_1, \Theta_2)= \Map^2(\theta_1, \theta_2)\,\frac{1}{A}\int_A \dd[2]{\eta}\; E_A(\etavec)\,\Map^{4,3}(\theta_4, \theta_5, \theta_6; \theta_3;\etavec) + \textrm{5 Perm.} - \big[2\,\TII(\Theta_1,\Theta_2)\big]\;,
\end{equation}
where the square brackets denote the terms caused by unconnected terms in $\Map^{4,3}$ and $\Map^{4,2}$.
The term $\TVII$ cannot be obtained from the aperture mass correlation functions independently of the other covariance terms. However, using Eq.~\eqref{eq: cov_from_estimator}, the full covariance $C_{\MapMapMapEst}$ is
\begin{equation}
\label{eq:Cov in terms of Map63}
    C_{\MapMapMapEst}(\Theta_1, \Theta_2)=\frac{1}{A}\int_A \dd[2]{\eta}\; E_A(\etavec)\, \Map^{6,3}(\theta_1, \theta_2, \theta_3;\theta_4,\theta_5, \theta_6; \etavec) -\MapMapMapEst(\Theta_1)\MapMapMapEst(\Theta_2)\;,
\end{equation}
so $\TVII$ can be estimated with
\begin{equation}
    \TVII(\Theta_1, \Theta_2) = C_{\MapMapMapEst}(\Theta_1, \Theta_2) - \TI(\Theta_1, \Theta_2) -\TII(\Theta_1, \Theta_2) - \TIV(\Theta_1, \Theta_2) - \TV(\Theta_1, \Theta_2) - \TVI(\Theta_1, \Theta_2)\;.
\end{equation}

The expressions of $C_{\MapMapMapEst}(\Theta_1, \Theta_2)$ in terms of the $\Map^{n,m}$ also hold if shape noise is present. Measuring the $\Map^{n,m}$ from a convergence field with a noise component $N$ is equivalent to replacing the $\kappa$ in Eq.~\eqref{eq: Map_nm in terms of kappa} by $K_N=\kappa+N$. Then, as shown in Appendix \ref{app: shape noise}, Eq.~\eqref{eq:Cov in terms of Map63} leads to
\begin{equation}
\expval{\MapMapMapEst\, \MapMapMapEst}(\Theta_1, \Theta_2)=\frac{1}{A}\int_A \dd[2]{\eta}\; E_A(\etavec)\, \Bigg[\prod_{i=1}^3\int \dd[2]{\vartheta_i}\, U_{\theta_i}(\vartheta_i)\Bigg]\,  \Bigg[\prod_{j=4}^6\int \dd[2]{\vartheta_j}\, U_{\theta_j}(|\varthetavec_j+\etavec|)\Bigg] \expval{K_N(\varthetavec_1)\dots K_N(\varthetavec_6)}
\end{equation}
which is equivalent to $\expval{\MapMapMapEst\, \MapMapMapEst}$ with shapenoise, as given by Eq.~\eqref{eq: Map3Map3 with shapenoise}. 

\subsection{Large-field approximation}
\label{sec: Derivation subsec: Large-field approximation}
In the previous sections, we have derived six components for the covariance, which might appear analogous to each other. However, we show here that the terms $\TII$ and $\TVI$ show significantly different behaviour with survey size than the other terms. For this, we consider the case of a large survey area, for which the window function $W_A$ can be approximated as one everywhere. In this case, the geometry factor $G_A$ becomes
\begin{equation}
    \label{eq: GA for infinite survey}
    G_A(\ellvec) \rightarrow \frac{(2\pi)^2}{A} \dirac(\ellvec)\;.
\end{equation}
In this approximation, the terms $\TI, \TIV, \TV$ and $\TVII$ become
\begin{align}
\label{eq: TI for infinite fields}
	\TI^\infty(\Theta_1, \Theta_2)
	&=  \frac{1}{A}\int \frac{\dd[2]{\ell_1}}{(2\pi)^2} \int \frac{\dd[2]{\ell_2}}{(2\pi)^2} \; P(\ell_1)\,P(\ell_2)\,P(|\ellvec_1+\ellvec_2|)\,\\
 &\notag \qquad \times \left[ \tilde{u}(\ell_1\,\theta_1)\,\tilde{u}(\ell_2\,\theta_2)\,\tilde{u}(|\ellvec_1+\ellvec_2|\,\theta_3)\,\tilde{u}(\ell_1\,\theta_4)\,\tilde{u}(\ell_2\,\theta_5)\,\tilde{u}(|\ellvec_1+\ellvec_2|\,\theta_6) + \textrm{5 Perm.} \right]\;,
\end{align}
\begin{align}
\label{eq: TIV for infinite fields}
	\TIV^\infty(\Theta_1, \Theta_2)
	&= \frac{1}{A} \int \frac{\dd[2]{\ell_1}}{(2\pi)^2} \int \frac{\dd[2]{\ell_2}}{(2\pi)^2}\int \frac{\dd[2]{\ell_3}}{(2\pi)^2} \;  B(\ellvec_1, \ellvec_2)\, B(\ellvec_1, \ellvec_3)\\
	&\notag\quad \times \left[\tilde{u}(\ell_1\,\theta_1)\,\tilde{u}(\ell_2\,\theta_2)\,\tilde{u}(|\ellvec_1+\ellvec_2|\,\theta_3)\,\tilde{u}(\ell_1\,\theta_4)\,\tilde{u}(\ell_3\,\theta_5)\,\tilde{u}(|\ellvec_1+\ellvec_3|\,\theta_6) + \textrm{8 Perm.}\right]\;,
\end{align}
\begin{align}
\label{eq: TV for infinite fields}
	\TV^\infty(\Theta_1, \Theta_2)
	&=\frac{1}{A}\int \frac{\dd[2]{\ell_1}}{(2\pi)^2}\int \frac{\dd[2]{\ell_2}}{(2\pi)^2}\int \frac{\dd[2]{\ell_3}}{(2\pi)^2}\; P(\ell_1)\, T(\ellvec_2, -\ellvec_1-\ellvec_2, \ellvec_3)\\
	&\notag\quad\times \left[ \tilde{u}(\ell_1\,\theta_1)\,\tilde{u}(\ell_2\,\theta_2)\,\tilde{u}(|\ellvec_1+\ellvec_2|\,\theta_3)\,\tilde{u}(\ell_1\,\theta_4)\,\tilde{u}(\ell_3\,\theta_5)\,\tilde{u}(|\ellvec_3-\ellvec_1|\,\theta_6) +  \textrm{8 Perm.}\right]\;,
\end{align}
and
\begin{align}
\label{eq: TVII for infinite fields}
	\TVII^\infty(\Theta_1, \Theta_2)
&= \frac{1}{A} \int\frac{\dd[2]{\ell_1}}{(2\pi)^2}\, \int\frac{\dd[2]{\ell_2}}{(2\pi)^2}  \int\frac{\dd[2]{\ell_3}}{(2\pi)^2} \int\frac{\dd[2]{\ell_4}}{(2\pi)^2}\;P_6(\ellvec_1, \ellvec_2, -\ellvec_1-\ellvec_2, \ellvec_3, \ellvec_4)\,\\
&\notag\quad\times  \tilde{u}(\ell_1\,\theta_1)\,\tilde{u}(\ell_2\,\theta_2)\,\tilde{u}(|\ellvec_1+\ellvec_2|\,\theta_3)\,\tilde{u}(\ell_3\,\theta_4)\,\tilde{u}(\ell_4\,\theta_5)\,\tilde{u}(|\ellvec_3+\ellvec_4|\,\theta_6)\;.
\end{align}
These terms scale with the inverse survey area, as is commonly expected for covariance. The terms $\TII$ and $\TVI$, though, behave differently. Applying the approximation leads to
\begin{align}
\label{eq: TII for infinite fields}
	\TII^\infty(\Theta_1, \Theta_2)
	&=  \frac{1}{A}\int \frac{\dd[2]{\ell_1}}{(2\pi)^2} \int \frac{\dd[2]{\ell_3}}{(2\pi)^2}\; P(\ell_1)\,P(0)\,P(\ell_3)\, \left[ \tilde{u}(\ell_1\,\theta_1)\,\tilde{u}(\ell_1\,\theta_2)\,\tilde{u}(0)\,\tilde{u}(0)\,\tilde{u}(\ell_3\,\theta_5)\,\tilde{u}(\ell_3\,\theta_6) + \textrm{8 Perm.}\right] = 0\;,
\end{align}
and
\begin{align}
\label{eq: TVI for infinite fields}
	\TVI^\infty(\Theta_1, \Theta_2)
	&= \frac{1}{A} \int \frac{\dd[2]{\ell_1}}{(2\pi)^2}\int \frac{\dd[2]{\ell_2}}{(2\pi)^2}\int \frac{\dd[2]{\ell_3}}{(2\pi)^2}\; P(\ell_1)\, T(0, \ellvec_2, \ellvec_3)\, \big[ \tilde{u}(\ell_1\,\theta_1)\,\tilde{u}(\ell_1\,\theta_2)\,\\
 &\notag\quad \times \tilde{u}(0)\,\tilde{u}(\ell_2\,\theta_4)\,\tilde{u}(\ell_3\,\theta_5)\,\tilde{u}(|\ellvec_2+\ellvec_3|\,\theta_6) + \textrm{5 Perm.} \big] = 0\;.
\end{align}
Both terms vanish because the filter function $\tilde{u}(0) = 0$ for a compensated filter. The same can be observed when writing $\TII$ in terms of the aperture mass correlation function (Eqs.~\ref{eq: TPPP2 from sims}). The large-field approximation implies $E(\etavec)=1$, so for a square survey with sidelength $\vartheta_\mathrm{max}$
\begin{align}
    \TII^\infty(\Theta_1, \Theta_2)=\Map^2(\theta_1, \theta_2)\, \Map^2(\theta_5, \theta_6)\, \frac{1}{A}\int_A \dd[2]{\eta} \Map^{2,1}(\theta_3; \theta_4;\etavec) +\textrm{ 8 Perm.}\;,
\end{align}
Since $\Map^{2,1}$ is a correlation function of a quantity with vanishing mean, the integral over $\etavec$ vanishes as $A\rightarrow \realspace^2$ and $\TII$ declines to zero faster than $1/A$. The same argument can be made for $\TVI$, which depends on the integral of the correlation function $\Map^{4,3}$. Consequently, these terms are directly connected to the finiteness of the survey area, which is why we refer to them as `finite-field terms' of the covariance. 

As shown in Eq.~\eqref{eq: G_A square}, the geometry factor $G_A$ is typically a strongly oscillating function. Consequently, integrals over this function are difficult to evaluate numerically. Therefore, in the following, we use the approximations $\TI^\infty$, $\TIV^\infty$ and $\TV^\infty$ unless otherwise noted. However, as we will show in Sect.~\ref{sec: MCMC} using $\TVII^\infty$ instead of $\TVII$ neglects a significant part of the covariance. As shown in Appendix \ref{app: TVII approximation}, $\TVII$ can be approximated as
\begin{equation}
    \TVII(\Theta_1, \Theta_2) \simeq \TVII^\infty(\Theta_1, \Theta_2) + T_{P_6,\mathrm{2h}}(\Theta_1, \Theta_2)\,,
\end{equation}
with $T_{P_6,\mathrm{2h}}$ given by Eq.~\eqref{eq:SSC_term}. The term $T_{P_6,\mathrm{2h}}$ is similar to the finite-field terms $\TII$ and $\TVI$ since it does not scale inversely with survey area and vanishes under the large-field approximation.

The approximations $\TI^\infty$,$\TIV^\infty$, and $\TV^\infty$ also neglect part of the covariance. However, since these terms are sub-dominant for the total covariance, as we will show in Sect.~\ref{sec: Validation}, we deem the approximation appropriate for these terms.

\section{\texorpdfstring{Attempt at alternative derivation of Gaussian $\MapMapMapEst$ covariance from bispectrum covariance}{Attempt at alternative derivation of Gaussian covariance from bispectrum covariance}}
\label{sec: Derivation From Bispec}
An alternative approach to finding an analytic covariance estimate of a real-space observable is via the covariance of the corresponding Fourier space quantity. As the $\MapMapMap$ are related to the bispectrum $B$ via Eq.~\eqref{eq: Map3 from Bispec}, it might appear natural to derive the covariance of $\MapMapMap$ from the covariance of $B$. However, this is impossible for finite survey areas without using the large-field approximation, even for Gaussian fields. To see this, we rewrite Eq.~\eqref{eq: Map3 from Bispec} as
\begin{equation}
   \MapMapMap(\theta_1, \theta_2, \theta_3) =  (2\pi)^2 \left[\prod_{i=1}^3  \int  \frac{\dd[2]{\ell_i}}{(2\pi)^2}\, \tilde{u}(\ell_i\, \theta_i) \right]\, \mathcal{P}_3(\ellvec_1, \ellvec_2, \ellvec_3) \, \dirac(\ellvec_1+\ellvec_2+\ellvec_3)\;,
\end{equation}
where we remind that $\mathcal{P}_3(\ellvec_1, \ellvec_2, \ellvec_3) = B(\ellvec_1, \ellvec_2)$. One could then assume that the covariance $C_{\MapMapMap}$ were given by
\begin{equation}
\label{eq: Cb-->Cmap2}
    C_{\MapMapMapEst}(\Theta_1, \Theta_2) =  (2\pi)^4  \left[\prod_{i=1}^6 \int \frac{\dd[2]{\ell_i}}{(2\pi)^2}\, \tilde{u}(\ell_i\, \theta_i) \right]\, C_B(\ellvec_1, \ellvec_2, \ellvec_3, \ellvec_4, \ellvec_5, \ellvec_6) \, \dirac(\ellvec_1+\ellvec_2+\ellvec_3)\, \dirac(\ellvec_4+\ellvec_5+\ellvec_6)\;,
\end{equation}
where $C_B$ is the covariance of the bispectrum. Then, it should be possible to infer $C_B$ from $C_{\MapMapMapEst}$. However, the Gaussian part of $C_{\MapMapMapEst}$ is
\begin{align} 
\label{eq: Gaussian Cov}
    \TI(\Theta_1, \Theta_2) + \TII(\Theta_1, \Theta_2) &=  (2\pi)^6  \left[\prod_{i=1}^6\int  \frac{\dd[2]{\ell_i}}{(2\pi)^2}\, \tilde{u}(\ell_i\, \theta_i) \right]\, P(\ell_1)\, P(\ell_2)\, P(\ell_3)\\
    &\notag \quad\times \Big\{[G_A(\ellvec_1+\ellvec_2+\ellvec_3)\,\dirac(\ellvec_1+\ellvec_4)\,\dirac(\ellvec_2+\ellvec_5)\,\dirac(\ellvec_3+\ellvec_6) + \textrm{5 Perm.} ]\\
    &\notag \qquad+ 
 [G_A(\ellvec_2)\,\dirac(\ellvec_1+\ellvec_2)\,\dirac(\ellvec_3+\ellvec_4)\,\dirac(\ellvec_5+\ellvec_6) + \textrm{8 Perm.}]\Big\}\;.
\end{align}
For Eq.~\eqref{eq: Cb-->Cmap2} to hold true, we need the expression in braces to equal $f(\ellvec_1, \ellvec_2, \ellvec_3, \ellvec_4, \ellvec_5, \ellvec_6)\,\dirac(\ellvec_1+\ellvec_2+\ellvec_3)\, \dirac(\ellvec_4+\ellvec_5+\ellvec_6)$ for some function $f$. However, this is impossible unless $G_A$ is proportional to a Dirac function. Therefore, Eq.~\eqref{eq: Cb-->Cmap2} cannot hold, and $C_{\MapMapMap}$ and $C_B$ are not related by simple integration.

Under the large-field approximation, though, Eq.~\eqref{eq: Cb-->Cmap2} can be used. We show this in Appendix~\ref{app: bispec cov}, where we start from the Gaussian bispectrum covariance from \citet{Joachimi2009} to recover $\TI^\infty$. 

\section{Validation of model covariance}
\label{sec: Validation}
We test the analytical model for $C_{\MapMapMapEst}$ by comparing it to numerical estimates from mock data. This section describes the used data, measurement method, and results.
\subsection{Validation data}
\label{sec: Validation subsec: Data}
We use two kinds of mock data: Shear maps for Gaussian density distributions and realistic shear and convergence maps from two cosmological $N$-body simulations, namely the Scinet LIghtcone Simulations (SLICS, \citealp{HarnoisDeraps2015}) and the suite by \citet[][T17 hereafter]{Takahashi2017}. The data from $N$-body simulations are the same as for \citetalias{Heydenreich2022}, which we describe here again for convenience. 

\subsubsection{SLICS}
\label{sec: Validation subsec: Data subsubsec: SLICS}
The SLICS contain $1536^3$ particles inside a $505\, h^{-1}\,$Mpc box. They are gravitationally evolved according to a flat $\Lambda$CDM cosmology with normalised Hubble constant $h=0.69$, matter clustering parameter $\sigma_8=0.83$, mater density parameter $\Omm=0.29$, baryon density parameter $\Omega_\mathrm{b}=0.047$, and primordial power spectrum spectral index  $n_\mathrm{s}=0.969$. We use two different sets of data products from these simulations. The first consists of mock shear catalogues from $924$  pseudo-independent lines of sight, each with a square area of $100\,\mathrm{deg}^2$. The source galaxies are distributed with a redshift distribution of
\begin{equation}
    n(z) \propto z^2\, \E^{-\left(z/z_0\right)^\beta}\;,
\end{equation}
with $z_0=0.637$, $\beta=1.5$ and normalisation such that the overall galaxy density is $30\, \mathrm{arcmin}^{-2}$. This redshift distribution and number density correspond to expectations for stage IV surveys.  The shear catalogues are infused with shape noise by adding random ellipticities from a Gaussian. We use a two-component ellipticity dispersion of $\sigma^2_\epsilon=(0.37)^2$.

The second set of data products from the SLICS are shape noise-free convergence maps. These convergence maps have the same area and cosmology as the shear lines of sight, but they were obtained with all source galaxies situated at redshift $z_\mathrm{s}=1$. 

\subsubsection{T17 Simulations}
\label{sec: Validation subsec: Data subsubsec: T17}
The T17 simulations are constructed from a series of boxes of side lengths $L$, $2L$, $3L$, with $L=450\,h^{-1}\,\mathrm{Mpc}$. The $2048^3$ particles in the box are evolved using GADGET2 \citep{Springel2001} and a flat $\Lambda$CDM cosmology with parameters $h=0.7$, $\sigma_8=0.82$, $\Omm=0.279$, $\Omega_\mathrm{b}=0.046$, and $n_\mathrm{s}=0.97$. Using \verb|GRayTrix|\footnote{\url{http://th.nao.ac.jp/MEMBER/hamanatk/GRayTrix/}}, light rays are traced through 108 realisations of the simulation, creating full-sky convergence shells at 38 redshifts.

\begin{figure}
\includegraphics[width=\columnwidth]{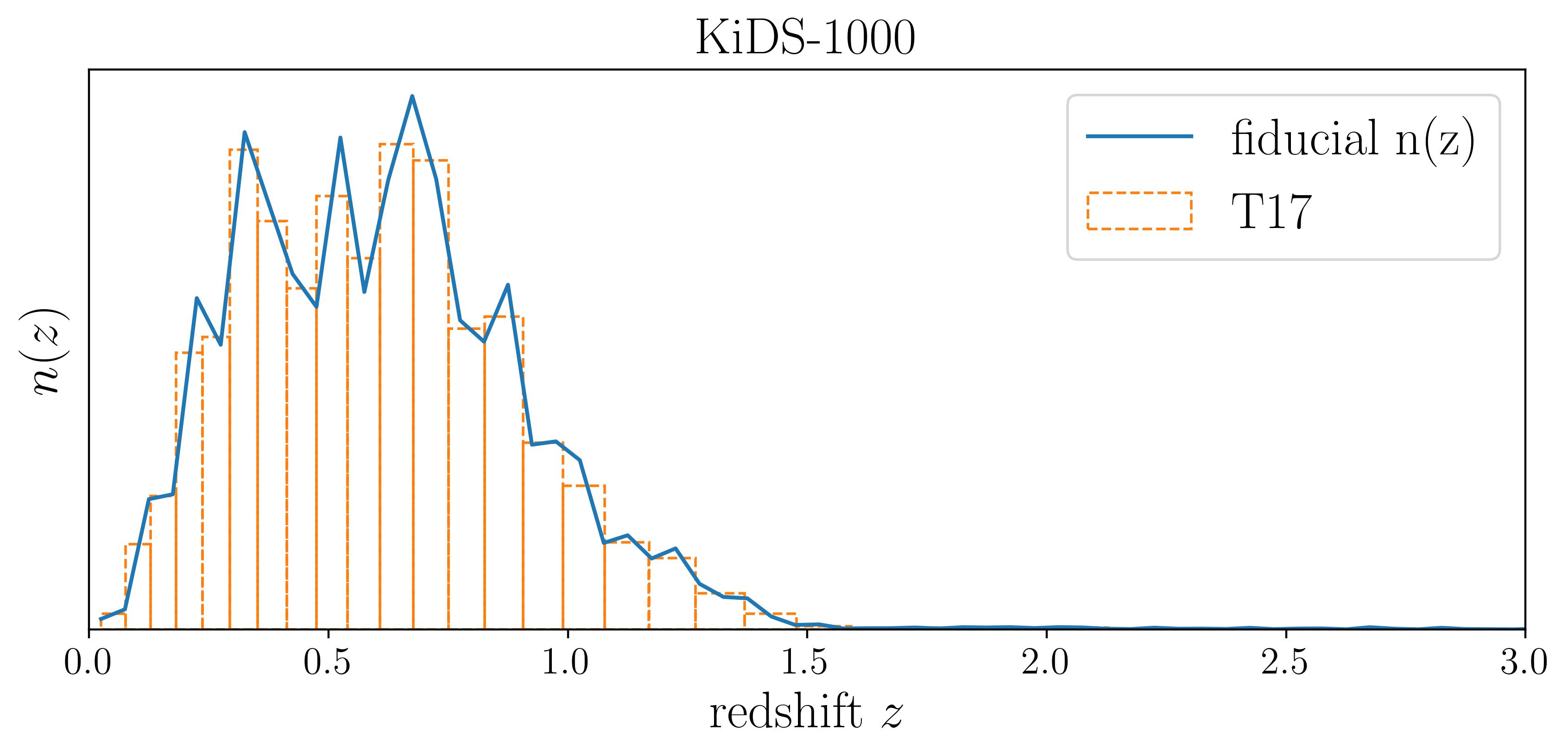}
\caption{Redshift distribution constructed from the T17 simulation given the fiducial $n(z)$ of the KiDS-1000 data.}
\label{fig:nofz}
\end{figure}

We create mock data similar to the KiDS-1000 data release of the Kilo Degree Survey (KiDS, \citealp{Kuijken2015}) from these convergence shells by cutting out an area of 859.4\,deg$^2$ from the convergence shells, combining them according to the KiDS-1000 $n(z)$ shown in Fig.~\ref{fig:nofz} \citep{Hildebrandt2021}, and adding a realistic shape noise with a standard deviation
\begin{equation}
    \sigma=\frac{\sigma_\epsilon}{\sqrt{2 n_\mathrm{gal}\, A_\mathrm{pix}}}\;,
\end{equation}
with the pixel area $A_\mathrm{pix}$, $n_\mathrm{gal}=6.17\, \mathrm{arcmin}^{-2}$, and $\sigma_\epsilon=0.375$ \citep{Giblin2021}. 

\subsubsection{Gaussian random fields (GRFs)}
\label{sec: Validation subsec: Data subsubsec: GRF}
We also generate shear maps for Gaussian density distributions, characterised solely by their power spectrum. We use a shape-noise-free cosmic shear power spectrum based on the non-linear matter power spectrum from the revised \verb|halofit| prescription \citep{Takahashi2012}. We use the cosmological parameters of the T17 simulations and the KiDS-1000 $n(z)$.

From the power spectrum, we create square maps of the shear $\gamma$ using the python library \texttt{lenstools} \citep{Petri2016} with the assumption of periodic boundary conditions. We create $4096$ realisations for each of the sidelengths $\varthetamax' \in \lbrace \astroang{10;;}, \astroang{15;;}, \astroang{20;;} \rbrace$. Each map contains $\Npix = 4096\times4096$ pixels.

\subsection{Covariance estimation from data}
\label{sec: Validation subsec: Measurement}
We estimate the covariance of $\MapMapMapEst$ in three ways from the validation data.  In the first method, we measure the aperture statistics $\MapMapMap$ on all simulation realisations using FFT and then use the sample covariance as an estimate. In the second method, we measure the third-order shear correlation functions $\Gamma_i$ and convert them to estimates of $\MapMapMap$ for all simulation realisations. Finally, in the third method, we measure the aperture mass correlation functions and convert them to the terms $\TI$ to $\TVII$ (see Sect.~\ref{sec: Derivation subsec: Correlation Functions}).

We choose the same aperture filter scales as used in \citetalias{Heydenreich2022} between \astroang{;4;} and \astroang{;32;}. There, we showed that $\MapMapMap$ at these scales contains cosmological information that is complementary to the information content of second-order statistics. Our choice of scales also is impacted by two practical concerns: At aperture filters larger than \astroang{;32;} the signal-to-noise for $\MapMapMap$ becomes small, and at scales smaller than \astroang{;4;}, the $\MapMapMap$-model disagrees with simulations (see Fig. 8 in \citetalias{Heydenreich2022}).

\subsubsection{Using FFT}
\label{sec: Validation subsec: Measurement subsubsec: FFT}
The measurement of $\MapMapMap$ in the simulations uses the same setup as the method described in \citetalias{Heydenreich2022}. We will summarise it here shortly. For the flat GRFs and the SLICS, we measure $C_{\MapMapMapEst}$ in the shear maps in three steps. 

At first, we measure ${\Map}_i(\varthetavec, \theta)$ for each realisation $i$, using scale radii $\theta \in \lbrace \astroang{;4;}, \astroang{;8;}, \astroang{;16;} \rbrace$ for the SLICS and  $\theta \in \lbrace \astroang{;4;}, \astroang{;8;}, \astroang{;16;}, \astroang{;32;} \rbrace$ for the GRFs. We use Eq.~\eqref{eq: map_from_gamma_via_convolution}, which defines aperture mass maps as convolutions of the filter function $Q$ and the tangential shear. We evaluate these convolutions using the convolution theorem, by first Fourier transforming $Q$ and $\gamma_\mathrm{t}$ with FFT, then multiplying them and inverse Fourier transforming the product. To remove edge effects, we cut off a border of 4 times the largest aperture radius, that is $4\times\astroang{;32;}= \astroang{;128;}$ for the GRFs and $4\times\astroang{;16;}=\astroang{;64;}$ for the SLICS, from each side of the ${\Map}_i$-maps. This large cut-off is needed because our filter function $Q_\theta$ has infinite support and is significantly non-zero even for $\vartheta>\theta$. The filter contains $99.9\%$ of its power within $4\theta$, so with the cut-off, boundary effects are negligible. The resulting ${\Map}_i$-maps have side lengths $ \varthetamax \in \lbrace\astroang{5.73;;}, \astroang{10.73;;}, \astroang{15.73;;}\rbrace$ for the GRFs and $\astroang{7.87;;}$ for the SLICS. 

Then, we estimate $\MapMapMap_i$ for each realisation $i$. For this, we multiply ${\Map}_i(\varthetavec, \theta_1)$, ${\Map}_i(\varthetavec, \theta_2)$, and ${\Map}_i(\varthetavec, \theta_3)$ and then average over $\varthetavec$.  Finally, we estimate the measured covariance $C^\mathrm{meas}_{\MapMapMapEst}$ with
\begin{align}
\label{eq: Cov measurement}
   C^\mathrm{sim}_{\MapMapMapEst}(\Theta_1, \Theta_2) &= \frac{1}{N-1} \sum_{i=1}^N \MapMapMap_i(\theta_1, \theta_2, \theta_3) \MapMapMap_i(\theta_4, \theta_5, \theta_6)\\
   &\notag \quad-\frac{1}{N (N-1)} \sum_{i=1}^N \MapMapMap_i(\theta_1, \theta_2, \theta_3) \sum_{j=1}^N \MapMapMap_j(\theta_4, \theta_5, \theta_6)\;.
\end{align}

For the curved-sky T17 convergence maps, we smooth the maps with the \texttt{healpy} function \texttt{smoothing} with a beam window function determined by the corresponding $U_\theta$ filter for the same aperture radii as for the GRFs, which are $\theta \in \lbrace \astroang{;4;}, \astroang{;8;}, \astroang{;16;}, \astroang{;32;} \rbrace$. Each filter yields a full-sky aperture mass map ${\Map}_i$, from which we extracted 18 tiles that are not adjacent. This gives in total 1944 almost independent realisations.

We compute uncertainty estimates with bootstrapping for both the SLICS and the T17 covariances from FFT. For this, we generate $10\,000$ vectors containing $N_\mathrm{real}$ randomly drawn integers between 1 and $N_\mathrm{real}$ where $N_\mathrm{real}$ is the number of realisations (924 for the SLICS and 1944 for the T17). The vectors can explicitly contain the same integer multiple times. For each vector, we take the $\MapMapMap_i$ for which $i$ is an element in the vector and use it to calculate a sample covariance with Eq.~\eqref{eq: Cov measurement}. In this way, we obtain $10\,000$ covariance estimates. We take the standard deviation of these estimates as uncertainty for the $C_{\MapMapMapEst}^\mathrm{sim}$.

\subsubsection{Using third-order shear correlation functions}
\label{sec: Validation subsec: Measurement subsubsec: Shear Correlation Functions}
As mentioned in Sect.~\ref{sec: Theory}, $\MapMapMap$ is related to the natural components $\Gamma_i$ of the third-order shear correlation function \citepalias[see also Section 5.3 in][]{Heydenreich2022}. Therefore, we can estimate the $\MapMapMap_i$ of each simulation realisation $i$ by first measuring the $\Gamma_i$ and then converting them to $\MapMapMap_i$, following Equations (69) and (71) of \citet{Schneider2005}. The covariance of $\MapMapMapEst$ can then be estimated as the sample variance of the $\MapMapMap_i$, using Eq.~\eqref{eq: Cov measurement}. 

We expect the covariance estimate from this approach to be smaller than the estimate from the FFT-based method described in the previous subsection. This is because, here, we do not cut off the border of the survey area. Consequently, more shear information is used to estimate $\MapMapMap$, leading to a smaller covariance.

The measurement of the $\Gamma_i$ is very time-intensive. Therefore, we restrict this approach to the SLICS data, where we measure the $\Gamma_i$ using \verb|treecorr| \citep{Jarvis2004}. We use ten bins each for the triangle parameters $u, v,$ and $r$ with $r$ between $\astroang{;0.1;}$ and $\astroang{;120;}$. As shown in \citetalias{Heydenreich2022}, binning the $\Gamma_i$ in this way yields accurate $\MapMapMap$ for scale radii between \astroang{;4;} and \astroang{;16;}. 

\subsubsection{Using aperture mass correlation functions}
\label{sec: Validation subsec: Measurement subsubsec: Aperture Mass Correlation Functions}
As shown in Sect.~\ref{sec: Derivation subsec: Correlation Functions}, the individual terms of the covariance can be estimated from the aperture mass correlation functions, which we use to validate the analytic expressions. 

For this, we measure the aperture mass maps ${\Map}_i(\varthetavec, \theta)$ according to Sect.~\ref{sec: Validation subsec: Measurement subsubsec: FFT}. From these maps, we first calculate the ${\Map}_i(\varthetavec, \theta_1){\Map}_i(\varthetavec, \theta_2)$ and ${\Map}_i(\varthetavec, \theta_1){\Map}_i(\varthetavec, \theta_2){\Map}_i(\varthetavec, \theta_3)$ by multiplying the aperture mass maps for different filter radii. Then, we estimate the correlation functions $\Map^{2,1}(\theta_1;\theta_2; \eta)$, $\Map^{3,1}(\theta_1,\theta_2; \theta_3; \eta)$, $\Map^{4,3}(\theta_1,\theta_2,\theta_3; \theta_4; \eta)$, and $\Map^{4,2}(\theta_1,\theta_2; \theta_3, \theta_4; \eta)$ by correlating the relevant aperture mass maps: We first extend the fields ${\Map}_i(\varthetavec, \theta)$ using zero-padding, meaning that we add pixel grids containing only zeros to the boundary of the map. Doing this, ${\Map}_i(\varthetavec, \theta)$ becomes $W_A(\varthetavec)\,{\Map}_i(\varthetavec, \theta)$. The correlation between two aperture mass maps then yields
\begin{equation}
    \int_A\dd[2]\vartheta\; W_A(\varthetavec)\,{\Map}_i(\varthetavec, \theta_1)\,W_A(\varthetavec+\etavec)\,{\Map}_i(\varthetavec+\etavec, \theta_2) = E(\etavec)\,\Map^{2}(\theta_1,\theta_2;\etavec) \; .
\end{equation}
We perform this correlation using the \verb|correlate2D| function from \verb|scipy|. Finally, we calculate the terms $\TI$ to $\TVII$ with the expressions in Sect.~\ref{sec: Derivation subsec: Correlation Functions}.

Our method to measure the correlation function uses the flat-sky approximation, so we did not apply it to the T17 simulations. Furthermore, the correlation function estimation of the terms $\TII$ and $\TVI$ is strongly affected by noise, which is why we only use it for the SLICS without shape noise.

\subsection{Results}
\label{sec: Validation subsec: Results}

\subsubsection{Gaussian random fields}
\label{sec: Validation subsec: Results subsubsec: GRF}
\begin{figure*}
\begin{minipage}[c]{12cm}
    \includegraphics[width=\linewidth, trim={1cm 1.5cm 1.5cm 2cm}, clip]{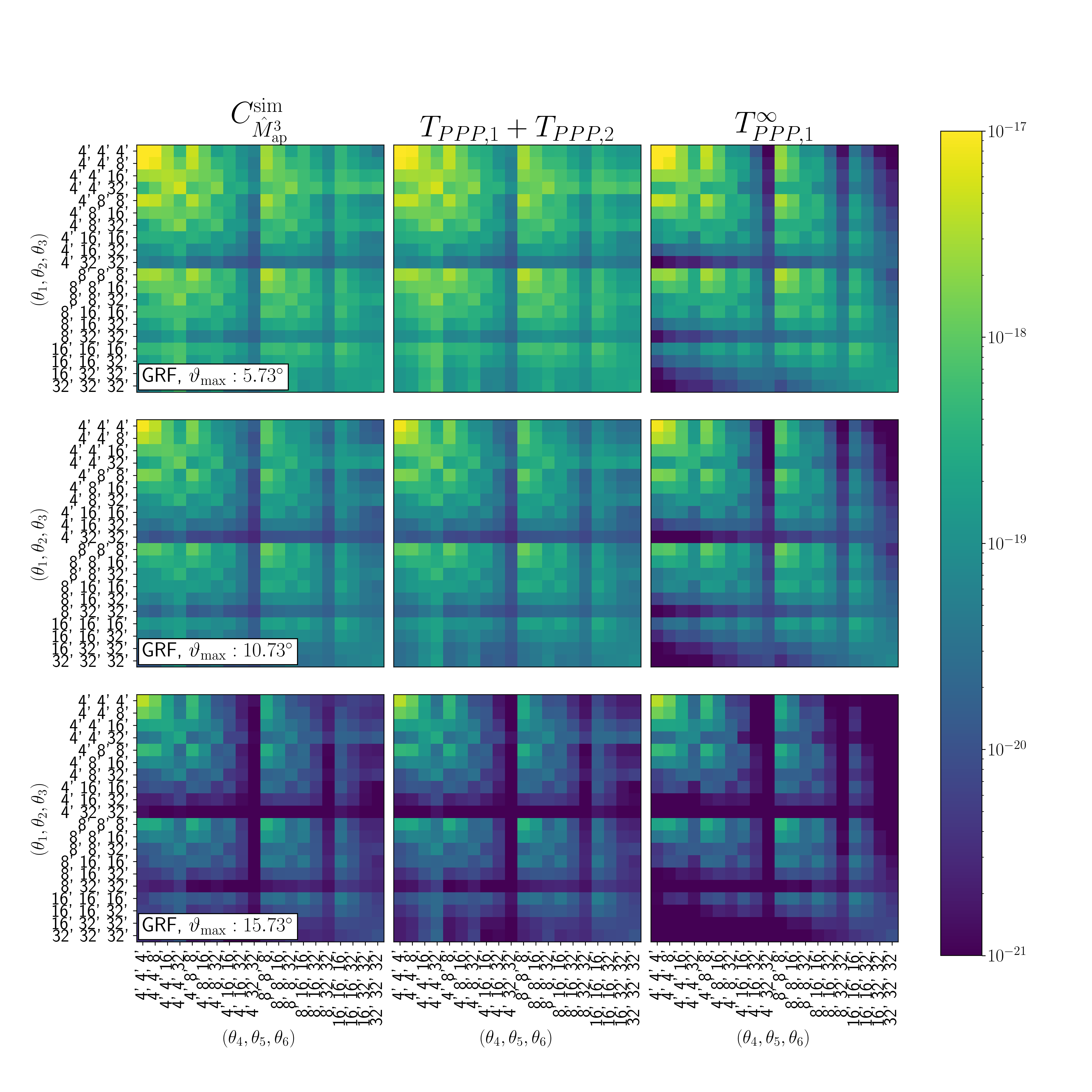}
\end{minipage}
\begin{minipage}[t]{\linewidth-12.5cm}
\caption{$C_{\MapMapMapEst}$ for the GRF, each row is showing a different field size. In the left column are the measured $C_{\MapMapMapEst}$ from the simulated GRF. The middle column shows the model prediction, including the finite-field term $\TII$. The covariances under the large-field approximation, for which $\TII$ vanishes,  are in the right column. }
    \label{fig: GRF_values}     
\end{minipage}
\end{figure*}

We compare the covariance for $\MapMapMap$ measured in the simulated GRFs with the analytical expression $\TI+\TII$ and the large-field approximation $\TI^\infty$ in Fig.~\ref{fig: GRF_values}. For the GRFs, all terms containing the bi-, tri-, or pentaspectrum vanish, so the middle column gives the full analytical prediction for the covariance. We see that, as expected, both simulated and analytical covariance decrease with increasing survey size. The full analytic covariance shows similar values to the simulated covariance for all combinations of aperture radii. The large-field approximation agrees only on and near the diagonal with the simulation and is much smaller for elements far from the diagonal.  

\begin{figure*}
\begin{minipage}[c]{12cm}
    \includegraphics[width=\linewidth, trim={1cm 1.5cm 1.5cm 2cm}, clip]{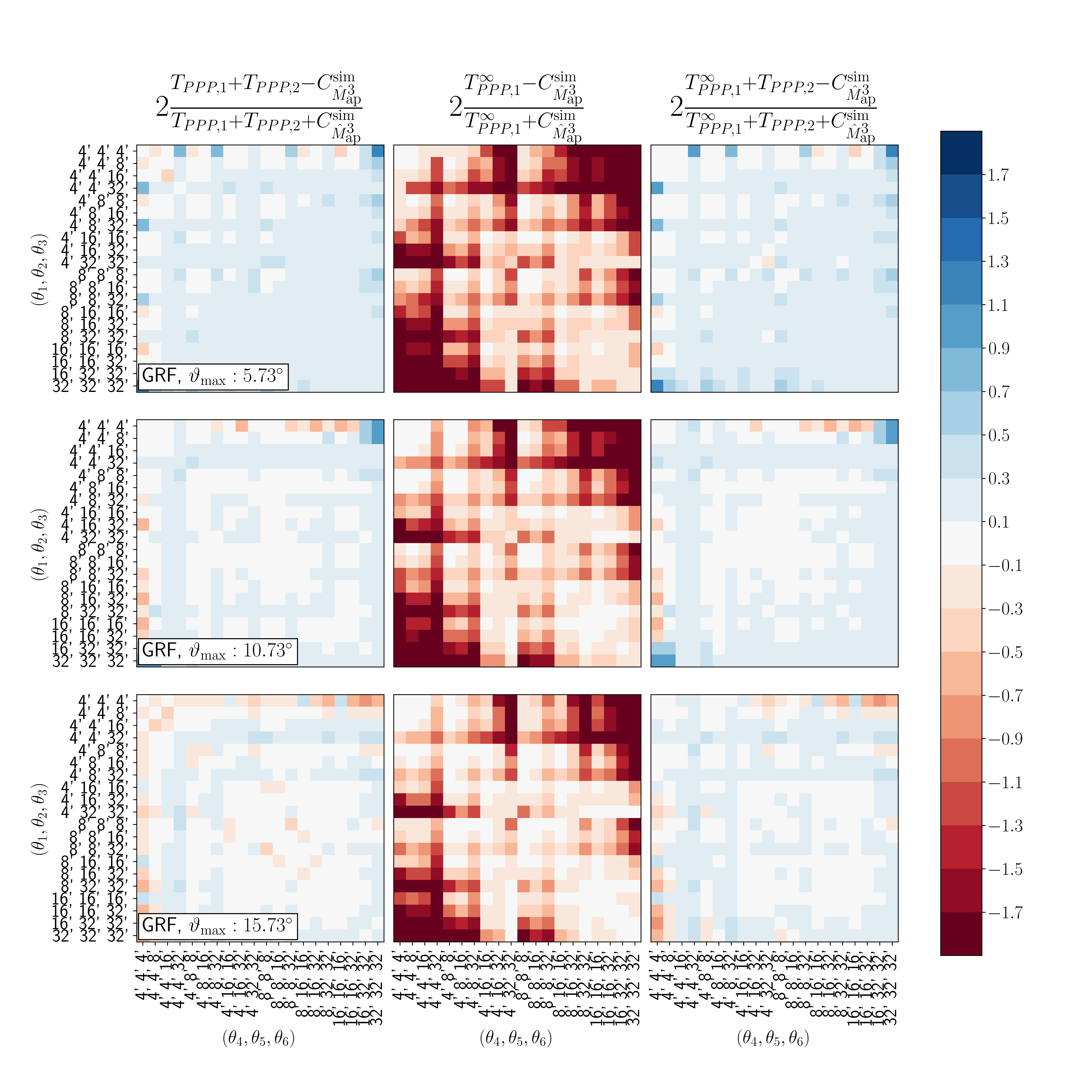}
\end{minipage}
\begin{minipage}[t]{\textwidth-12.5cm}
\caption{Fractional differences of the model covariance to the measured $C_{\MapMapMapEst}$ for the GRF, each row showing a different field size. In the left column are the fractional differences between the full analytic model and the simulated GRF. The middle column shows the difference between the large-field approximation, neglecting the finite field term, and the simulated covariance. In the right column, the difference of the model using the large-field approximation for $\TI$, but keeping $\TII$ as it is, is shown. }
    \label{fig: GRF_fracErrs}   
\end{minipage}
\end{figure*}

This observation is quantified in Fig.~\ref{fig: GRF_fracErrs}, where we show the fractional differences between the simulated and analytically calculated covariances. The full analytical expression shows deviations of less than 30\% to the simulation, while the large-field approximation is almost a factor of two too small for the non-diagonal elements. While the deviations decrease with field size, they are still significant for the largest side length of $\astroang{15.73}$.  Consequently, neglecting $\TII$ is inappropriate, even for surveys of a size of $(\astroang{15.73})^2 = 247.43\, \mathrm{deg}^2$.  
 
\begin{figure*}
    \centering
    \includegraphics[width=\linewidth, trim={1cm 0 3cm 0cm}, clip]{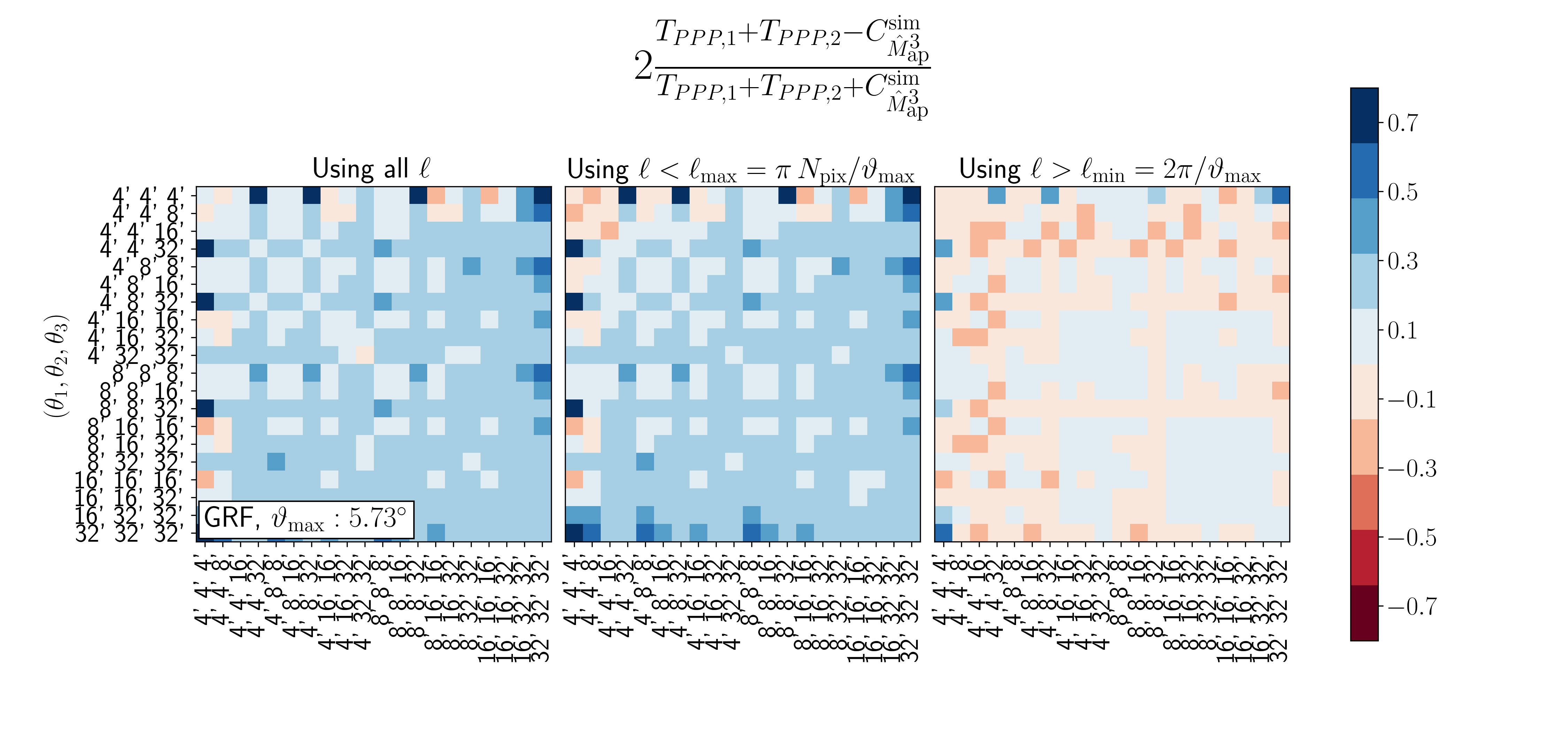}
    \caption{Fractional differences between covariance model and estimate from GRFs with $\vartheta_\mathrm{max}=\astroang{5.78;;}$ for different choices of the boundaries of the $\ell$-integrals. Left: Fiducial calculation of model, where $P(\ell)$ is used for all $\ell$. Middle: $\ell$-modes larger than $\ell_\mathrm{max}$ are cut-off. This removes modes not present in the simulation due to the finite pixel size. Right: $\ell$-modes smaller than $\ell_\mathrm{min}$ are cut-off. This removes modes not present in the simulation due to the finite field size.}
   \label{fig: l boundaries}
\end{figure*}

The remaining disagreement between the full analytic covariance model and the simulations can be explained by the discreteness of the shear maps. Due to the finite pixel number, the shear maps contain only modes which are $\ell=2\pi\sqrt{i^2+j^2}/\vartheta'_\mathrm{max}$, where $i$ and $j$ are integers between 0 and $N_\mathrm{pix}/2$, where $N_\mathrm{pix}$ is the pixel number and $\vartheta'_\mathrm{max}$ is the side length of the shear maps (before boundary cut-off). Therefore, while the model assumes a smooth power spectrum $P(\ell)$, the actual power spectrum of the GRFs is
\begin{equation}
    P^\mathrm{GRF}(\ellvec)=\left(\frac{2\pi}{\vartheta'_\mathrm{max}}\right)^2\,\sum_{i=1}^{N_\mathrm{pix}/2}\sum_{j=1}^{N_\mathrm{pix}/2} P\left(\frac{2\pi\sqrt{i^2+j^2}}{\vartheta'_\mathrm{max}}\right)\, \dirac\left(\ell_1-\frac{2\pi i}{\vartheta'_\mathrm{max}}\right)\, \dirac\left(\ell_2-\frac{2\pi j}{\vartheta'_\mathrm{max}}\right)\;,
\end{equation}
with $\ellvec=(\ell_1, \ell_2)$. Due to the large pixel number, it is not feasible to directly calculate $\TI$ and $\TII$ for this power spectrum. However, using $P^\mathrm{GRF}$ is similar to assuming that $P$ is zero for $\ell$ outside of $[\ell_\mathrm{min},\ell_\mathrm{max}]$, with $\ell_\mathrm{min}=2\pi/\vartheta'_\mathrm{max}$, and $\ell_\mathrm{max}=\pi \, N_\mathrm{pix}/\vartheta'_\mathrm{max}$. This is the same as increasing the lower and decreasing the upper integration boundary for the $\ell$-integrals in $\TI$ and $\TII$. We show the impact of these integration borders in Fig.~\ref{fig: l boundaries}. The upper boundary $\ell_\mathrm{max}$, and consequently the pixel number, has only a small impact on the model covariance. This is because the aperture filters $\hat{u}$ decrease with $\ell$ and are already small at $\ell\simeq \ell_\mathrm{max}$. The lower boundary $\ell_\mathrm{min}$ has a stronger impact. By increasing the lower integration border to $\ell_\mathrm{min}$, we decrease the model covariance. This decrease significantly improves the agreement between the simulated GRFs and the model.

As the field size (and $\vartheta'_\mathrm{max}$) increases, $\ell_\mathrm{min}$ decreases so that smaller $\ell$-modes are included in the GRFs. Therefore, the difference between the model, evaluated for all $\ell$, and the covariance from the GRFs shrinks with increasing field size, as seen in the right column of Fig.~\ref{fig: GRF_fracErrs}. 

\begin{figure*}
    \centering
    \includegraphics[width=\linewidth, trim={1cm 0 7cm 1cm}, clip]{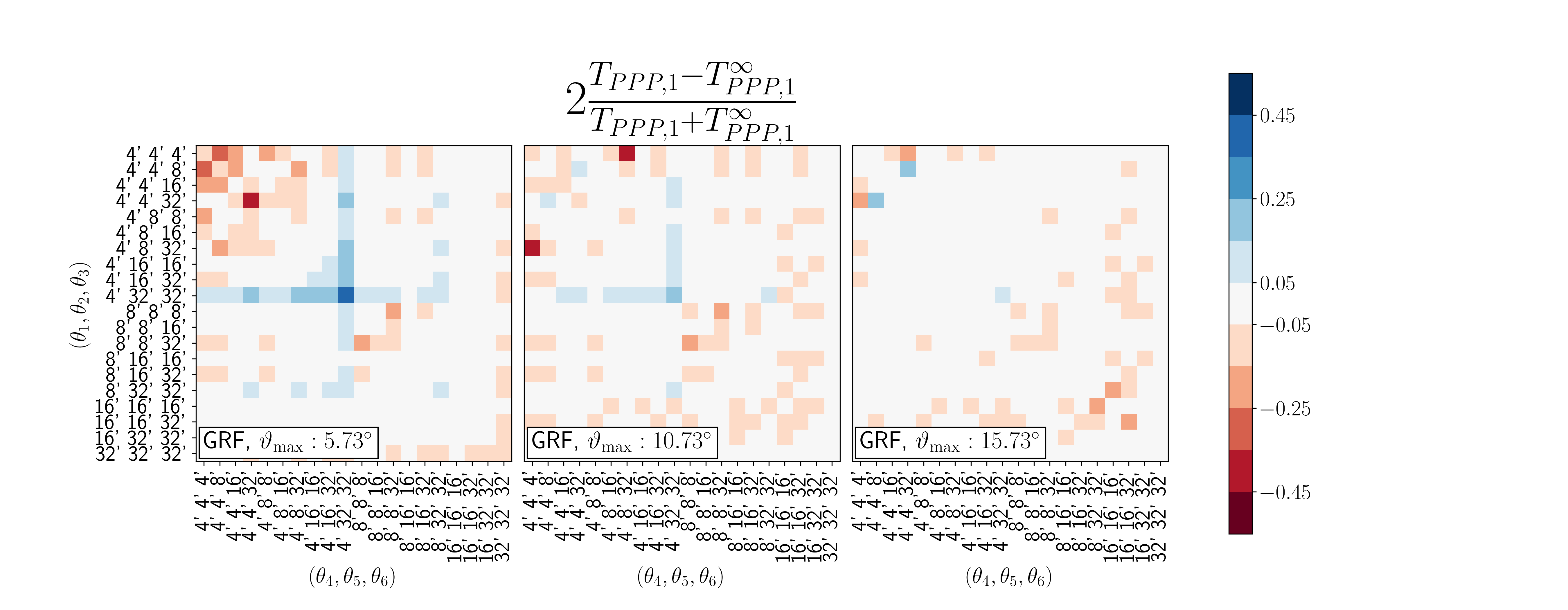}
    \caption{Fractional differences of $\TI$ with and without large-field approximation for different field sizes}
    \label{fig: T1 vs T1 inf}
\end{figure*}

We also show in Fig.~\ref{fig: GRF_fracErrs} the fractional difference between $\TI^\infty + \TII$ and the simulated covariance. This approximated analytic expression shows a similar level of agreement with the simulated covariance as the full expression. The cause of this good agreement is the similarity between $\TI$ and $\TI^\infty$, shown in Fig.~\ref{fig: T1 vs T1 inf}. Even for the smallest field size with side length $\vartheta_\mathrm{max}=\astroang{5.73}$, most elements of $\TI$ and $\TI^\infty$ differ by less than 5\%, with stronger deviations mostly at very large aperture radii ($\astroang{;32;}$) or very small aperture radii ($\astroang{;4;}$). For the largest field, all elements agree at better than 15\%.  

The agreement between $\TI$ and $\TI^\infty$ encourages a practical simplification, which is substituting $\TI$ by the large-field approximation. This significantly simplifies the calculation: Equation \eqref{eq: TI for infinite fields} to obtain $\TI^\infty$ contains only three integrals, while Eq.~\eqref{eq: T1 final} for $\TI$ contains six integrals, two of which involve oscillating functions. Consequently, the numerical integration for $\TI^\infty$ is much simpler and faster. 

\begin{figure*}
\begin{minipage}[c]{0.49\linewidth}
    \includegraphics[width=\linewidth, trim={0 0 0 0}, clip]{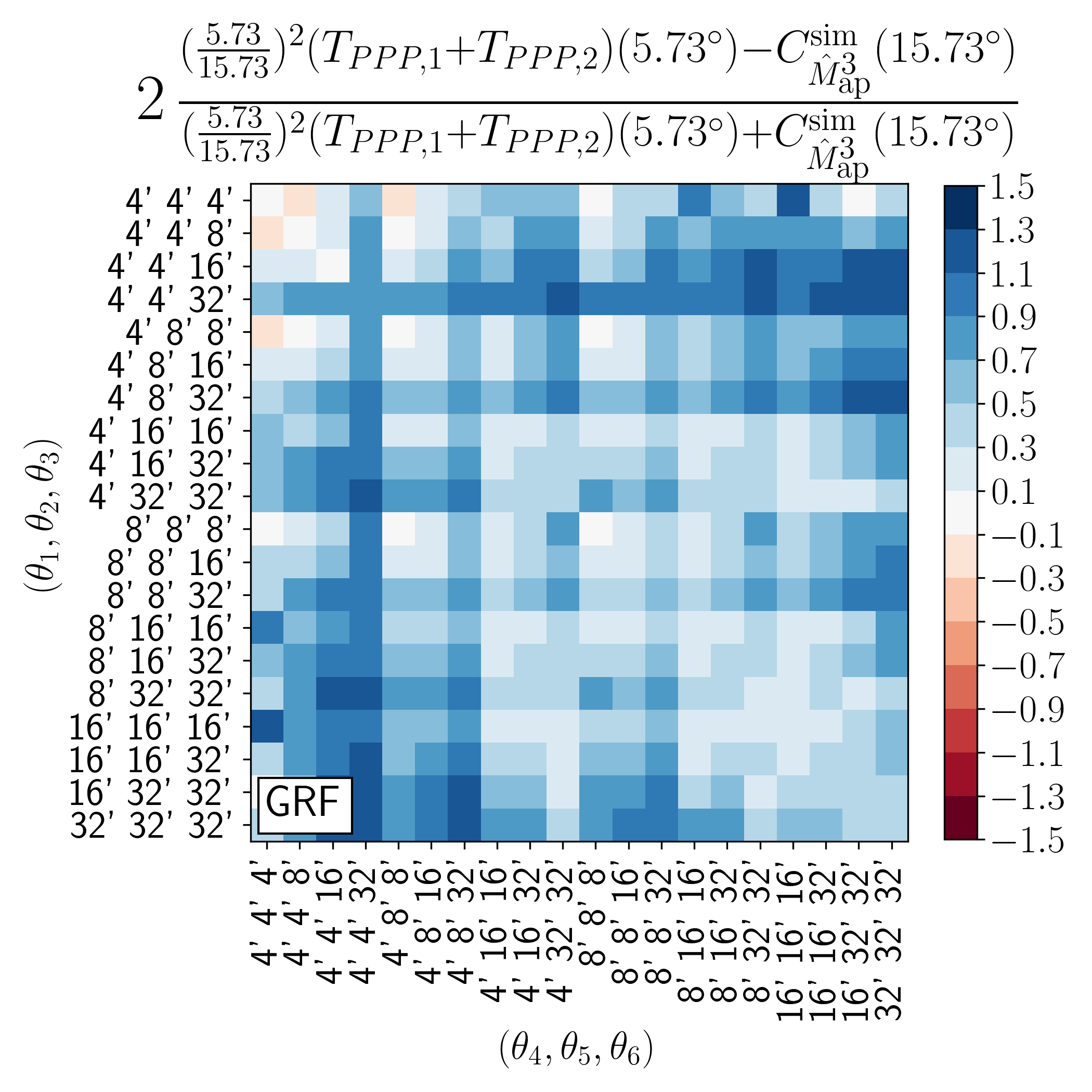}
\end{minipage}
\begin{minipage}[t]{0.49\linewidth}
    \caption{Fractional difference of model covariance, rescaled from a field size of $\astroang{5.73}^2$ to $\astroang{15.73}^2$ to the covariance from a simulated GRF of size $\astroang{15.73}^2$. The rescaling is performed under the assumption that the covariance scales inversely proportional to the survey area, which is not true for the finite field term $\TII$.}
    \label{fig: GRF rescaling}
\end{minipage}
\end{figure*}

The finite field effect, which gives rise to $\TII$, can be important for covariance estimates derived from simulations. Such estimates are sometimes performed on simulations with a smaller area than the survey to which the covariance is applied. To account for the different areas, the covariances are rescaled by a factor $A_\mathrm{sim}/A_\mathrm{survey}$, where $A_\mathrm{sim}$ and $A_\mathrm{survey}$ are the simulation and survey areas, respectively. This rescaling assumes that the covariance is inversely proportional to the survey area, which is only true under the large-field approximation. While $\TI^\infty$ indeed scales with the inverse survey area, $\TII$ decreases faster, as shown in Sect.~\ref{sec: Derivation subsec: Large-field approximation}. We demonstrate the failure of the rescaling approach in Fig.~\ref{fig: GRF rescaling}, where we show the fractional difference between the covariance for the simulated GRF with a side length of $\astroang{15.73}$ and the analytical estimate for the side length $\astroang{5.73}$, rescaled by the factor $(5.73/15.73)^2$. The rescaled covariance is too large by a factor of up to 4.7.  

\subsubsection{\texorpdfstring{$N$}{N}-body simulations}
\label{sec: Validation subsec: Results subsubsec: N-body}
\begin{figure*}
    \centering
    \includegraphics[width=\linewidth, trim={0 0 0 0}, clip]{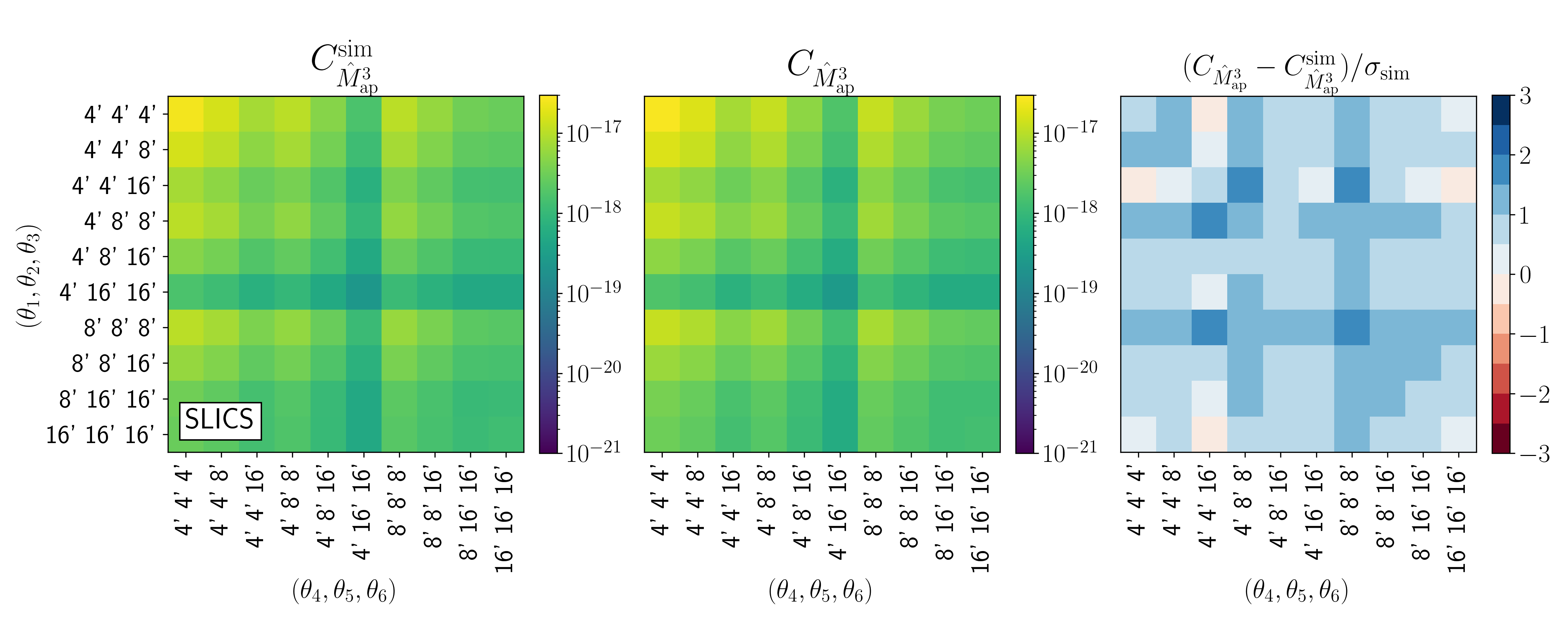}
    \includegraphics[width=\linewidth]{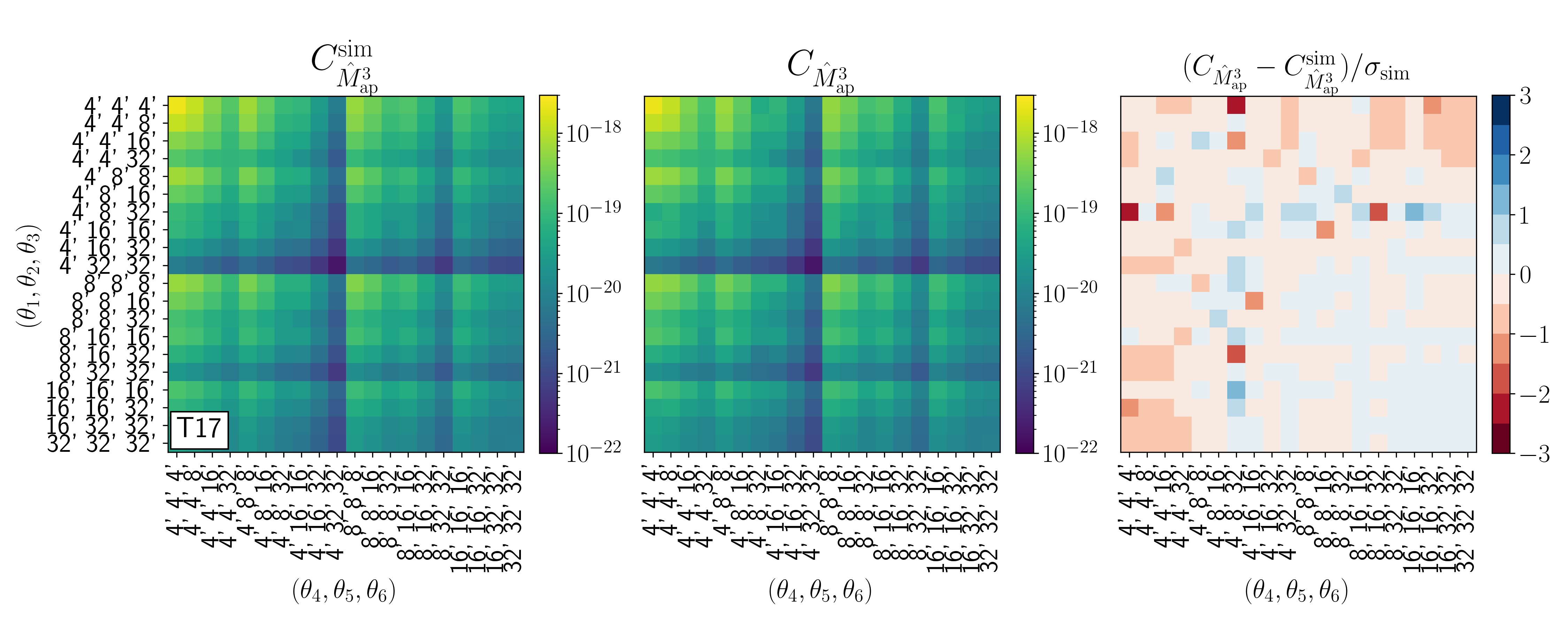}
    \caption{Comparison of measured and modelled covariance for the SLICS (top) and T17 simulations (bottom). Left are the measured covariance, middle are the model predictions; right are the differences between model and simulation, normalised by the simulation bootstrap error.}
    \label{fig: Nbody comparison}
\end{figure*}

We now compare the full analytic  $C_{\MapMapMapEst}$, including the non-Gaussian terms, to the covariances from the $N$-body simulations. Figure~\ref{fig: Nbody comparison} compares the $C_{\MapMapMapEst}$ and $C_{\MapMapMapEst}^\mathrm{sim}$ for the SLICS and the T17 simulations. For the SLICS, we see deviations of 1 to 2 times the simulation uncertainty, with the model being larger than the simulated covariance at most scales. These deviations could indicate that the power-, tri-, and pentaspectrum of the SLICS do not correspond to the model polyspectra. This could be caused by the finite resolution and smoothing in the simulations, the need for additional halo terms for the tri- and pentaspectrum in the model, or the general inaccuracy of the halo model. However, for the T17, we generally see a better agreement between simulation and model. Most elements agree within the simulation accuracy, with a few elements outside the $2\sigma$ range. 

\begin{figure*}
    \centering
    \includegraphics[width=0.49\linewidth]{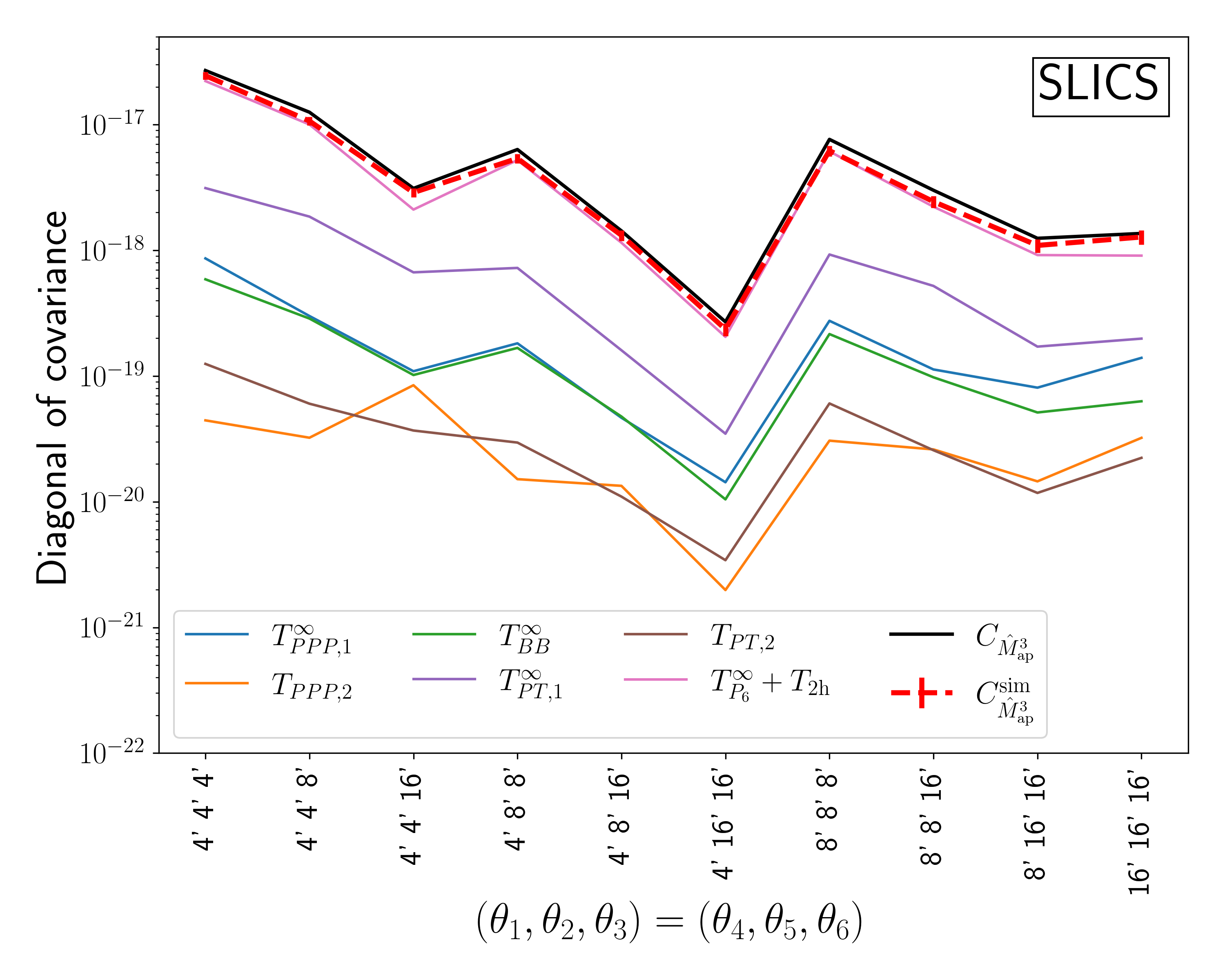}
     \includegraphics[width=0.49\linewidth]{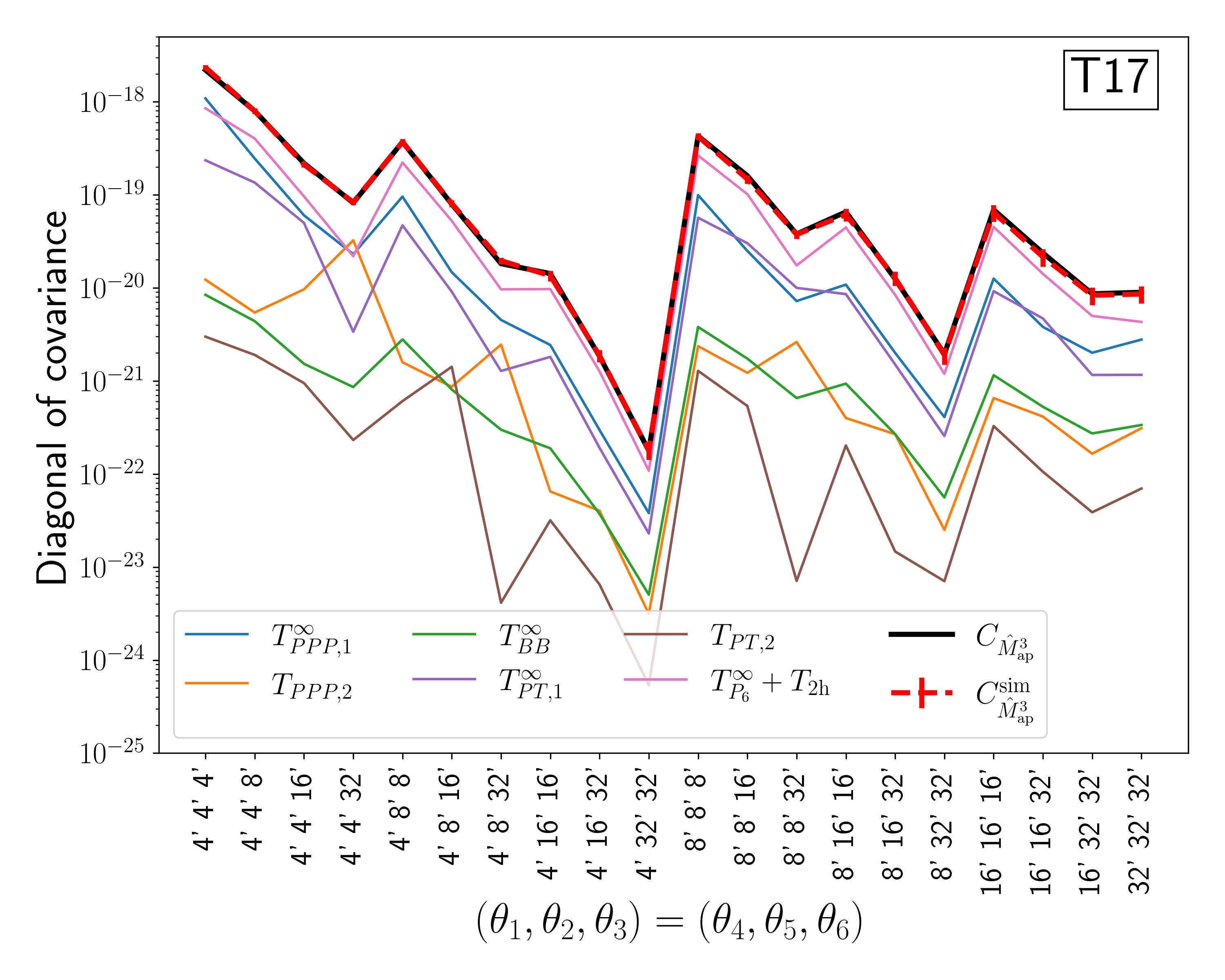}
      \caption{Diagonal of $C_{\MapMapMapEst}$ and the individual covariance terms for the SLICS (left) and the T17 (right). The measurement from the simulation is shown as a red dashed line, while the full modelled $C_{\MapMapMapEst}$ is the bold black line. The other lines show individual terms. Error bars on the simulated covariances originate from bootstrapping}
    \label{fig: individual terms diagonal}
\end{figure*}

\begin{figure}
\begin{minipage}[c]{0.49\linewidth}
    \includegraphics[width=\linewidth]{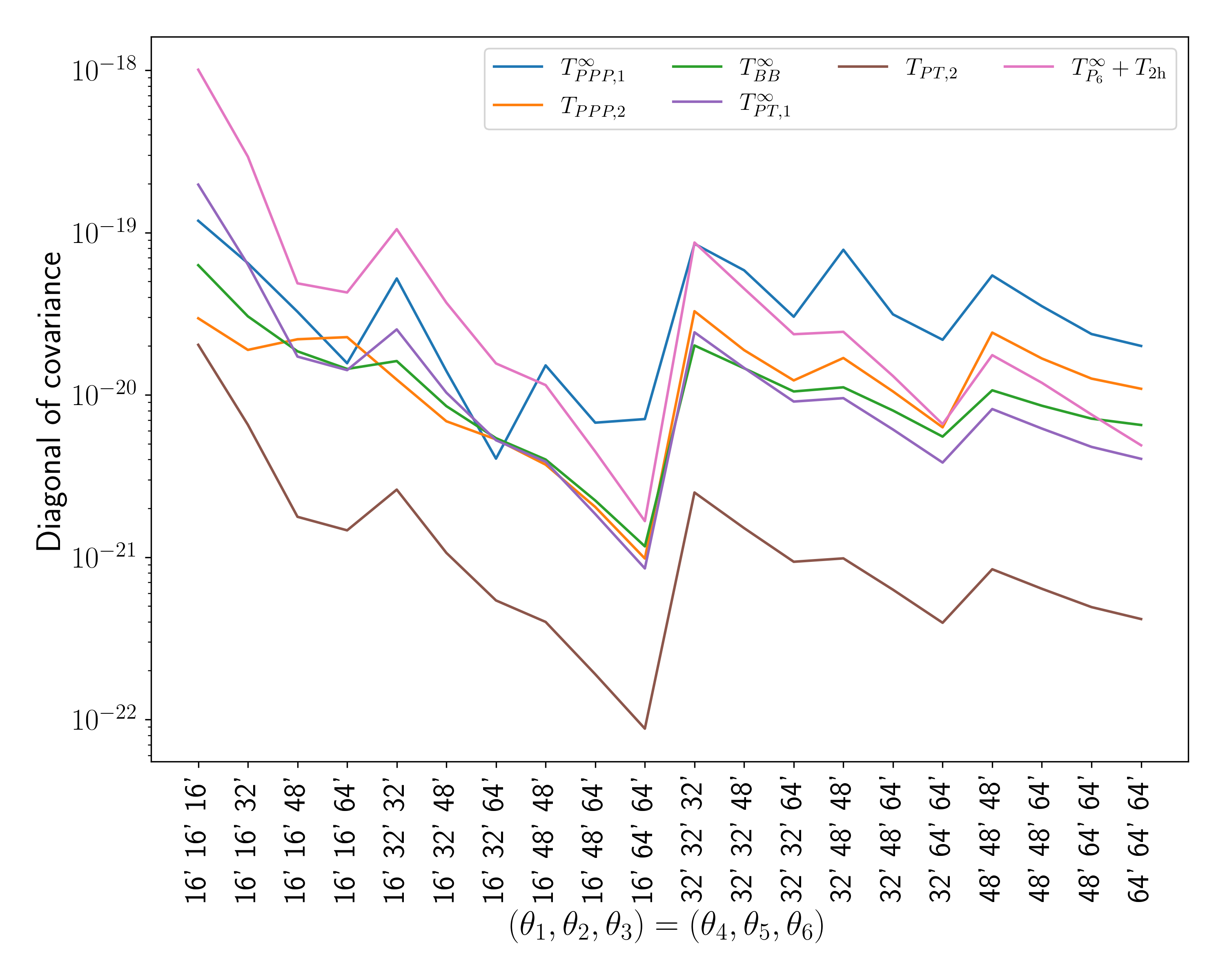}
\end{minipage}
\begin{minipage}[t]{0.49\linewidth}
    \caption{Model predictions for the individual covariance terms for the SLICS-like setup. }
    \label{fig: SLICS Cov larger Scales}
\end{minipage}
\end{figure}

We compare the diagonals of the individual terms of the analytic covariance in Fig.~\ref{fig: individual terms diagonal}. We see that for both the SLICS and T17 setup, the non-Gaussian terms are dominant. For the SLICS,  $\TVII$ dominates on the scales considered here, whereas for the T17, the first Gaussian term $\TI$ is of a similar magnitude as $\TVII$. This difference is caused by the different levels of shape noise in the two simulations. For the SLICS, we use a Stage-IV-like setup, where the galaxy number density is $n_\mathrm{g}=30\, \mathrm{arcmin}^{-2}$, while the T17 use the KiDS-like number density of $n_\mathrm{g}=6.17\, \mathrm{arcmin}^{-2}$. Therefore, the Gaussian covariance term for the T17 has a larger impact than for the SLICS. The relative importance of the Gaussian and non-Gaussian terms is also related to the considered aperture radii. For larger scales, the importance of the Gaussian terms increases, while the contribution of the pentaspectrum-term decreases, as shown in Fig.~\ref{fig: SLICS Cov larger Scales} for the SLICS setup and aperture radii between \astroang{;16;} and \astroang{;64;}. However, we stress that we consider these large scales suboptimal for cosmological analyses with $\MapMapMap$ since, as mentioned in Sect.~\ref{sec: Validation subsec: Measurement}, the matter field is more Gaussian and the signal-to-noise of third-order statistics is small.

\begin{figure}
\centering
\begin{minipage}[t]{0.49\linewidth}
 \includegraphics[width=\linewidth]{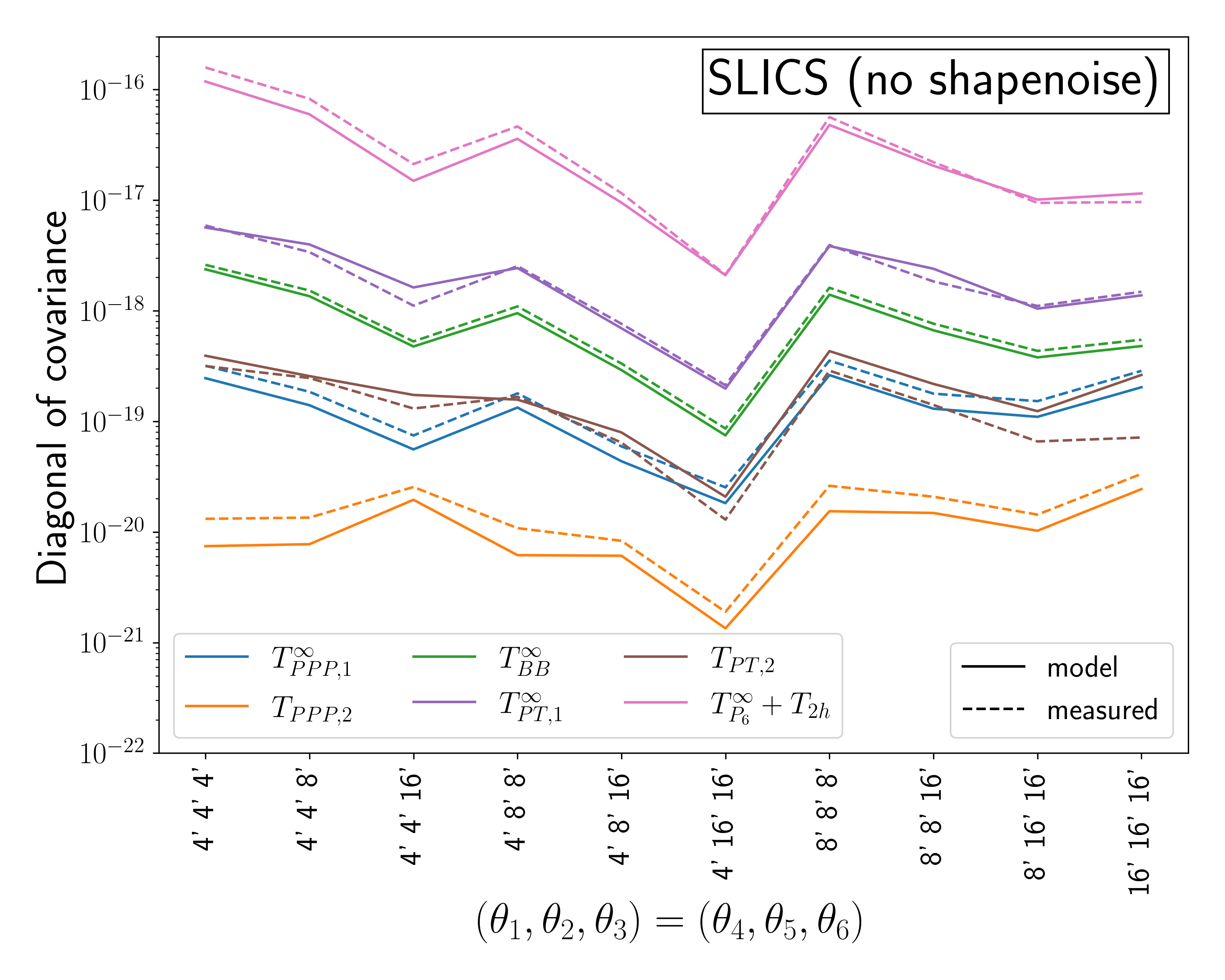}
    \caption{Comparison of individual covariance terms modelled (solid) and measured (dashed) in the SLICS without shape noise and with all sources at redshift 1}.
    \label{fig: SLICS individual terms measured}
\end{minipage}
\begin{minipage}[t]{0.49\linewidth}
    \includegraphics[width=\linewidth]{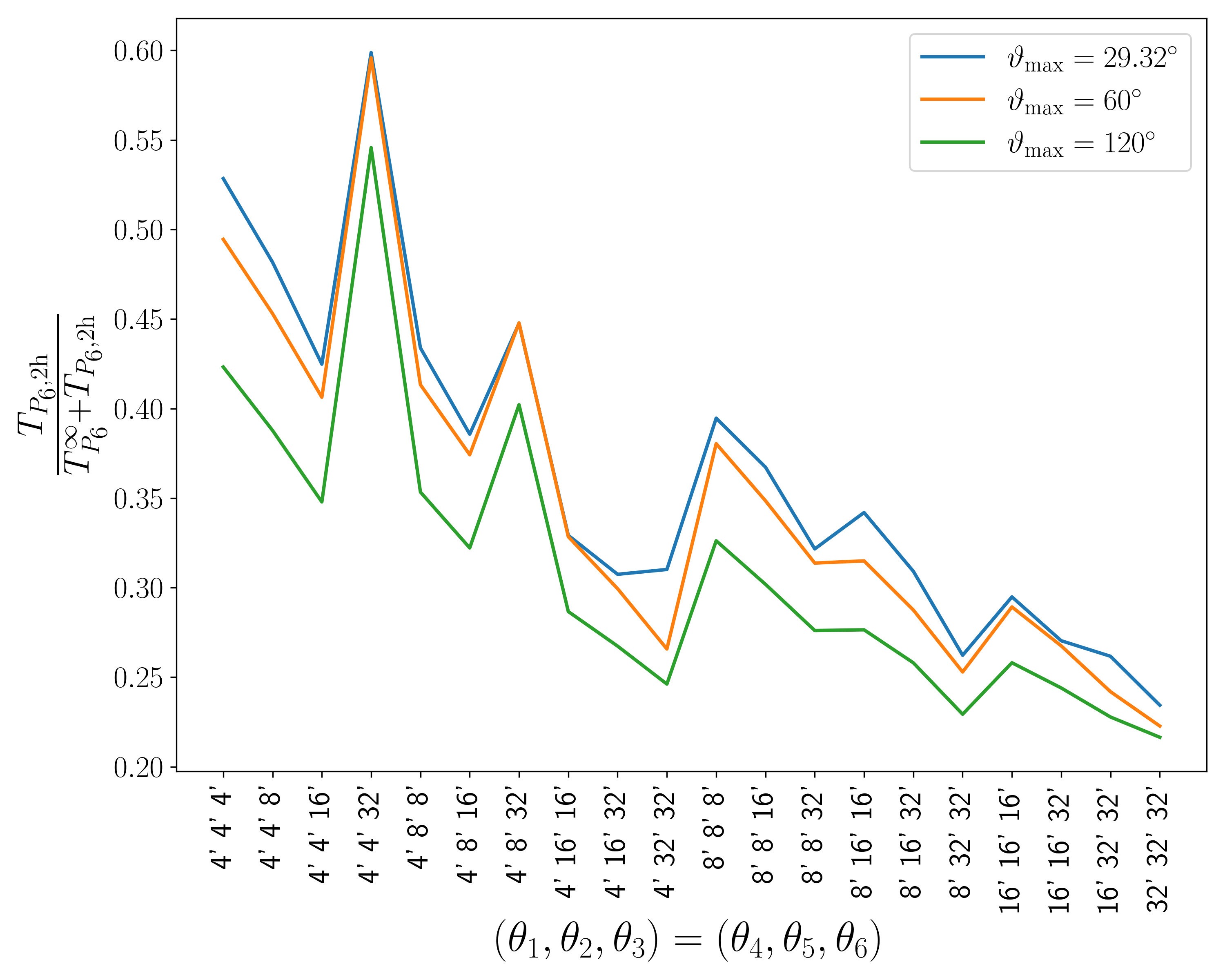}
    \caption{Ratio of two-halo contribution $T_{P_6,\mathrm{2h}}$ to full $T_{P_6}$ term for various field sizes with T17 cosmology. The blue line corresponds to the KiDS1000-like field size.}
    \label{fig: T2h various areas}
\end{minipage}
\end{figure}

As mentioned before, we can also directly measure the individual covariance terms and compare them to the analytic prediction in Fig.~\ref{fig: SLICS individual terms measured} using aperture mass correlation functions. {Unfortunately, measuring the finite-field terms $\TII$ and $\TVI$ is challenging: In Eqs.~\eqref{eq: TPPP2 from sims} and \eqref{eq: TPT2 from sims}, one can see that, when neglecting the function $E_A(\etavec)$, the integral over the correlation function is, by definition, zero. So the only part that leads to $\TII$ and $\TVI$ being non-zero is multiplication with the slowly varying function $E_A(\etavec)$. This amplifies the effect of noise in the measured correlation function. Thus, the measurements of $\TII$ and $\TVI$ are very sensitive to the noise level}, and we can only perform them in the SLICS without shape noise. We note that the source galaxies in this data set are all situated at redshift $z=1$ and do not follow the same redshift distribution as the data used for Fig.~\ref{fig: individual terms diagonal}. The measured and modelled covariance terms show similar dependencies on the aperture scale radii. The model for the dominating term $\TVII$ agrees within a few per cent with the measurement. For $\TI$, $\TII$, and $\TVI$, though, the model underpredicts the measurements. This difference is likely due to differences between the model polyspectra and the actual polyspectra of the simulation. Since our trispectrum model contains only the 1-halo term, we do not expect a perfect match between the modelled $\TV$ and $\TVI$ and the measurement. Furthermore, the \verb|BiHalofit| bispectrum model only agrees with the data at the $10\%$ level, \citepalias[compare to Appendix A of][]{Heydenreich2022}, so deviations at the $20\%$ level are expected for $\TIV$, which depends on the square of the bispectrum.  

The term $T_{P_6,\mathrm{2h}}$ gives a significant contribution to $T_{P_6}$. We show this in Fig.~\ref{fig: T2h various areas}, where the blue line shows the diagonal of the ratio between $T_{P_6,\mathrm{2h}}$ and $T_{P_6}^\infty+T_{P_6,\mathrm{2h}}$ for the T17 setup. The two-halo term is substantial for small apertures and the combinations of small and large apertures. The figure also shows the same ratio for larger field sizes. The importance of $T_{P_6,\mathrm{2h}}$ decreases with field size. This is as expected, since $T_{P_6}^\infty$ scales with the inverse survey area, while $T_{P_6,\mathrm{2h}}$ decreases faster. 

\begin{figure}
\begin{minipage}[c]{0.49\linewidth}
    \includegraphics[width=\linewidth]{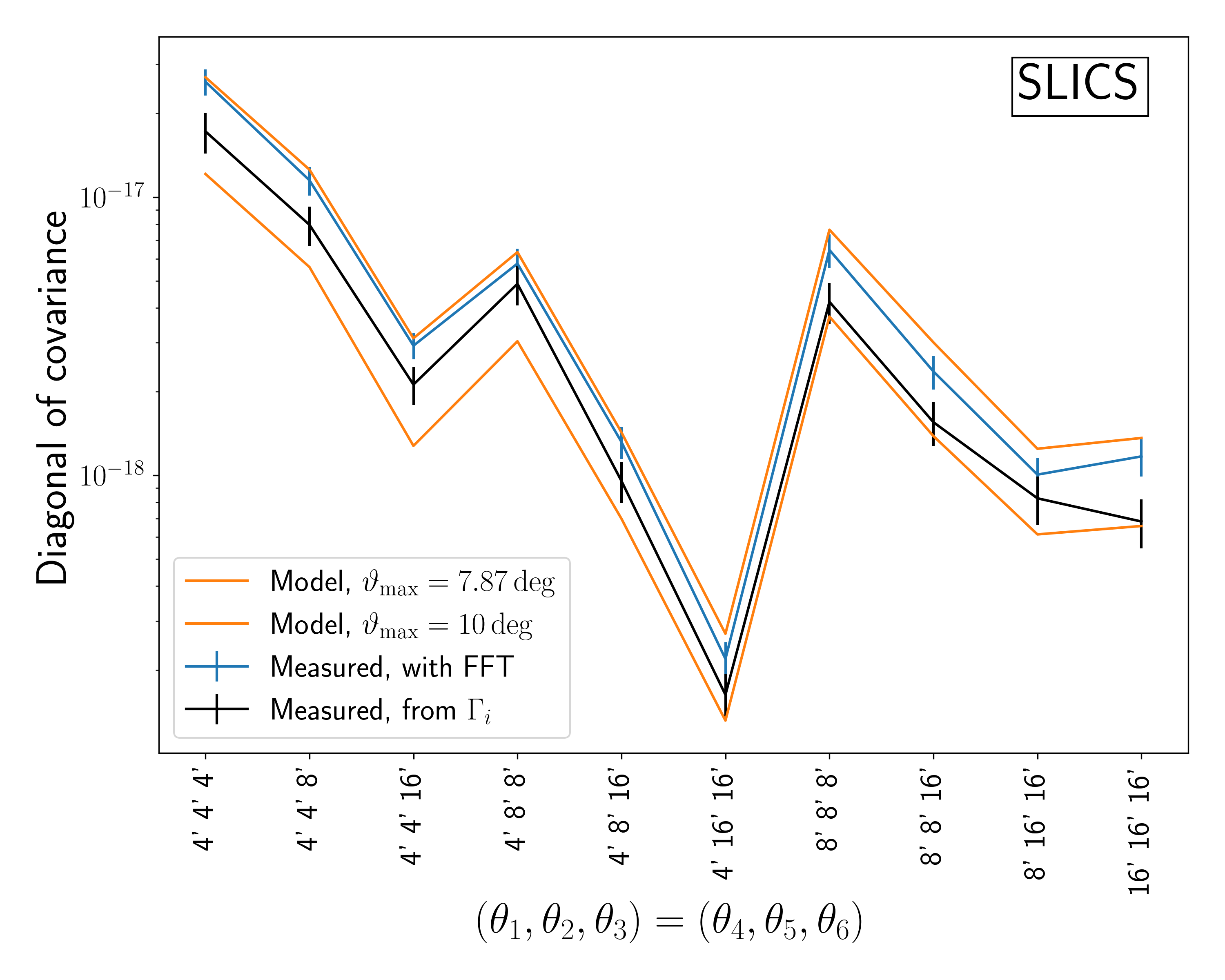}
\end{minipage}
\begin{minipage}[t]{0.49\linewidth}
        \caption{Covariance in SLICS, measured with shear correlation functions $\Gamma_i$ (black) and FFT (blue), as well as the model covariance for the full survey area of $\astroang{10;;}\times\astroang{10;;}$ (orange, dashed) and for the effective survey area of $\astroang{7.87;;}\times\astroang{7.87;;}$.}
    \label{fig: SLICS Cov from CorrFunc}
\end{minipage}
\end{figure}

Finally, we compare the covariance estimated in the SLICS using the shear three-point correlation function $\Gamma_i$ to the FFT-based estimate in Fig.~\ref{fig: SLICS Cov from CorrFunc}. We also show the model prediction for two different survey areas: either the entire survey area of $\astroang{10}\times \astroang{10}$ or the survey area of $\astroang{7.87}\times \astroang{7.87}$ after boundary removal. As expected, the FFT-based covariance estimate is larger than the $\Gamma_i$-based estimate. This is a direct consequence of the boundary removal for the FFT - the effective survey area shrinks.

The $\Gamma_i$-based covariance is itself larger than the full-survey model covariance. This is because, as mentioned before, estimates of $\MapMapMap$ for the whole survey area $A'$ require shear information outside of $A'$. Consequently, unbiased estimates of $\MapMapMap$ on $A'$ contain more information than estimates of $\Gamma_i$, which are restricted to $A'$. Therefore, the covariance of the $\Gamma_i$-based estimator lies in between the model covariance for the total area $A'$ and cut-off survey area $A$. Since the $\Gamma_i$-based estimator is required to analyse survey data, the covariance model gives an upper and lower bound to the expected survey constraints.

\section{Influence of covariance terms on cosmological parameter estimation}
\label{sec: MCMC}

We have shown in the previous section that the analytical covariance estimates agree with estimates from the simulations within 1--2 times the simulations' statistical uncertainty. However, it is unclear whether this level of agreement is sufficient for a cosmological parameter analysis. To test this, in this section, we perform mock MCMC analyses using $C_{\MapMapMapEst}$ and $C_{\MapMapMapEst}^\mathrm{sim}$, and compare the resulting parameter constraints. We also test the impact of the individual covariance terms on the cosmological analysis.

\subsection{Analysis setup}
\label{sec: MCMC subsec: Setup}
We perform mock cosmological analyses with the same setup as in \citetalias{Heydenreich2022}. In this setup, we use the neural network emulator \verb|CosmoPower| \citep{SpurioMancini2022} to quickly evaluate the $\MapMapMap$ model and sample the parameter likelihood. With the trained emulator, we evaluate the parameter likelihoods for different choices for the covariance: the estimate from the simulations, the full analytic expressions, and various combinations of the individual terms of $C_{\MapMapMapEst}$. Since the $\MapMapMap$ are not sensitive to $w_0$ and $h$ in a non-tomographic setting, we evaluate the likelihood for fixed $w_0=-1$ and $h=0.69$. For the T17 validation, we evaluated the $\MapMapMap$-model at the aperture scale radii $\astroang{;4;}, \astroang{;8;}, \astroang{;16;},$ and $\astroang{;32;}$, whereas for the SLICS setup, we used the aperture scale radii $\astroang{;4;}, \astroang{;8;},$ and $\astroang{;16;}$. The reference data vector is measured at the corresponding cosmology of the respective simulation.

\subsection{Results}
\label{sec: MCMC subsec: Results}
\begin{figure}
    \centering
    \begin{subfigure}{0.49\linewidth}
    \includegraphics[width=\linewidth]{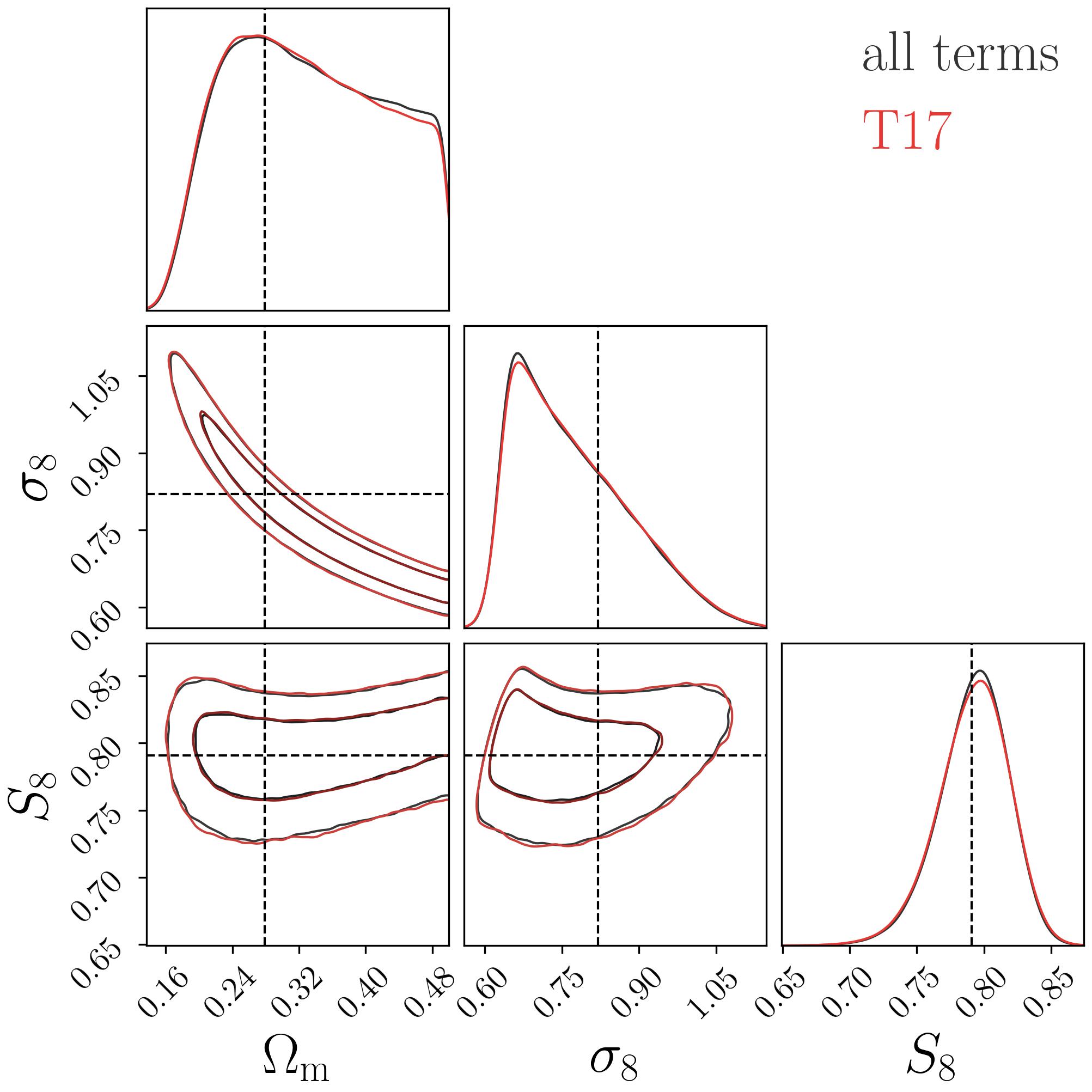}
    \label{fig: MCMC Takahashi}            
    \end{subfigure}
    \begin{subfigure}{0.49\linewidth}
    \includegraphics[width=\linewidth]{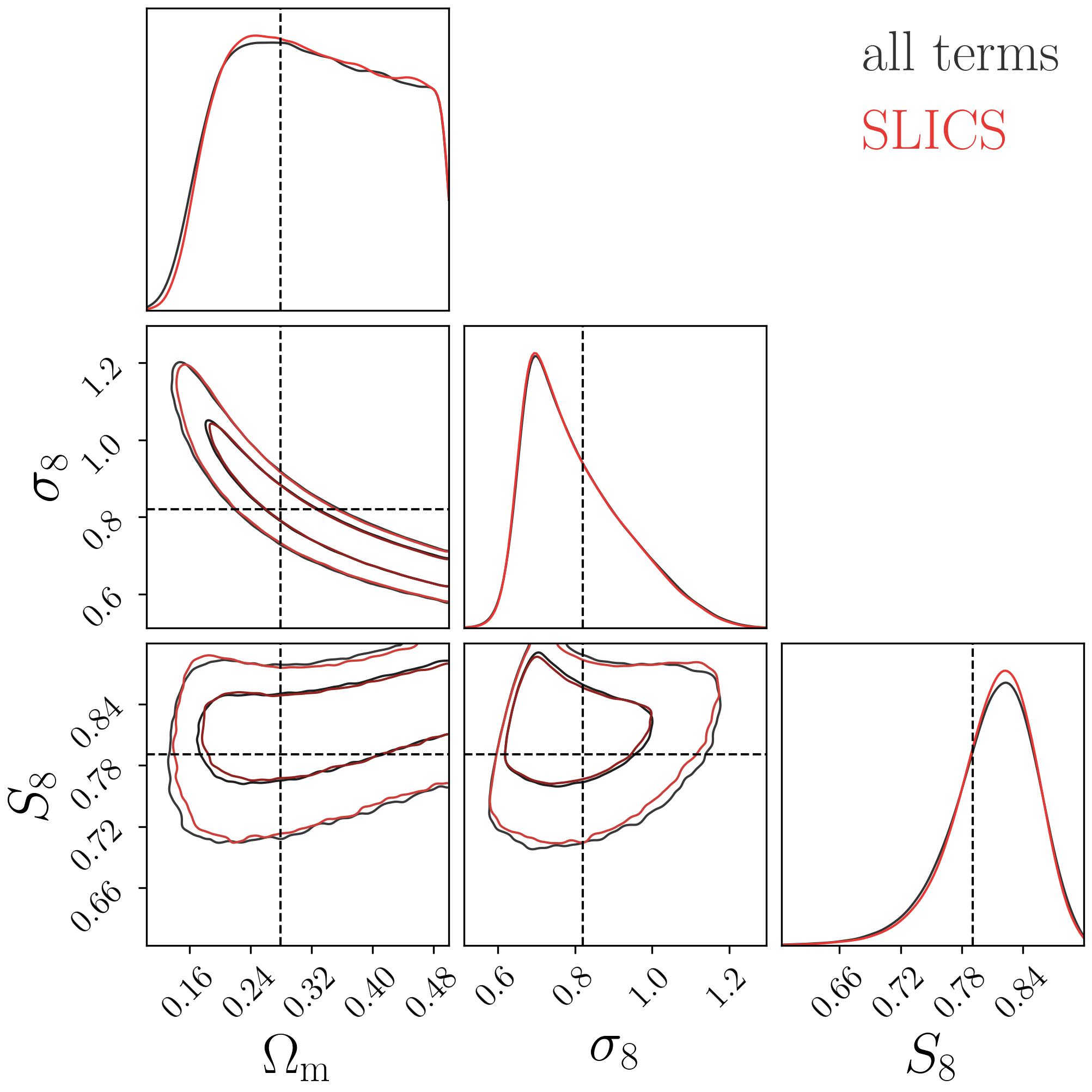}
    \label{fig: MCMC analytical SLICS}
    \end{subfigure}
    \caption{Parameter constraints, using either the covariance from the simulations (red) or the analytic model (black). Left are the constraints for a KiDS-1000-like survey, and right are the constraints for the SLICS setup, which uses a stage IV-like $n(z)$ and shape noise, but a small survey area of $\astroang{7.87;;}\times \astroang{7.87;;}$}
    \label{fig: MCMC}
\end{figure}

In the left panel of Fig.~\ref{fig: MCMC}, we show the T17 results for the constraints on $\Omm$, $S_8$, and $\sigma_8$, while varying only $\Omm$ and $S_8$. In Table \ref{table:MAP_values}, we report the resulting marginalised parameter constraints together with the figure of merit (FoM) estimated as 
\begin{equation}
    \mathrm{FoM} = \frac{1}{\sqrt{\mathrm{det}(C)}}\, ,
\end{equation}
where $C$ is the parameter covariance matrix resulting from the MCMC process. The T17 covariances are tailored to a KiDS-like survey area and shape noise. The constraints from the simulated and analytic covariance coincide. The FoM of $\Omega_\mathrm{m}$--$S_8$ differs by less than 3\%. Consequently, the analytic covariance is ideal for an unbiased cosmological analysis in our setup.  We display in the right panel of Fig.~\ref{fig: MCMC} the comparison for the SLICS setup. The posteriors for both the simulated and modelled covariance matrix clearly agree. This supports the robustness of the modelled covariance for varying survey areas, shape noise and redshift distribution.

\begin{table*}
\centering
\caption{Overview of the marginalised MAP values and $68\%$ confidence intervals resulting from MCMC chains where $\Omega_\mathrm{m}$, $S_8$ are varied and $\sigma_8=S_8\sqrt{0.3/\Omega_\mathrm{m}}$. We fixed $h=0.7$ and $w_0=-1$.}
\begin{tabular}{c|ccc}
\hline
\hline
 MAP & $\Omega_\mathrm{m}$ & $\sigma_8$ & $S_8$  \\
\hline
T17 &  $0.272^{+0.148}_{-0.065}$ & $0.665^{+0.173}_{-0.038}$ & $0.798^{+0.023}_{-0.029}$  \\
analytic model (all terms) & $0.259^{+0.166}_{-0.048}$ & $0.661^{+0.172}_{-0.035}$ & $0.798^{+0.023}_{-0.027}$ \\
analytic model (all terms, without 2-halo term) & $0.260^{+0.170}_{-0.047}$ & $0.663^{+0.166}_{-0.033}$ & $0.798^{+0.021}_{-0.021}$ \\
analytic model (neglecting finite-field terms) &  $0.271^{+0.150}_{-0.061}$ & $0.661^{+0.171}_{-0.033}$ & $0.797^{+0.023}_{-0.025}$ \\
analytic model (only Gaussian)  & $0.271^{+0.132}_{-0.052}$ & $0.664^{+0.164}_{-0.019}$ & $0.795^{+0.015}_{-0.013}$ \\
\hline
\hline
 FoM & $\Omega_\mathrm{m}$-$\sigma_8$ & $\Omega_\mathrm{m}$-$S_8$ & $\sigma_8$-$S_8$  \\
\hline
T17 &  293 & 429 & 341  \\
analytic model (all terms) & 299 & 441 & 353 \\
analytic model (all terms, without 2-halo term) & 334 & 561 & 445 \\
analytic model (neglecting finite-field terms) &  321 & 482 & 388  \\
analytic model (only Gaussian)  & 553 & 1216 & 966 \\
\end{tabular}
\label{table:MAP_values}
\end{table*}

\begin{figure}
\begin{minipage}[c]{0.49\linewidth}
    \includegraphics[width=\linewidth]{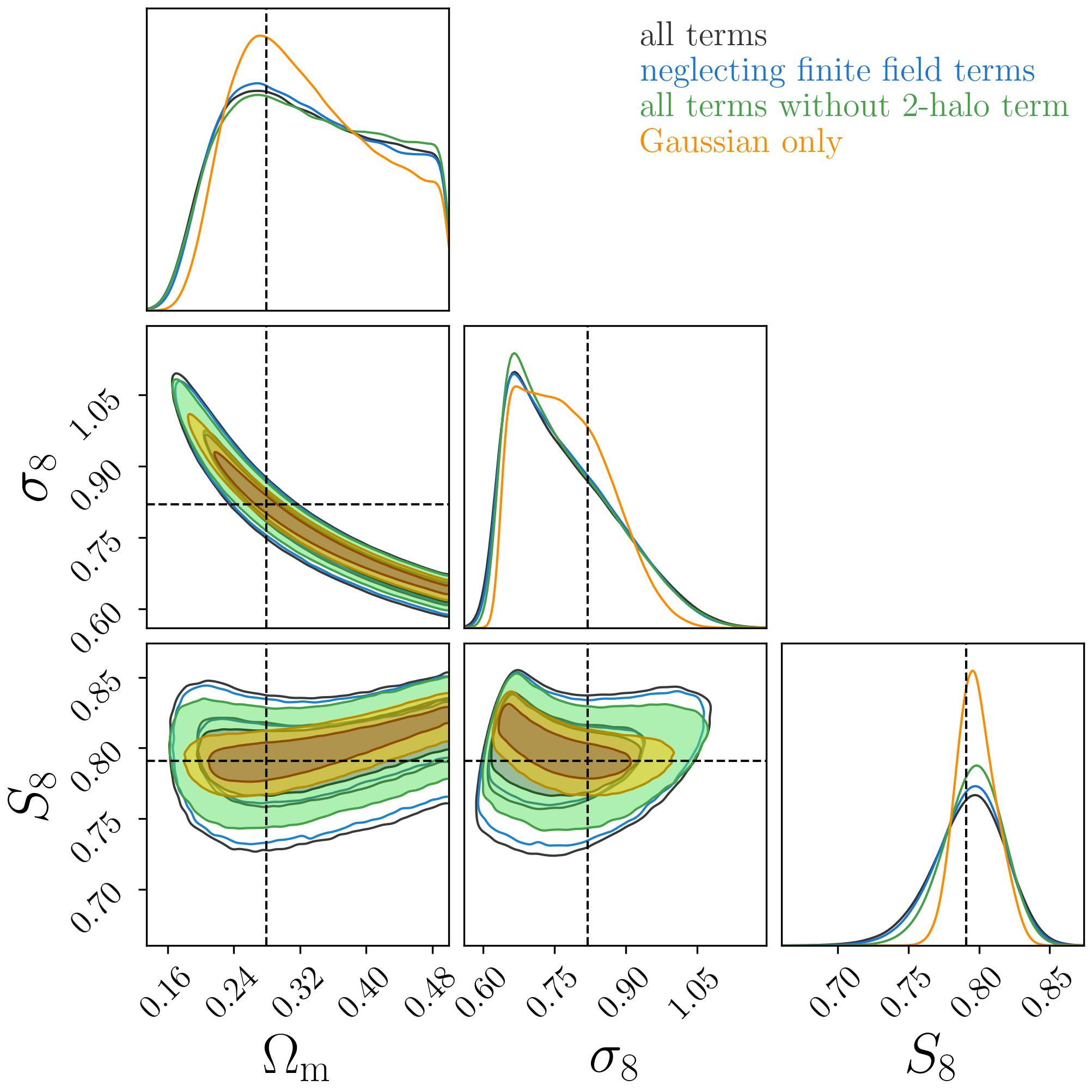}
\end{minipage}
\begin{minipage}[t]{0.49\linewidth}
\caption{Parameter constraints for a KiDS-1000-like survey, using the full model covariance (black), neglecting the finite field terms (blue), neglecting the 2-halo contribution to $T_{P_6}$ (green) or using only the Gaussian covariance (orange). The FoMs of $\Omega_\mathrm{m}$-$S_8$ are given in Table~\ref{table:MAP_values}. }
    \label{fig: MCMC analytical}  
\end{minipage}
\end{figure}

In Fig.~\ref{fig: MCMC analytical}, we compare the parameter constraints obtained when using the total analytic covariance estimate, only the Gaussian terms $\TI$ and $\TII$, when neglecting the finite-field terms $\TII$ and $\TVI$, and when neglecting the term $T_{P_6, \mathrm{2h}}$. The results demonstrate the importance of the non-Gaussian covariance terms, as the FoM of $\Omega_\mathrm{m}$--$S_8$ approximately triples if only the Gaussian terms are used (see Table~\ref{table:MAP_values}). Consequently, for the application to a Stage III survey, non-Gaussian covariance terms for $\MapMapMap$ cannot be neglected.
 
For the KiDS-like survey area used here, the finite-field terms $\TII$ and $\TVI$ play a secondary role (see Table~\ref{table:MAP_values}). The FoM of $\Omega_\mathrm{m}$-$S_8$ increases by 12\% if these terms are neglected. Consequently, depending on the desired accuracy of the modelled covariance, these terms could be neglected for a real parameter analysis. Since calculating $\TII$ and $\TVI$ for a real survey requires evaluating Eq.~\eqref{eq: G_A} for the true, most likely complicated, survey geometry, bypassing them could significantly decrease the computational complexity of the covariance calculation. Ignoring $T_{P_6, \mathrm{2h}}$, the 2-halo contribution to $\TVII$, has a stronger effect. Neglecting $T_{P_6, \mathrm{2h}}$ increases the FoM of $\Omega_\mathrm{m}$-$S_8$ by 27\% and has a significant impact on the inferred constraints on the individual parameters.

\section{Discussion}
\label{sec: Discussion}
In this work, we derived an analytic model for the covariance of the third-order aperture statistics $\MapMapMap$. We make our modelling code publically available \footnote{ \url{https://github.com/sheydenreich/threepoint/releases}}.

We performed three tests for this analytic model. First, we compared  the Gaussian part of the covariance to simulated GRFs. We found that all elements of the modelled covariance agree with the GRFs within 30\%, with most elements agreeing better than 10\% with the simulated data. Second, we compared the full model to two independent $N$-body simulations, the SLICS and T17 simulations. These comparisons showed that the model agrees within 1--2 times the statistical uncertainty on the simulations with the simulated covariance estimates. Third, we compared the expected constraints from a KiDS-1000-like survey for the cosmological parameters $\Omm$ and $S_8$ using the covariance from the T17 simulations and the analytic expression. The constraints were remarkably similar, with only a 3\% deviation in the FoM on $\Omm$-$S_8$. These three tests confirmed that our covariance model is sufficiently accurate for analysing a stage III survey.

Additionally, we found that all terms of the $\MapMapMapEst$ covariance can be estimated by measuring aperture mass correlation functions. Using this approach, we compared the individual terms of the model to measurements in the SLICS and confirmed that the model agrees with the simulations.

In the derivation of the covariance model, we found terms that decrease faster than the inverse survey area and vanish entirely under the large-field approximation, which assumes an infinitely broad survey window function. These finite-field terms show a complex dependence on the survey geometry, while the other terms mainly scale with the inverse survey area. The finite field term $\TII$ is already present in the Gaussian case. Comparing the model to GRFs showed that it is the dominating contribution for non-diagonal elements, and neglecting it leads to a severe bias.
However, the effect of the finite-field term is small for the realistic covariances, including non-Gaussianity, as they are dominated by the pentaspectrum term $\TVII$. Ignoring the finite-field terms leads to an increment of only 12\% in the FoM on $\Omm$-$S_8$.

We showed that it is incorrect to derive the covariance for the real-space statistic $\MapMapMapEst$ from a bispectrum covariance. Using an integral over a bispectrum covariance to model the covariance of $\MapMapMapEst$ is only possible under the large-field approximation. However, in this case, one automatically neglects the finite-field terms. 

We note that our covariance model assumes the flat-sky approximation. Nevertheless, the favourable comparison to the T17 simulation indicates that this assumption is valid at the relatively small scales considered here. This is not a surprising finding since we consider $\MapMapMap$ at small, sub-degree scales. At these small scales, the flat-sky approximation introduces no significant bias. Our model should therefore be applicable to actual observations. 

Our covariance model is based on an estimator for $\MapMapMap$ that is easily applicable to simulations but should not be applied to realistic survey data, which include masks, because it requires a continuous convergence (or shear) map. Instead, one must use an estimator based on the third-order shear correlation functions. However, we have shown that our model can quantify the covariance of this estimator. The model covariance for the entire survey area gives a lower bound to the expected covariance, while the model covariance for a smaller effective area after boundary removal gives an upper bound. For a square survey with an area of $1000\,\mathrm{deg}^2$ and a boundary cut-off of $4\times \astroang{;16;}$, the upper bound is around $15\%$ above the lower bound, which gives relatively tight constraints on the true covariance. For even larger stage IV surveys, the bound will be tighter, as the boundary is a smaller fraction of the overall area.

In conclusion, we have presented and validated an analytic model for the covariance of third-order aperture statistics. Together with the analytical model for $\MapMapMap$ presented in \citetalias{Heydenreich2022}, our covariance model paves the way for the cosmological analysis of third-order shear statistics of stage III and stage IV weak lensing surveys.

We have deliberately avoided the term super-sample covariance (SSC) to describe parts of $C_{\MapMapMapEst}$. This term, originally devised for parts of the covariance of the power spectrum \citep{Takada2013}, is used to describe covariance parts depending on $\ell$-modes larger than a given survey area. The term $T_{P_6,\mathrm{2h}}$ corresponds to what is commonly called SSC for the bispectrum \citep{Chan2018, Pyne2021}. However, as we will show in a forthcoming paper, the SSC can be defined as the difference between the exact form of a covariance and its large-field approximation. By this definition, the finite-field terms $\TII$ and $\TVI$ are also part of the SSC for $\MapMapMapEst$.

\begin{acknowledgements}
Funded by the TRA Matter (University of Bonn) as part of the Excellence Strategy of the federal and state governments. This work has been supported by the Deutsche Forschungsgemeinschaft through the project SCHN 342/15-1 {and DFG SCHN 342/13}. PAB and SH acknowledge support from the German Academic Scholarship Foundation. {We would like to thank Joachim Harnois-D\'eraps for making public the SLICS mock data, which can be found at \url{http://slics.roe.ac.uk/}.} We thank Benjamin Joachimi, Susan Pyne, Lucas Porth and Niek Wielders for many helpful comments and discussions.
\end{acknowledgements}

\bibliographystyle{aa}
\bibliography{cite}
\onecolumn
\begin{appendix}
    \section{List of all permutations for covariance terms}
\label{app: permutations}
    In Table~\ref{tab: permutations}, we list all permutations of aperture scale radii for the terms $\TI$ to $\TVI$ from Sect.~\ref{sec: Derivation}. The first permutation corresponds to the expressions in Sect.~\ref{sec: Derivation}. The vertical dashes separate groups of scale radii, in which the terms are symmetric. These symmetries can be easily seen in the expressions in terms of the aperture mass correlation functions in Sect.~\ref{sec: Derivation subsec: Correlation Functions}.

\begin{table}[]
        \centering
        \caption{Permutations of aperture scale radii for covariance terms.}
\begin{tabular}{|ll|ll|ll|}
\hline
Term              & permutations & Term              & permutations & Term              & permutations \\ \hline
\multirow{6}{*}{$\TI$} & $\theta_1\theta_4|\theta_2\theta_5|\theta_3\theta_6$      & \multirow{9}{*}{$\TIV$} & $\theta_2\theta_3|\theta_4|\theta_5\theta_6|\theta_1$             & \multirow{9}{*}{$\TV$} & $\theta_1\theta_4|\theta_2\theta_3|\theta_5\theta_6$             \\
                  & $\theta_1\theta_4|\theta_2\theta_6|\theta_3\theta_5$               &   &  $\theta_2\theta_3|\theta_5|\theta_4\theta_6|\theta_1$            &                   & $\theta_1\theta_5|\theta_2\theta_3|\theta_4\theta_6$              \\
                  & $\theta_1\theta_5|\theta_2\theta_4|\theta_3\theta_6$               &   &  $\theta_2\theta_3|\theta_6|\theta_4\theta_5|\theta_1$            &                   & $\theta_1\theta_6|\theta_2\theta_3|\theta_4\theta_5$              \\
                  & $\theta_1\theta_5|\theta_2\theta_6|\theta_3\theta_4$               &   &  $\theta_1\theta_3|\theta_4|\theta_5\theta_6|\theta_2$            &                   & $\theta_2\theta_4|\theta_1\theta_3|\theta_5\theta_6$              \\
                  & $\theta_1\theta_6|\theta_2\theta_4|\theta_3\theta_5$               &   & $\theta_1\theta_3|\theta_5|\theta_4\theta_6|\theta_2$             &                   & $\theta_2\theta_5|\theta_1\theta_3|\theta_4\theta_6$              \\
                  & $\theta_1\theta_6|\theta_2\theta_5|\theta_3\theta_4$               &   & $\theta_1\theta_3|\theta_6|\theta_4\theta_5|\theta_2$             &                   & $\theta_2\theta_6|\theta_1\theta_3|\theta_4\theta_5$              \\ \cline{1-2}
\multirow{9}{*}{$\TII$} & $\theta_1\theta_2|\theta_3\theta_4|\theta_5\theta_6$         &   & $\theta_1\theta_2|\theta_4|\theta_5\theta_6|\theta_3$              &                   & $\theta_3\theta_4|\theta_1\theta_2|\theta_5\theta_6$              \\
                  & $\theta_1\theta_2|\theta_3\theta_5|\theta_4\theta_6$              &    &  $\theta_1\theta_2|\theta_5|\theta_4\theta_6|\theta_3$            &                   & $\theta_3\theta_5|\theta_1\theta_2|\theta_4\theta_6$              \\
                  & $\theta_1\theta_2|\theta_3\theta_6|\theta_4\theta_5$             &     & $\theta_1\theta_2|\theta_6|\theta_4\theta_5|\theta_3$              &                   & $\theta_3\theta_6|\theta_1\theta_2|\theta_4\theta_5$              \\ \cline{3-6} 
                  & $\theta_1\theta_3|\theta_2\theta_4|\theta_5\theta_6$             &                   &              & \multirow{6}{*}{$\TVI$} & $\theta_1\theta_2|\theta_4\theta_5\theta_6|\theta_3$             \\
                  & $\theta_1\theta_3|\theta_2\theta_5|\theta_4\theta_6$             &                   &              &                   & $\theta_1\theta_3|\theta_4\theta_5\theta_6|\theta_2$             \\
                  & $\theta_1\theta_3|\theta_2\theta_6|\theta_4\theta_5$             &                   &              &                   & $\theta_2\theta_3|\theta_4\theta_5\theta_6|\theta_1$             \\
                  & $\theta_2\theta_3|\theta_1\theta_4|\theta_5\theta_6$             &                   &              &                   & $\theta_4\theta_5|\theta_1\theta_2\theta_3|\theta_6$             \\
                  & $\theta_2\theta_3|\theta_1\theta_5|\theta_4\theta_6$             &                   &              &                   & $\theta_4\theta_6|\theta_1\theta_2\theta_3|\theta_5$             \\
                  & $\theta_2\theta_3|\theta_1\theta_6|\theta_4\theta_5$             &                   &              &                   & $\theta_5\theta_6|\theta_1\theta_2\theta_3|\theta_4$             \\ \cline{1-2} \cline{5-6}
\end{tabular}
        \label{tab: permutations}
\end{table}

\section{Impact of shape noise on aperture mass covariance}
\label{app: shape noise}

In this section, we show that the impact of shape noise on $C_{\MapMapMapEst}$ can be estimated by replacing the power spectrum $P(\ell)$ in the expressions derived in Sect.~\ref{sec: Derivation} by $P(\ell)+\sigma_\epsilon^2/2n$, where $\sigma_\epsilon^2$ is the two-component intrinsic ellipticity dispersion and $n$ is the galaxy number density. For this, we consider the smoothed convergence $K$, which is
\begin{equation}
    K(\varthetavec) = \frac{1}{n} \sum_{i=1}^N F_\sigma(|\varthetavec-\varthetavec_i|)\, \kappa(\varthetavec_i)\;,
\end{equation}
where $F_\sigma$ is a smoothing kernel of width $\sigma$, the sum runs over the $N$ galaxies and $\varthetavec_i$ is the position of the $i$th galaxy. We assume that the galaxies are distributed uniformly within the survey area $A'$. The kernel $F_\sigma$ needs to be normalised, such that $\int_{A'} \dd[2]{\vartheta} F_\sigma(\varthetavec) = 1$. As noted in \citet{VanWaerbeke2000},  the smoothed convergence field $K_\mathrm{N}$ including shape noise can be written as
\begin{equation}
    K_{\mathrm{N}}(\varthetavec) = K(\varthetavec) + \mathcal{N}(\varthetavec)\quad \textrm{with}\quad  \mathcal{N}(\varthetavec) = \frac{1}{n} \sum_{i=1}^N\int \frac{\dd[2]{\ell}}{(2\pi)^2} \Tilde{F}_\sigma(\ellvec)\, \E^{-\I\ellvec\cdot(\varthetavec-\varthetavec_i)}\left[ \cos{2\phi_\ell}\, \epsilon_1^\mathrm{s}(\varthetavec_i)+\sin{2\phi_\ell}\, \epsilon_2^\mathrm{s}(\varthetavec_i)\right]\;,
\end{equation}
where $\mathcal{N}$ is the noise contribution, $\phi_\ell$ is the polar angle of $\ellvec$ and $ \epsilon_{1/2}^\mathrm{s}(\varthetavec_i)$ are the components of the intrinsic ellipticity $\epsilon^\mathrm{s}$ of the galaxy at $\varthetavec_i$. In general, $K_\mathrm{N}$ has an imaginary component. However, aperture statistics, the quantity we are ultimately interested in,  do not depend on the imaginary parts, which is why we concentrate on the real part here.

We now introduce a filter function $U'_\theta$, defined such that
\begin{equation}
    \int \dd[2]{\vartheta} U'_\theta(|\alphavec-\varthetavec|)\, K(\varthetavec) = \int \dd[2]{\vartheta} U_\theta(|\alphavec-\varthetavec|) \, \kappa(\varthetavec)\;.
\end{equation}
This indicates that the aperture filter $U$ is a convolution of $U'$ and the smoothing kernel $F_\sigma$. Therefore, we can write Eq.~\eqref{eq: cov_from_estimator} including shape noise with $K_\mathrm{N}$ as
\begin{align}
\label{eq: Map3Map3 with shapenoise}
\notag\expval{\MapMapMapEst\, \MapMapMapEst}(\Theta_1, \Theta_2)&= \frac{1}{A^2}\int \dd[2]{\alpha_1} \int \dd[2]{\alpha_2} W_A(\alphavec_1)\,W_A(\alphavec_2)\,\Bigg[\prod_{i=1}^3 \int \dd[2]{\vartheta_i} U'_{\theta_i}(|\alphavec_1-\varthetavec_i|)\Bigg]\,\Bigg[\prod_{j=4}^6 \int \dd[2]{\vartheta_j} U'_{\theta_j}(|\alphavec_2-\varthetavec_j|)\Bigg]\\
&\quad\times\,\expval{K_\mathrm{N}(\varthetavec_1)\,K_\mathrm{N}(\varthetavec_2)\,K_\mathrm{N}(\varthetavec_3)\,K_\mathrm{N}(\varthetavec_4)\,K_\mathrm{N}(\varthetavec_5)\,K_\mathrm{N}(\varthetavec_6)}\\
&=\notag \frac{1}{A^2}\int \dd[2]{\alpha_1} \int \dd[2]{\alpha_2} W_A(\alphavec_1)\,W_A(\alphavec_2)\,\Bigg[\prod_{i=1}^3 \int \dd[2]{\vartheta_i} U'_{\theta_i}(|\alphavec_1-\varthetavec_i|)\Bigg]\,\Bigg[\prod_{j=4}^6 \int \dd[2]{\vartheta_j} U'_{\theta_j}(|\alphavec_2-\varthetavec_j|)\Bigg]\\
&\notag 
\quad\times\,\Big\lbrace\expval{K(\varthetavec_1)\,K(\varthetavec_2)\,K(\varthetavec_3)\,K(\varthetavec_4)\,K(\varthetavec_5)\,K(\varthetavec_6)}\\
&\notag \qquad + \left[\expval{\mathcal{N}(\varthetavec_1)\,\mathcal{N}(\varthetavec_2)}\expval{K(\varthetavec_3)\,K(\varthetavec_4)\,K(\varthetavec_5)\,K(\varthetavec_6)} + \textrm{14 perm.}\right]\\
&\notag \qquad + \left[\expval{\mathcal{N}(\varthetavec_1)\,\mathcal{N}(\varthetavec_2)\,\mathcal{N}(\varthetavec_3)\,\mathcal{N}(\varthetavec_4)}\,\expval{K(\varthetavec_5)\,K(\varthetavec_6)} + \textrm{14 perm.}\right]\\
&\notag \qquad + \expval{\mathcal{N}(\varthetavec_1)\,\mathcal{N}(\varthetavec_2)\,\mathcal{N}(\varthetavec_3)\,\mathcal{N}(\varthetavec_4)\,\mathcal{N}(\varthetavec_5)\,\mathcal{N}(\varthetavec_6)}\Big\rbrace\;,
\end{align}
where we used that all odd moments of the noise vanish and that $\mathcal{N}$ and $K$ are uncorrelated.

The expectation values are taken by averaging over all galaxy positions inside $A'$ and taking the ensemble average. For the second-order moment of $\mathcal{N}$, \citet{VanWaerbeke2000} showed that this leads to
\begin{equation}
    \expval{\mathcal{N}(\varthetavec)\,\mathcal{N}(\varthetavec')} = \frac{\sigma^2_\epsilon}{2n}\int \frac{\dd[2]{\ell}}{(2\pi)^2}\E^{\I\ellvec\cdot(\varthetavec-\varthetavec')}\, |\Tilde{F}_\sigma(\ellvec)|^2\;.
\end{equation}
For $K$ we find
\begin{align}
    \expval{K(\thetavec)\,K(\thetavec')} &= \left[\prod_{g=1}^N \frac{1}{A'} \int \dd[2]{\vartheta_g}\right] \sum_{i=0}^N \sum_{j=0}^N F_\sigma(|\thetavec-\varthetavec_i|)\,F_\sigma(|\thetavec'-\varthetavec_j|)\, \expval{\kappa_i\, \kappa_j}\\
    &\notag= \frac{N(N-1)}{n^2\,A'^2}\int \dd[2]{\vartheta_1}\, \int \dd[2]{\vartheta_2} F_\sigma(|\thetavec-\varthetavec_1|)\,F_\sigma(|\thetavec'-\varthetavec_2|) \expval{\kappa_1\, \kappa_2} \\
    &\notag \quad + \frac{N}{n^2\,A'}\int \dd[2]{\vartheta}\, F_\sigma(|\thetavec-\varthetavec|)\,F_\sigma(|\thetavec'-\varthetavec|) \expval{\kappa^2(\varthetavec)}\\
    &\notag \simeq \int \dd[2]{\vartheta_1}\, \int \dd[2]{\vartheta_2} F_\sigma(|\thetavec-\varthetavec_1|)\,F_\sigma(|\thetavec'-\varthetavec_2|) \expval{\kappa_1\, \kappa_2}  + \frac{\expval{\kappa^2}}{n}\,\int \dd[2]{\vartheta}\, F_\sigma(|\thetavec-\varthetavec|)\,F_\sigma(|\thetavec'-\varthetavec|)\;,
\end{align}
where we assumed that $N\gg 1$ and used that $\expval{\kappa^2(\varthetavec)}$ is the $\kappa$-correlation function at vanishing separation and thus independent of $\varthetavec$. The second summand describes the noise contribution due to the finite number of galaxies \citep{Schneider1998}. However, since $\expval{\kappa^2}$ is much smaller than $\sigma^2_\epsilon$ for realistic shape noise, this term is small compared to $ \expval{\mathcal{N}(\varthetavec)\,\mathcal{N}(\varthetavec')}$ and we neglect it in the following. With the same approximation,
\begin{align}
    \expval{K(\thetavec_1)\,K(\thetavec_2)\, \,K(\thetavec_3)\, \,K(\thetavec_4)} &\simeq \left[ \prod_{i=1}^4 \int \dd[2]{\vartheta_i}\,  F_\sigma(|\thetavec_i-\varthetavec_i|)\,\right]\expval{\kappa_1\, \kappa_2\, \kappa_3\, \kappa_4}\\
    \expval{K(\thetavec_1)\,K(\thetavec_2)\, \,K(\thetavec_3)\, \,K(\thetavec_4)\,K(\thetavec_5)\,K(\thetavec_6)} &\simeq \int \left[ \prod_{i=1}^6 \int \dd[2]{\vartheta_i}\,  F_\sigma(|\thetavec_i-\varthetavec_i|)\,\right]\,\expval{\kappa_1\, \kappa_2\, \kappa_3\, \kappa_4\,\kappa_5\, \kappa_6}\;.
\end{align}

We now assume that the width $\sigma$ of the smoothing kernel is much smaller than the aperture radii $\theta$. In that case, we can approximate the $F_\sigma$ with Dirac functions, so
\begin{equation}
    U_\theta(\vartheta) = \int \dd[2]{\vartheta'} U'_\theta(|\varthetavec-\varthetavec'|)\, F_\sigma(\varthetavec')   = \int \dd[2]{\vartheta'} U'_\theta(|\varthetavec-\varthetavec'|)\, \dirac(\varthetavec') = U'_\theta(\vartheta)\;,
\end{equation}
and
\begin{align}
\expval{\MapMapMapEst\, \MapMapMapEst}(\Theta_1, \Theta_2)
&=\notag \frac{1}{A^2}\int \dd[2]{\alpha_1} \int \dd[2]{\alpha_2} W_A(\alphavec_1)\,W_A(\alphavec_2)\,\Bigg[\prod_{i=1}^3 \int \dd[2]{\vartheta_i} U_{\theta_i}(|\alphavec_1-\varthetavec_i|)\Bigg]\,\Bigg[\prod_{j=4}^6 \int \dd[2]{\vartheta_j} U_{\theta_j}(|\alphavec_2-\varthetavec_j|)\Bigg]\\
& \quad\times\,\Bigg\lbrace\expval{\kappa_1\,\kappa_2\,\kappa_3\,\kappa_4\,\kappa_5\,\kappa_6} + \left[\frac{\sigma^2_\epsilon}{2n}\, \dirac(\varthetavec_1-\varthetavec_2)\,\expval{\kappa_3\, \kappa_4\,\kappa_5\, \kappa_6} + \textrm{14 perm.}\right]\\
&\notag \qquad + \left[ \left(\frac{\sigma^2_\epsilon}{2n}\right)^2\, \dirac(\varthetavec_1-\varthetavec_2)\,\dirac(\varthetavec_3-\varthetavec_4)\,\expval{\kappa_5\,\kappa_6} + \textrm{44 perm.}\right]\\
&\notag \qquad + \left[ \left(\frac{\sigma^2_\epsilon}{2n}\right)^3\, \dirac(\varthetavec_1-\varthetavec_2)\,\dirac(\varthetavec_3-\varthetavec_4)\,\dirac(\varthetavec_5-\varthetavec_6) + \textrm{14 perm.}\right]\Bigg\rbrace\;,
\end{align}
By decomposing the six- and four-point function into its connected components and considering all permutations, one can show that this is equal to the right-hand side of Eq.~\eqref{eq: cov_from_estimator_decomposed} after replacing $\expval{\kappa_i\, \kappa_j}$ by $\expval{\kappa_i\, \kappa_j}+\sigma_\epsilon^2\, \dirac(\varthetavec_i-\varthetavec_j)/2n$. This implies a power spectrum $P'$ given by
\begin{align}
    P'(\ell)\, (2\pi)^2\dirac(\ellvec+\ellvec') &= \int \dd[2]{\vartheta_1}\, \int \dd[2]{\vartheta_2} \left[ \expval{\kappa_1\, \kappa_2} + \frac{\sigma_\epsilon^2}{2n}\,\dirac(\varthetavec_1-\varthetavec_2) \right]\, \E^{\I(\ellvec\cdot\varthetavec_1 + \ellvec'\cdot\varthetavec_2)}\\
    &\notag = \left[ P(\ell) + \frac{\sigma_\epsilon^2}{2n}\right]\, (2\pi)^2\dirac(\ellvec+\ellvec')\;,
\end{align}
where $P$ is the power spectrum without shape noise. Consequently, replacing $P(\ell)$ by $P(\ell)+\sigma_\epsilon^2/2n$ in the covariance expressions from Sect.~\ref{sec: Derivation} gives the correct covariance for $\MapMapMap$ in the presence of shape noise.

\section{\texorpdfstring{Approximation of $T_{P_6}$}{Approximation of TP6}}
\label{app: TVII approximation}

We here derive an approximation for $T_{P_6}$, which is necessary to reduce the computational complexity of the covariance model. For this, we rewrite Eq.~\eqref{eq: T7 final} as
\begin{align}
	\TVII(\Theta_1, \Theta_2)
&=  \int\frac{\dd[2]{\ell_1}}{(2\pi)^2}\, \int\frac{\dd[2]{\ell_2}}{(2\pi)^2}  \int\frac{\dd[2]{\ell_3}}{(2\pi)^2}  \int\frac{\dd[2]{\ell_4}}{(2\pi)^2} \int\frac{\dd[2]{s}}{(2\pi)^2}\;P_6(\ellvec_1, \ellvec_2, \svec-\ellvec_1-\ellvec_2, \ellvec_3, \ellvec_4)\,\\
&\notag\quad\times  \tilde{u}(\ell_1\,\theta_1)\,\tilde{u}(\ell_2\,\theta_2)\,\tilde{u}(|\svec-\ellvec_1-\ellvec_2|\,\theta_3)\,\tilde{u}(\ell_3\,\theta_4)\,\tilde{u}(\ell_4\,\theta_5)\,\tilde{u}(|\svec+\ellvec_3+\ellvec_4|\,\theta_6)\, G_A(\svec)\;.
\end{align}
According to the Limber approximation (Eq.~\ref{eq: limber}),
\begin{align}
    P_6(\ellvec_1, \ellvec_2, \svec-\ellvec_1-\ellvec_2, \ellvec_3, \ellvec_4)&=\left(\frac{3H_0^2\Omm}{2c^2}\right)^6\,\int_0^\infty \dd{\chi}\; \frac{q^6(\chi)}{\chi^{5}\, a^6(\chi)}\, \\
    &\notag\quad\times \mathcal{P}_6^\mathrm{(3d)}\left[\ellvec_1/\chi, \ellvec_2/\chi, (\vec{s}-\ellvec_1-\ellvec_2)/\chi, \ellvec_3, \ellvec_4, (-\vec{s}-\ellvec_3-\ellvec_4)/\chi; \chi\right]\;.
\end{align}
We now approximate $\mathcal{P}_6^\mathrm{(3d)}$ under the assumption that $s \ll \ell_1, \ell_2, \ell_3, \ell_4$. This approximation is valid if $G_A$ varies on larger scales (smaller $s$) than $P_6$. Then,  as shown by \citet{Chan2018},
\begin{align}
\label{eq: P6 decomposition}
    \mathcal{P}_6^\mathrm{(3d)}(\vec{k}_1, \vec{k}_2, \vec{q}-\vec{k}_1-\vec{k}_2, \vec{k}_3, \vec{k}_4, -\vec{q}-\vec{k}_3-\vec{k}_4; \chi) = 
    \mathcal{P}_6^\mathrm{(3d)}(\vec{k}_1, \vec{k}_2, -\vec{k}_1-\vec{k}_2, \vec{k}_3, \vec{k}_4, -\vec{k}_3-\vec{k}_4; \chi)  \textrm{ + additional terms}\;,
\end{align}
where the additional terms can be calculated from the halo model. We approximate the first part of Eq.~\eqref{eq: P6 decomposition} with the 1-halo term, which is $I_6^0({k}_1, {k}_2, {k}_3, {k}_4, k_6, \chi)$, with \begin{align}
\label{eq: halomodel Is}
    I_n^i(k_1, \dots, k_n; \chi) &= \int \dd{m} n[m, z(\chi)]\, b^i[m, z(\chi)]\, \left(\frac{m}{\bar{\rho}}\right)^n\, \tilde{u}_\mathrm{NFW}(\vec{k}_1, m)\dots \tilde{u}_\mathrm{NFW}(\vec{k}_n, m)\;, 
\end{align}
with comoving mean density $\bar{\rho}$, halo mass function $n(m,z)$ and halo bias $b(m,z)$. The $\tilde{u}_\mathrm{NFW}$ is defined as Fourier-transformation of the normalised NFW-halo profile $u_\mathrm{NFW}$, which is
\begin{equation}
    u_\mathrm{NFW}(r, m)= \frac{1}{m} \rho(r, m)\;,
\end{equation}
where $\rho$ is the halo density profile.

\citet{Pyne2021} showed that the dominant additional term in Eq.~\eqref{eq: P6 decomposition} for $k>0.3\,h\,\mathrm{Mpc}^{-1}$ is given by a part of the 2-halo term of the pentaspectrum, so that
\begin{align}
    \mathcal{P}_6^\mathrm{(3d)}(\vec{k}_1, \vec{k}_2, \vec{q}-\vec{k}_1-\vec{k}_2, \vec{k}_3, \vec{k}_4, -\vec{q}-\vec{k}_3-\vec{k}_4;\chi)
    &\simeq I_6^0({k}_1, {k}_2, |\vec{q}-\vec{k}_1-\vec{k}_2|, {k}_3, {k}_4, |-\vec{q}-\vec{k}_3-\vec{k}_4|;\chi)\\
    &\notag \quad + P_\mathrm{L}(q)\, I_3^1(k_1, k_2, |\qvec-\vec{k}_1-\vec{k}_2|;\chi)\, I_3^1(k_3, k_4, |-\qvec-\vec{k}_3-\vec{k}_4|;\chi)\;, 
\end{align}
where $P_\mathrm{L}$ is the linear matter power spectrum.
With this pentaspectrum, the last covariance part becomes
\begin{align}
\label{eq:SSC_term}
\TVII(\Theta_1, \Theta_2)
&\notag= \left(\frac{3H_0^2\Omm}{2c^2}\right)^6\,\int_0^\infty \dd{\chi}\; \frac{q^6(\chi)}{\chi^{4}\, a^6(\chi)} \int\frac{\dd[2]{\ell_1}}{(2\pi)^2}\, \int\frac{\dd[2]{\ell_2}}{(2\pi)^2}  \int\frac{\dd[2]{s}}{(2\pi)^2}  \int\frac{\dd[2]{\ell_3}}{(2\pi)^2} \int\frac{\dd[2]{\ell_4}}{(2\pi)^2} \,\;\,\\
&\notag\quad\times  \tilde{u}(\ell_1\,\theta_1)\,\tilde{u}(\ell_2\,\theta_2)\,\tilde{u}(|\svec-\ellvec_1-\ellvec_2|\,\theta_3)\,\tilde{u}(\ell_3\,\theta_4)\,\tilde{u}(\ell_4\,\theta_5)\,\tilde{u}(|\svec+\ellvec_3+\ellvec_4|\,\theta_6)\, G_A(\svec)\\
&\notag \quad \times \Big[ I_6^0(\ell_1/\chi, \ell_2/\chi, |\svec-\ellvec_1-\ellvec_2|/\chi, \ell_3/\chi, \ell_4/\chi, |-\svec-\ellvec_3-\ellvec_4|/\chi; \chi)\\
&\notag \qquad + P_\mathrm{L}(s/\chi)\, I_3^1(\ell_1/\chi, \ell_2/\chi, |\svec-\ellvec_1-\ellvec_2|/\chi; \chi)\, I_3^1(\ell_3/\chi, \ell_4/\chi, |-\svec-\ellvec_3-\ellvec_4|/\chi; \chi)\Big]\\
&\simeq \left(\frac{3H_0^2\Omm}{2c^2}\right)^6\,\int_0^\infty \dd{\chi}\; \frac{q^6(\chi)}{\chi^{4}\, a^6(\chi)} \int\frac{\dd[2]{\ell_1}}{(2\pi)^2}\, \int\frac{\dd[2]{\ell_2}}{(2\pi)^2}  \int\frac{\dd[2]{s}}{(2\pi)^2}  \int\frac{\dd[2]{\ell_3}}{(2\pi)^2} \int\frac{\dd[2]{\ell_4}}{(2\pi)^2} \\
&\notag\quad\times  \tilde{u}(\ell_1\,\theta_1)\,\tilde{u}(\ell_2\,\theta_2)\,\tilde{u}(|\ellvec_1+\ellvec_2|\,\theta_3)\,\tilde{u}(\ell_3\,\theta_4)\,\tilde{u}(\ell_4\,\theta_5)\,\tilde{u}(|\ellvec_3+\ellvec_4|\,\theta_6)\, G_A(\svec)\\
&\notag \quad \times \Big[ I_6^0(\ell_1/\chi, \ell_2/\chi, |\ellvec_1+\ellvec_2|/\chi, \ell_3/\chi, \ell_4/\chi, |\ellvec_3+\ellvec_4|/\chi; \chi)\\
&\notag \qquad + P_\mathrm{L}(s/\chi)\, I_3^1(\ell_1/\chi, \ell_2/\chi, |\ellvec_1+\ellvec_2|/\chi; \chi)\, I_3^1(\ell_3/\chi, \ell_4/\chi, |\ellvec_3+\ellvec_4|/\chi; \chi)\Big]\\
&\notag= \TVII^\infty(\Theta_1, \Theta_2) + \left(\frac{3H_0^2\Omm}{2c^2}\right)^6\,\int_0^\infty \dd{\chi}\; \frac{q^6(\chi)}{\chi^{4}\, a^6(\chi)} \left[\int\frac{\dd[2]{s}}{(2\pi)^2} G_A(\svec)\, P_\mathrm{L}(s/\chi) \right]\\
&\notag \quad \times \int\frac{\dd[2]{\ell_1}}{(2\pi)^2}\, \int\frac{\dd[2]{\ell_2}}{(2\pi)^2}   \int\frac{\dd[2]{\ell_3}}{(2\pi)^2} \int\frac{\dd[2]{\ell_4}}{(2\pi)^2} I_3^1(\ell_1/\chi, \ell_2/\chi, |\vec{\ell}_1+\vec{\ell}_2|/\chi; \chi)\, I_3^1(\ell_3/\chi, \ell_4/\chi, |\vec{\ell}_3+\vec{\ell}_4|/\chi; \chi)\\
&\notag \quad \times  \tilde{u}(\ell_1\,\theta_1)\,\tilde{u}(\ell_2\,\theta_2)\,\tilde{u}(|\ellvec_1+\ellvec_2|\,\theta_3)\,\tilde{u}(\ell_3\,\theta_4)\,\tilde{u}(\ell_4\,\theta_5)\,\tilde{u}(|\ellvec_3+\ellvec_4|\,\theta_6)\;,
\end{align}
where we used $s \ll \ell_1, \ell_2, \ell_3, \ell_4$. We refer to the last summand in this equation as $T_{P_6, \mathrm{2h}}$. Similar to $\TII$ and $\TVI$, $T_{P_6, \mathrm{2h}}$ vanishes under the large-field-approximation. This can be seen by replacing $G_A(\svec)$ with $2\pi\dirac(\svec)/A$. In this approximation, the integral in square brackets becomes zero, since $P_\mathrm{L}(0)=0$. Consequently, all of $T_{P_6, \mathrm{2h}}$ vanishes.

\section{\texorpdfstring{Derivation of Gaussian $\MapMapMap$ covariance from bispectrum covariance}{Derivation of Gaussian covariance from bispectrum covariance}}
\label{app: bispec cov}
In this appendix, we try to use the Gaussian bispectrum covariance derived by \citet{Joachimi2009} to obtain $C_{\MapMapMap}$. We will show that this approach only recovers the large-field approximation $\TI^\infty$. In this appendix (and only here), we parameterise the bispectrum by the lengths of three $\ellvec$, namely using
\begin{equation}
   \expval{\hat{\kappa}(\ellvec_1)\,\hat{\kappa}(\ellvec_2)\,\hat{\kappa}(\ellvec_3)}  = (2\pi)^2\, B(\ell_1, \ell_2, \ell_3)\, \dirac(\ellvec_1+\ellvec_2+\ellvec_3)\;.
\end{equation}

The bispectrum covariance has been derived for Gaussian fields by \citet[][see also references therein]{Joachimi2009}. It is
	\begin{align}
		\label{eq: Bispec_Covariance_Gaussian}
		\mathrm{C}_{B}(\ell_1, \ell_2, \ell_3; \ell_4, \ell_5, \ell_6) 
		&=\frac{(2\pi)^3 \Lambda^{-1}(\ell_1, \ell_2, \ell_3)}{A\, \,A_R(\ell_1)\, A_R(\ell_2)\, A_R(\ell_3)}\, \left[\kronecker{1}{4}\,\kronecker{2}{5}\,\kronecker{3}{6} + \text{5 permutations}\right]\, {P}(\ell_1)\, {P}(\ell_2)\, {P}(\ell_3)\;,
	\end{align}
	where $A_R(\ell)$ is the size of the bin of $\ellvec$, defined as	$A_R(\ell)=2\pi\, \ell\, \Delta \ell$, the $\kronecker{i}{j}$ denote Kronecker-deltas 
	and $\Lambda$ is defined by
		\begin{align}
		\label{eq: Definition Lambda}
\left[\prod_{i=1}^3\,\int \frac{\dd[2]{\ell_i}}{(2\pi)^2} \right]\, \dirac(\ellvec_1+\ellvec_2+\ellvec_3) 
		= \int_0^\infty \frac{\dd{\ell_1}}{(2\pi)^2} \int_0^\infty \frac{\dd{\ell_2}}{(2\pi)^2} \int_0^\infty \frac{\dd{\ell_3}}{(2\pi)^2} \; \ell_1\, \ell_2\, \ell_3\,  2\pi\, \Lambda(\ell_1, \ell_2, \ell_3)\;.   
	\end{align}
We note that Eq.~\eqref{eq: Bispec_Covariance_Gaussian} was derived under the assumption that the Fourier transform $\tilde{\kappa}(\ellvec)$ of the convergence field is known. However, this assumption cannot be fulfilled if $\tilde{\kappa}$ is derived from the $\kappa$ on only a finite survey area, as the Fourier transform formally requires $\kappa$ on all of $\realspace^2$. Nevertheless, we try to derive the covariance of $\MapMapMapEst$ from $C_B$. Using Eq.~\eqref{eq: Map3 from Bispec} and Eq.~\eqref{eq: Definition Lambda}, the aperture statistics are
	\begin{equation}
		\label{eq: Map3 in radial bins}
		\MapMapMap(\theta_1, \theta_2, \theta_3) = \left[\prod_{i=1}^3\,\int {\frac{\dd{\ell_i}}{2\pi}}\, \ell_i\, \tilde{u}(\ell_i\, \theta_i) \right]\, B(\ell_1, \ell_2, \ell_3) \, \Lambda(\ell_1,\ell_2,\ell_3)\;.
	\end{equation}
	We go from the continuous integration to a discrete sum in the $\ell_i$, so
	\begin{equation}
		\label{eq: Map3 discrete}
		\MapMapMap(\theta_1, \theta_2, \theta_3) = \frac{1}{(2\pi)^3} \sum_{ijk} \Delta \ell_i\, \Delta \ell_j\, \Delta \ell_k\, \ell_i\, \ell_j\, \ell_k\,  \tilde{u}(\ell_i\, \theta_1)\, \tilde{u}(\ell_j\, \theta_2)\, \tilde{u}(\ell_k\, \theta_3)\, B(\ell_i, \ell_j, \ell_k) \, (2\pi)^3 \Lambda(\ell_i,\ell_j,\ell_k)\;.
	\end{equation}
	Here, the $\Delta \ell_i$ are the bin sizes along the $\ell_i$. 
	Using Eq.~\eqref{eq: Map3 discrete}, we derive $C_{\MapMapMapEst}$ from $C_B$ with
	\begin{align}
		\label{eq: Map3 Cov General}
		C_{\MapMapMapEst}(\Theta_1, \Theta_2) &=  \frac{1}{(2\pi)^{6}}\sum_{ijk} \sum_{lmn} \Delta \ell_i\, \Delta \ell_j\, \Delta \ell_k\,  \Delta \ell_l\, \Delta \ell_m\, \Delta \ell_n\, \ell_i\, \ell_j\, \ell_k\, \ell_l\, \ell_m\, \ell_n\,  \tilde{u}(\ell_i\, \theta_1)\, \tilde{u}(\ell_j\, \theta_2)\, \tilde{u}(\ell_k\, \theta_3)\\
		&\notag \quad \times  \tilde{u}(\ell_l\, \theta_4)\, \tilde{u}(\ell_m\, \theta_5)\, \tilde{u}(\ell_n\, \theta_6)\, C_B(\ell_i, \ell_j, \ell_k; \ell_l, \ell_m, \ell_n) \, \Lambda(\ell_i,\ell_j,\ell_k)\, \Lambda(\ell_l,\ell_m,\ell_n)\;.
	\end{align}	
	So the Gaussian covariance of the aperture statistics is
	\begin{align}
		\label{eq: Map3 Cov Gauss}
		C_{\MapMapMapEst}(\Theta_1, \Theta_2) 
		&= \frac{1}{(2\pi)^{6}} \sum_{ijk} \sum_{lmn} \Delta \ell_i\, \Delta \ell_j\, \Delta \ell_k\,  \Delta \ell_l\, \Delta \ell_m\, \Delta \ell_n\, \ell_i\, \ell_j\, \ell_k\, \ell_l\, \ell_m\, \ell_n\,  \tilde{u}(\ell_i\, \theta_1)\, \tilde{u}(\ell_j\, \theta_2)\, \tilde{u}(\ell_k\, \theta_3)\\
		&\notag \quad \times  \tilde{u}(\ell_l\, \theta_4)\, \tilde{u}(\ell_m\, \theta_5)\, \tilde{u}(\ell_n\, \theta_6)\, \frac{(2\pi)^3 \Lambda^{-1}(\ell_i, \ell_j, \ell_k)}{A\,\,\Delta \ell_i\, \Delta \ell_j\, \Delta \ell_k\, \ell_i\, \ell_j\, \ell_k}\, \left[\kronecker{i}{l}\,\kronecker{j}{m}\,\kronecker{k}{n} + \text{5 Perm.}\right]\,\\
		&\notag \quad \times P(\ell_i)\, P(\ell_j)\, P(\ell_k) \, \Lambda(\ell_i,\ell_j,\ell_k)\, \Lambda(\ell_l,\ell_m,\ell_n)\\\
		&\notag=  \frac{1 }{(2\pi)^3\,A\, }\, \sum_{ijk} \sum_{lmn} \Delta \ell_l\, \Delta \ell_m\, \Delta \ell_n\,\ell_l\, \ell_m\, \ell_n\, \tilde{u}(\ell_i\,\theta_1)\,\tilde{u}(\ell_j\, \theta_2)\, \tilde{u}(\ell_k\, \theta_3)\, \tilde{u}(\ell_l\, \theta_4)\, \tilde{u}(\ell_m\, \theta_5)\,\tilde{u}(\ell_n\,\theta_6)\\
		&\notag \quad \times \left[\kronecker{i}{l}\,\kronecker{j}{m}\,\kronecker{k}{n} + \text{5 Perm.}\right]\, P(\ell_i)\, P(\ell_j)\, P(\ell_k) \,  \Lambda(\ell_l,\ell_m,\ell_n)\;.
	\end{align}	
We evaluate the sums over the $l$, $m$, and $n$ using the Kronecker-Deltas, so
	\begin{align}
		C_{\MapMapMapEst}(\Theta_1, \Theta_2) 
		&=  \frac{1}{(2\pi)^3\, A\, }\, \sum_{ijk} \Delta \ell_i\, \Delta \ell_j\, \Delta \ell_k\,\ell_i\, \ell_j\, \ell_k\,\tilde{u}(\ell_i\,\theta_1)\, \tilde{u}(\ell_j\, \theta_2)\, \tilde{u}(\ell_k\, \theta_3)\\
		&\notag \quad \times \left[  \tilde{u}(\ell_l\, \theta_4)\, \tilde{u}(\ell_m\, \theta_5)\, \tilde{u}(\ell_n\,\theta_6) + \text{5 Perm.}\right]\, P(\ell_i)\, P(\ell_j)\, P(\ell_k) \,  \Lambda(\ell_i,\ell_j,\ell_k)\;.
	\end{align}		
Finally, we go from the discrete sum back to continuous integrals (assuming $\Delta \ell_i \rightarrow 0$), so
	\begin{align}
		C_{\MapMapMapEst}(\Theta_1, \Theta_2) 
		&= \frac{1}{A}\left[\prod_{i=1}^3\,\int_0^\infty \frac{\dd{\ell_i}}{(2\pi)^2}\,\ell_i\, \tilde{u}(\ell_i\,\theta_i) \right]\, P(\ell_1)\, P(\ell_2)\, P(\ell_3) \, (2\pi)^3\, \Lambda(\ell_1,\ell_2,\ell_3)\,\left[  \tilde{u}(\ell_1\, \theta_4)\, \tilde{u}(\ell_2\, \theta_5)\, \tilde{u}(\ell_3\, \theta_6) + \text{5 Perm.}\right]\,\;.
	\end{align}	
This can also be written as
\begin{align}
\label{eq: FinalCov_general}
	C_{\MapMapMapEst}(\Theta_1, \Theta_2) 
	&=  \frac{1}{A}\left[\prod_{i=1}^3\,\int \frac{\dd[2]{\ell_i}}{(2\pi)^2}\, \tilde{u}(\ell_i\,\theta_i) \right]\, P(\ell_1)\, P(\ell_2)\, P(\ell_3) \,(2\pi)^2  \dirac(\ellvec_1+\ellvec_2+\ellvec_3)\, \left[  \tilde{u}(\ell_1\, \theta_4)\, \tilde{u}(\ell_2\, \theta_5)\, \tilde{u}(\ell_3\,\theta_6) + \text{5 Perm.}\right]\,\\
	&\notag=  \frac{1 }{A}\, \int \frac{\dd[2]{\ell_1}}{(2\pi)^2} \int \frac{\dd[2]{\ell_2}}{(2\pi)^2} \tilde{u}(\ell_1\,\theta_1)\,\tilde{u}(\ell_2\, \theta_2)\, \tilde{u}(|\ellvec_1+\ellvec_2|\, \theta_3)\,P(\ell_1)\, P(\ell_2)\, P(|\ellvec_1+\ellvec_2|)\\
		&\notag \quad \times \left[  \tilde{u}(\ell_1\, \theta_4)\, \tilde{u}(\ell_2\, \theta_5)\, \tilde{u}(|\ellvec_1+\ellvec_2|\, \theta_6) + \text{5 Perm.}\right]\,  \;.
\end{align}
By comparing Eq.~\eqref{eq: FinalCov_general} with Eq.~\eqref{eq: TI for infinite fields}, we see that this expression corresponds to $\TI^\infty$, which is the Gaussian covariance of $\MapMapMapEst$ in the limiting case of large survey areas $A$. The term $\TII$ is not recovered from the bispectrum covariance. This is a direct consequence of the assumption that $\tilde{\kappa}(\ellvec)$ can be reconstructed from the $\kappa$ on the finite survey window $A$ in the derivation of Eq.~\eqref{eq: Bispec_Covariance_Gaussian}. This assumption is equivalent to assuming a window function $W_A$, which is one on the whole $\realspace^2$, leading to the large-field approximation for $G_A$ in Eq.~\eqref{eq: GA for infinite survey}. This approximation directly reduces $\TI$ to $\TI^\infty$ and $\TII$ to zero.

\end{appendix}

\end{document}